\documentclass[twocolumn]{aastex631}
\usepackage{natbib}
\usepackage{url}
\usepackage{colortbl}
\usepackage{float}
\usepackage{multirow}
\usepackage{graphicx}
\usepackage{lipsum}
\usepackage{subfigure}
\usepackage{amsmath}

\DeclareUnicodeCharacter{2212}{-}

\newcommand{\msun}{\mbox{$\,{\rm M}_\odot$}}
\newcommand{\lsun}{\mbox{$\,{\rm L}_\odot$}}

\newcommand{\oversim}[2]{\protect{\mbox{\lower0.5ex\vbox{%
  \baselineskip=0pt\lineskip=0.2ex
  \ialign{$\mathsurround=0pt #1\hfil##\hfil$\crcr#2\crcr\sim\crcr}}}}}
\DeclareRobustCommand{\VAN}[3]{#2}
\let\VANthebibliography\thebibliography
\def\thebibliography{\DeclareRobustCommand{\VAN}[3]{##3}\VANthebibliography}

\begin{document}

%\title{Long-period variable stars in NGC 6822 – II. Their dust production}
\title{SEDust: A DUSTY-Based SED-Fitting Pipeline for Dust and Mass-Loss from AGB and LPV Stars}

%\correspondingauthor{Hedieh Abdollahi}
\email{mahani@ipm.ir}
\email{atefeh@ipm.ir}

\author[0009-0002-9957-5818]{Hamidreza Mahani}
\affiliation{School of Astronomy, Institute for Research in Fundamental Sciences (IPM), P.O. Box 1956836613, Tehran, Iran}
\affiliation{Iranian National Observatory (INO), Institute for Research in Fundamental Sciences (IPM), P.O. Box 1956836613, Tehran, Iran}

\author[0009-0006-5762-8327]{Narges Rostami}
\affiliation{Department of Physics, Faculty of Physics, Alzahra University, Vanak, 1993891176, Tehran, Iran}

\author[0000-0001-8392-6754]{Atefeh Javadi}
\affiliation{School of Astronomy, Institute for Research in Fundamental Sciences (IPM), P.O. Box 1956836613, Tehran, Iran}
\affiliation{Iranian National Observatory (INO), Institute for Research in Fundamental Sciences (IPM), P.O. Box 1956836613, Tehran, Iran}

 \author[0000-0002-7823-7169] {Hedieh Abdollahi}
 \affiliation{School of Astronomy, Institute for Research in Fundamental Sciences (IPM), P.O. Box 1956836613, Tehran, Iran}
 \affiliation{Konkoly Observatory, HUN-REN Research Centre for Astronomy and Earth Sciences, MTA Centre of Excellence, Konkoly Thege M. {\'u}t 15-17, Budapest, H–1121, Hungary}

\author[0000-0002-1272-3017]{Jacco Th. van Loon}
\affiliation{Lennard-Jones Laboratories, Keele University, ST5 5BG, UK}

 \author[0000-0003-2743-8240]{Francisca Kemper}
 \affiliation{Institut de Ci{\'e}ncies de l’Espai (ICE, CSIC), Can Magrans, s/n, E-08193 Cerd anyola del Vall{\'e}s, Barcelona, Spain}
 \affiliation{ICREA, Pg. Llu{\'i}s Companys 23, E-08010 Barcelona, Spain}
 \affiliation{Institut d'Estudis Espacials de Catalunya (IEEC), E-08860 Castelldefels Barcelona, Spain}

 \author[0000-0002-3078-9482]{{\'A}lvaro S{\'a}nchez-Monge}
 \affiliation{Institut de Ci{\'e}ncies de l’Espai (ICE, CSIC), Can Magrans, s/n, E-08193 Cerd anyola del Vall{\'e}s, Barcelona, Spain}
 \affiliation{Institut d'Estudis Espacials de Catalunya (IEEC), E-08860 Castelldefels Barcelona, Spain}

%@@@@@@@@@@@@@@@@@@@@@@@@@@@@@@@@@@@@@@@@@@@@@@@@@@@@@@@@@@@@@@@@@@@@@@@@@@@@@@@@@@@@@@@@@@

\begin{abstract}

We present \texttt{SEDust}, an automated pipeline for spectral energy distribution (SED) fitting and the estimation of stellar mass-loss and dust-production rates. We apply this pipeline to quantify the mass-loss and dust injection rates in the dwarf galaxy NGC~6822, focusing on long-period variables (LPVs) and pulsating asymptotic giant branch (AGB) stars as major contributors to dust feedback into the interstellar medium. The total stellar mass-loss rate is found to be $(3.70 \pm 1.16) \times 10^{-3}\ \msun \mathrm{yr}^{-1}$. Two alternative methods (an LPV-based estimate and \texttt{DESK}) give totals that differ from this value by factors of $\sim$2.3 (lower) and $\sim$2.4 (higher), respectively, which we adopt as an estimate of the method-dependent systematic uncertainty. A positive correlation between stellar luminosity, intrinsic reddening caused by circumstellar dust self-extinction, and mass-loss rate is observed. The specific mass-return rate from AGB stars is $(7.40 \pm 2.30) \times 10^{-11}\ \mathrm{yr}^{-1}$, indicating that mass return from AGB stars alone is insufficient to sustain the current star formation rate of NGC~6822. This result implies the need for external gas accretion or additional mass-return mechanisms.

\end{abstract}


%@@@@@@@@@@@@@@@@@@@@@@@@@@@@@@@@@@@@@@@@@@@@@@@@@@@@@@@@@@@@@@@@@@@@@@@@@@@@@@@@@@@@@@@@@@

\section{Introduction} 

%\textcolor{red}{
%\begin{itemize}
%\item , in dimensions
%\item Errors of values
%\item Caption of plots 
%\item $-$ in colors 
%\end{itemize}
%}

The mass-loss processes of evolved stars play a fundamental role in regulating the chemical enrichment and evolution of galaxies \citep{Jura89,Jura90,Matsuura09,Boyer11}. During their late evolutionary stages, asymptotic giant branch (AGB) stars and red supergiants (RSGs) lose substantial fractions of their mass through dust-driven stellar winds, returning gas and heavy elements to the interstellar medium (ISM) \citep{vanLoon25}. Among evolved stars, AGB stars are major contributors to the dust and gas budget of intermediate-age stellar populations. This mass return shapes subsequent star formation, drives dust evolution, and contributes to the overall baryonic cycle within galaxies \citep{Habing13,Hofner18}. Therefore, quantifying the dust and gas injection rates from evolved stars is essential for understanding the mechanisms governing galactic evolution across a wide range of environments, including low-metallicity galaxies \citep{Srinivasan09,Javadi13_III,Abdollahi23}.

Dwarf galaxies provide key laboratories for such studies, as their simple structures, low masses, and diverse star-formation histories enable investigations of stellar feedback across a wide range of environmental conditions \citep{Mahani25}. At the same time, because dwarf galaxies contain relatively small stellar populations, their total dust and gas return may be dominated by only a few highly evolved stars undergoing brief episodes of intense mass-loss. Consequently, these phases may not always be well sampled in observational studies \citep{Jones15}. In particular, dwarf irregular galaxies are of special interest because they are often regarded as transitional systems bridging gas-rich irregulars and gas-poor dwarf spheroidals \citep{Mateo98,Tolstoy09}. As such, they offer valuable insights into the interplay among stellar mass-loss processes, interstellar medium evolution, and star formation \citep{Javadi13_III,Boyer17,Mahani23,Gholami25,Abdollahi26}.

NGC 6822 (Barnard’s Galaxy) is an isolated member of the Local Group located at a distance of approximately 490 kpc, corresponding to the TRGB distance modulus of $(m-M)_0=23.46\pm0.20$~mag reported by \citet{lee1993tip,Mateo98}. Propagating the quoted uncertainty in the distance modulus gives $D=490 \pm 47$~kpc.

 Classified as a barred irregular galaxy, it exhibits a complex structure consisting of an optically bright bar, an extended HI envelope, and a spheroidal halo populated by older stars \citep{hodge1991cosmos,de2000evidence}. With a stellar mass of roughly 5 $\times 10^{7} \msun$ and metallicity ranging from [Fe/H] $\sim$ $-$1.0 to $-$0.7 dex \citep{Gallart96_III,davidge2003metallicity}, NGC 6822 serves as a nearby, low-metallicity proxy of star-forming dwarf systems in the early universe. Its active and prolonged star-formation history \citep{gallart1996local,wyder2001star, Kinson21} makes it a particularly suitable target for examining the ongoing enrichment of the ISM through stellar mass-loss.

In this work, we investigate the mass-loss and dust-production rates of evolved stars in NGC~6822. To achieve this, we developed \texttt{SEDust}\footnote{The \texttt{SEDust} pipeline is publicly available at \url{https://github.com/hmahani/SEDust} and archived at Zenodo with DOI \url{https://doi.org/10.5281/zenodo.22858441}.}, an automated pipeline designed to derive best-fit spectral energy distributions (SEDs) from radiative-transfer models and to estimate stellar mass-loss and dust-production rates in a uniform and reproducible manner. The pipeline combines multi-wavelength photometry with precomputed \texttt{DUSTY} model grids \citep{Ivezic97}, allowing reliable characterization of circumstellar dust properties for large stellar samples. We apply \texttt{SEDust} to long-period variables (LPVs) and AGB stars in NGC~6822, to systematically assess their contribution to the dust and gas budget of a low-metallicity dwarf galaxy.

%This paper is organized as follows: Section~\ref{sec:data} describes the observational data, Section~\ref{sec:model} outlines the \texttt{SEDust} methodology, Section~\ref{sec:result} presents the results, Section~\ref{sec:Discussion} discusses their implications for the dust budget of NGC~6822, , and Section~\ref{sec:Conclusions} summarizes the main conclusions.

%@@@@@@@@@@@@@@@@@@@@@@@@@@@@@@@@@@@@@@@@@@@@@@@@@@@@@@@@@@@@@@@@@@@@@@@@@@@@@@@@@@@@@@@@@@

\section{Data}
\label{sec:data}

\subsection{Near- and Mid-Infrared Photometry}

\subsubsection{Asymptotic Giant Branches}

Our analysis is primarily based on the catalog presented by \citet{Hirschauer20}, which provides one of the most comprehensive multiwavelength datasets of dusty and evolved stellar populations in NGC 6822. This catalog combines near- and mid-infrared photometry obtained with the United Kingdom Infrared Telescope (UKIRT; $JHK$ bands \citep{Hewett06}) and the \textit{Spitzer} Space Telescope (IRAC 3.6, 4.5, 5.8, and 8.0~$\mu$m, and MIPS 24~$\mu$m \citep{Fazio04, Rieke04}). The typical photometric uncertainties range from 0.02–0.30~mag at 3.6 and 4.5~$\mu$m, 0.01–0.30~mag at 5.8, 8.0, and 24~$\mu$m, and 0.04–0.85, 0.03–0.94, and 0.03–0.84~mag in the $J$, $H$, and $K$ bands, respectively. The combined dataset offers excellent sensitivity over a broad wavelength range, enabling accurate identification and characterization of evolved and dust-producing stars in this metal-poor galaxy. The catalog includes 560 carbon-rich (C-type) and 1050 oxygen-rich (M-type) AGB stars with reliable classifications.

The photometry used in our analysis was corrected for Galactic foreground extinction using the extinction map of \citet{Schlegel98}, following the procedure described in \citet{Hirschauer20}. No correction was applied for internal reddening within NGC 6822, given its significant spatial variation across the galaxy.

The completeness of the \citet{Hirschauer20} catalog is non-uniform, as it is based on the $1''$ cross-match of 210979 UKIRT and 30745 \textit{Spitzer} sources, yielding 14534 sources with both near- and mid-IR photometry. Source crowding, the coarser MIPS $24~\mu$m resolution, and increasing photometric uncertainties toward the faint end further limit source recovery. Moreover, the larger distance of NGC~6822 makes these selection effects more severe for faint and heavily obscured evolved stars. %The derived integrated mass-loss and dust-production rates should therefore be regarded as likely lower estimates of the total AGB contribution.

In addition to the AGB candidates, \citet{Hirschauer20} identify 1292 red supergiant (RSG) candidates in NGC~6822 via the intersection of three color--magnitude diagrams, with an estimated contamination rate of $52.93\%$. Since the present study focuses on AGB and LPV stars, RSGs are not included in the analysis that follows. Their potential contribution to the mass and dust budget of NGC~6822 is discussed in Sect.~\ref{Sec:RSG}.

\subsubsection{Long-Period Variables}

The study by \citet{Whitelock13} presented a multi-epoch near-infrared monitoring survey of the central regions of NGC~6822 with the aim of identifying and characterizing large-amplitude AGB variables. The survey spanned 3.5 years and covered three overlapping fields centered on the optical bar of NGC~6822. The fields were observed over 16, 18, and 19 epochs in the $K_S$, $H$, and $J$ bands, respectively, providing sufficient temporal coverage to determine pulsation periods. Observations were obtained with the SIRIUS camera on the Japanese--South African Infrared Survey Facility (IRSF), which provides simultaneous imaging in all three filters \citep{Nagashima99}.

The published catalog reports photometric uncertainties smaller than 0.1~mag in all bands. However, bolometric magnitudes are more uncertain for stars with very faint $J$-band measurements, particularly for $J > 20.0$ mag, due to increased photometric errors and source crowding in the central regions of the galaxy.

From this catalog, we select a subsample of large-amplitude stars listed in Table 4 of \citet{Whitelock13} with pulsation periods longer than 100~days, treating such sources as LPVs as commonly adopted in studies such as \citet{Javadi13_III, Gholami25}. These large-amplitude, long-period pulsators correspond to Mira-type variables, representing the most evolved and vigorously pulsating stars along the AGB. Cross-matching these sources with the AGB catalog of \citet{Hirschauer20} yielded a final sample of 59 LPVs, comprising 50 carbon-rich and 9 oxygen-rich stars. These objects constitute the LPV catalog used throughout the present analysis.

\begin{figure*}
\centering
\subfigure[]{
    \includegraphics[width=0.45\textwidth]{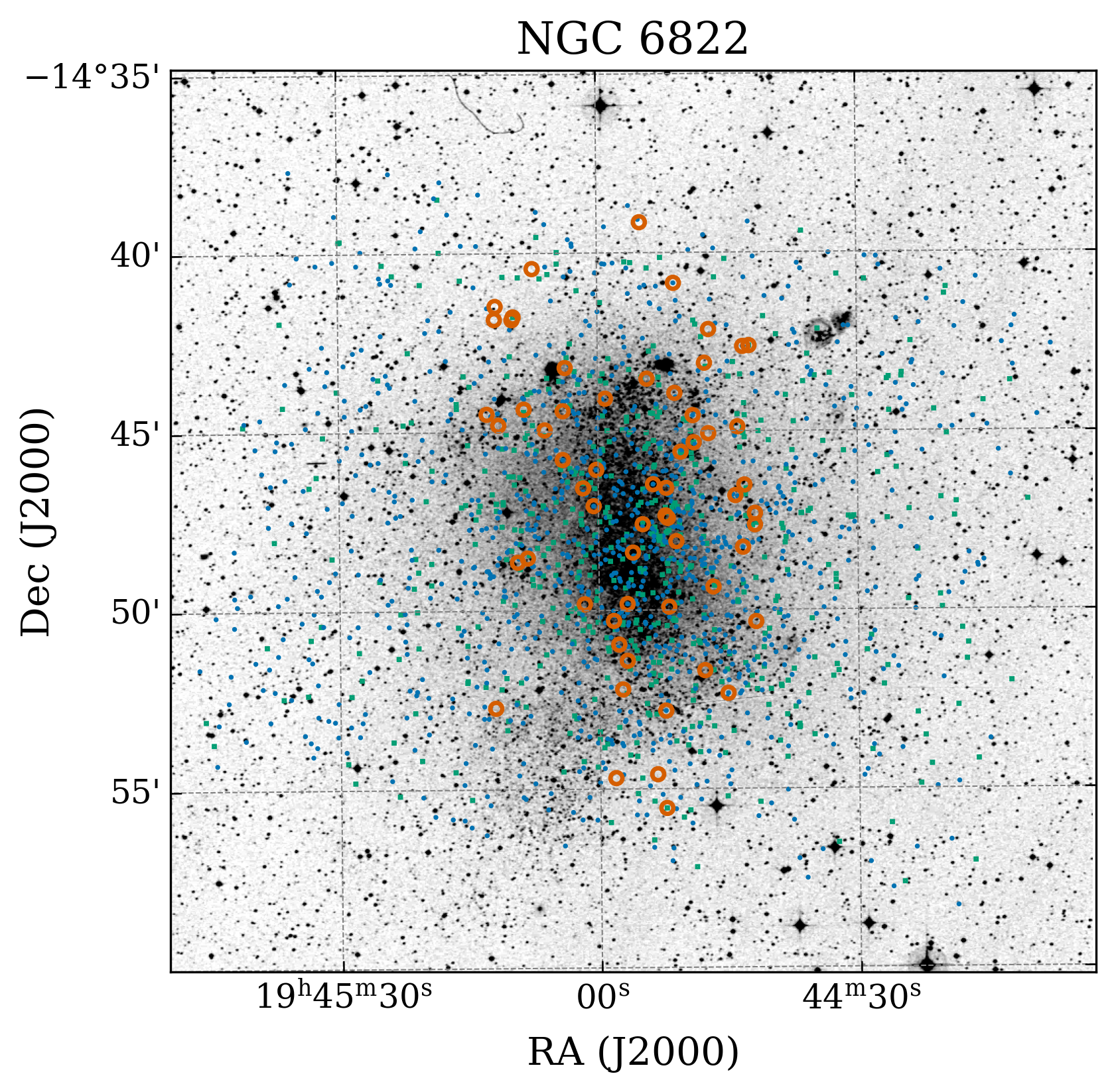}
    \label{fig:DS9}
}
\hfill
\subfigure[]{
    \includegraphics[width=0.45\textwidth]{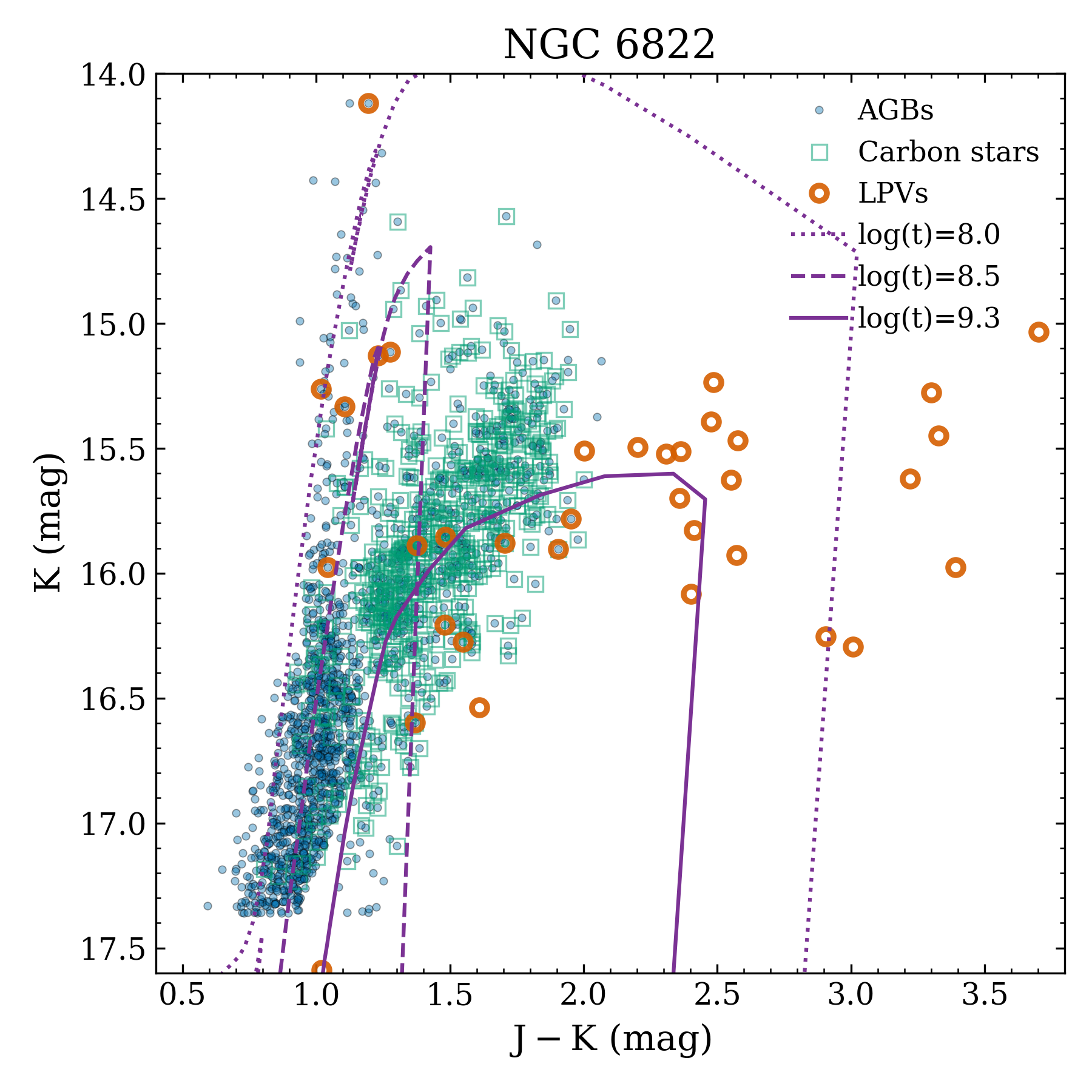}
    \label{fig:CMD}
}
\caption{
(\textit{Left}) Image of NGC~6822 showing the spatial distribution of evolved stars.
AGB stars from the catalog of \citet{Hirschauer20} are marked by blue circles, while LPVs from \citet{Whitelock13} are indicated by orange circles. Green squares identify carbon stars from \citet{Letarte02} that are also present in the \citet{Hirschauer20} catalog.
(\textit{Right}) $K$ versus $(J-K)$ color--magnitude diagram of NGC~6822.
AGB stars and LPVs are shown using the same color and symbol coding as in the left panel.
Theoretical isochrones with ages $\log(t/\mathrm{yr}) = 8.0$, 8.5, and 9.3 are overplotted as dotted, dashed, and solid purple lines, respectively.
}
\label{fig:NGC6822_overview}
\end{figure*}

\subsection{Optical Carbon-Star Catalog}

\citet{Letarte02} reported the presence of an extensive and homogeneous population of carbon stars in NGC~6822. Their primary observations were obtained in 2000 September with the CFH12K wide--field camera on the Canada--France--Hawaii Telescope \citep{Cuillandre01}, covering an area of $42\farcm0 \times 28\farcm0$. Imaging was secured in the $R$ and $I$ broadband filters together with CN and TiO narrowband filters. Complementary data from 1999 October were collected with the Las Campanas Swope Telescope in Kron--Cousins $R_{\mathrm{KC}}$, CN (810\,nm), and TiO (770\,nm), mainly to pre-select carbon star candidates for spectroscopic follow--up. From the CFH12K dataset, a total of 904 carbon stars were identified. Stars were classified as carbon stars if they satisfied $(R-I)_0 > 0.90$ mag and $(\mathrm{CN}-\mathrm{TiO}) > 0.30$. Reported uncertainties for the brightest stars ranged from $0.004 \leq \Delta I \leq 0.052$ mag and $0.007 \leq \Delta (R-I) \leq 0.100$ mag. Cross-matching our final catalog with that of \citet{Letarte02} identified 410 carbon stars in common.

Figure~\ref{fig:DS9} shows the spatial distribution of the 1610 AGB stars in the final catalog of NGC~6822, including the LPVs and carbon stars identified by cross-matching with the catalogs of \citet{Whitelock13} and \citet{Letarte02}, respectively. The final catalog was constructed using the AGB catalog of \citet{Hirschauer20} as the reference catalog.  Figure~\ref{fig:CMD} presents the locations of these stars in the near-infrared color--magnitude diagram.

\subsection{JWST Data}

Our analysis also makes use of the catalog presented by \citet{Nally24}, which provides the most detailed infrared view of the evolved stellar populations in the central bar of NGC~6822. This study utilized the \textit{James Webb Space Telescope} (JWST) with both the Near-Infrared Camera (NIRCam) and the Mid-Infrared Instrument (MIRI), achieving a depth up to seven magnitudes greater than previous $JHK$ surveys \citep{Rieke23, Dicken24}. The primary goal was to characterize the infrared stellar content, with particular emphasis on evolved stars such as oxygen-rich and carbon-rich AGBs.

We find a small number of outliers when comparing the JWST photometry with the closest corresponding bands in the \citet{Hirschauer20} catalog, as discussed in Appendix~\ref{app:JWST}. Since the JWST and \citet{Hirschauer20} filters are not exactly equivalent, the observed differences cannot be straightforwardly interpreted as systematic changes in the underlying SED. We therefore did not adopt the full JWST catalog in our analysis. Instead, we selected a subset of the JWST photometry from \citet{Nally24} based on the procedure described in Appendix~\ref{app:JWST}. The selected JWST measurements are included in the final photometric catalog and incorporated into the SED fitting.

The final catalog is obtained by cross-matching the reference catalog of \citet{Hirschauer20} with the LPV catalog of \citet{Whitelock13}, the optical carbon-star catalog of \citet{Letarte02} (410 common stars), and the JWST photometry of \citet{Nally24} (340 F115W and 111 F1500W common stars, the representative NIRCam and MIRI bands used in the quality-control comparison of Appendix~\ref{app:JWST}). The resulting 1610-source combined catalog is used throughout this analysis. \footnote{The final catalog is publicly available in the \texttt{SEDust} repositories mentioned above.}

\section{SEDust}
\label{sec:model}

To derive mass-loss rates for evolved stars in a consistent manner, we developed an automated spectral energy distribution fitting pipeline, hereafter referred to as \texttt{SEDust}. The pipeline converts multi-band photometric measurements into fluxes and compares them with synthetic SEDs generated using the \texttt{DUSTY} radiative transfer code \citep{Ivezic97}. Model fitting is performed through a minimum $\chi^{2}$ approach, with separate grids constructed for carbon-rich and oxygen-rich chemistries \citep{Draine84, Hanner88, Pegourie88}. Once the best-fitting model is identified, the corresponding \texttt{DUSTY} outputs are used to infer the physical properties of the circumstellar envelope, including the stellar mass-loss rate. In the following, we briefly outline the chemical distinction between carbon- and oxygen-rich AGB stars and summarize the relevant features of the \texttt{DUSTY} code that underpin the \texttt{SEDust} methodology.

Following the third dredge-up, the carbon abundance in AGB envelopes increases even when oxygen initially dominates \citep{Mowlavi99, Uttenthaler19}. Based on the carbon-to-oxygen ratio, these stars are broadly classified as carbon-rich (C/O $>1$) or oxygen-rich (C/O $<1$) \citep{Nowotny03,Ferrarotti06}. Accurately constraining the atmospheric composition is therefore essential for reproducing the spectral energy distributions of AGBs using radiative transfer tools such as \texttt{DUSTY}.

\texttt{DUSTY}, introduced by \citet{Ivezic97}, is a flexible radiative transfer package developed to model how dust interacts with radiation across a wide range of astrophysical environments. It solves the transfer equation while incorporating absorption, scattering, and thermal re-emission by dust grains, allowing synthetic SEDs to be produced and compared with photometric or spectroscopic measurements. A major advantage of \texttt{DUSTY} is the ability to adopt diverse dust compositions, grain size distributions, and optical properties, which is essential for representing the variety of dusty media in the cosmos.

\begin{figure}
\begin{center}
\includegraphics[width=0.48\textwidth]{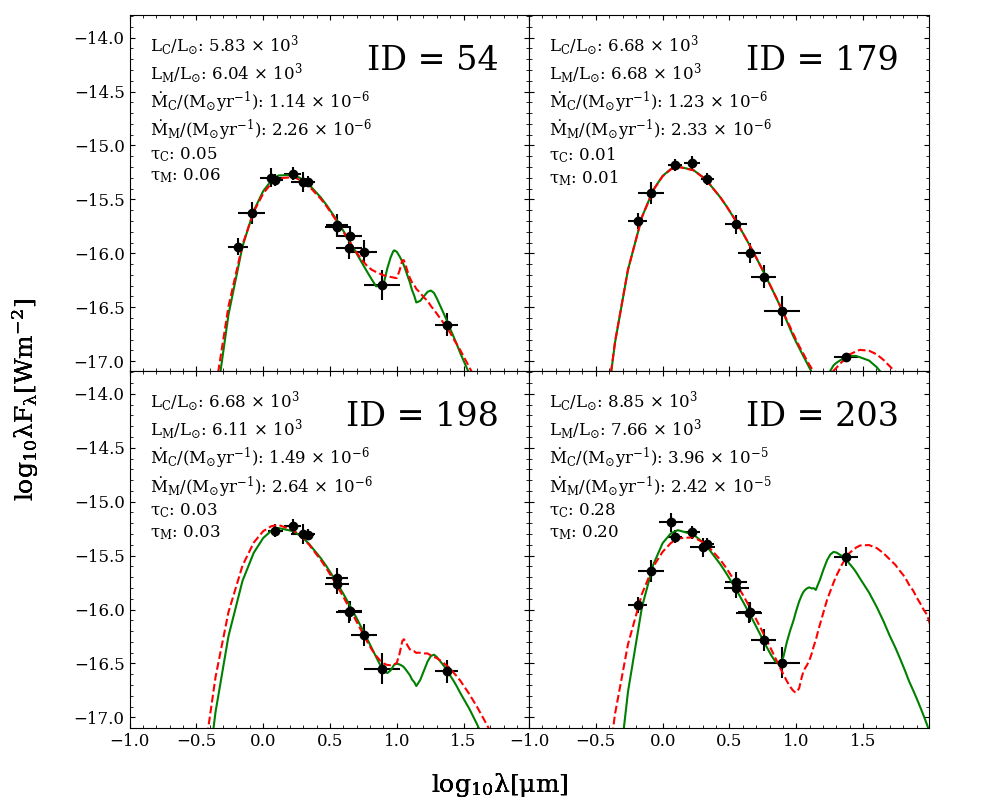}
\caption{SEDs for a representative subset of stars. Observed photometry is shown as black points with error bars. The best-fitting oxygen-rich (M-type) dust model is indicated by the solid green line, while the carbon-rich dust model is shown by the red dashed line. Each panel lists the source identifier and the derived stellar and circumstellar parameters, including luminosity, mass-loss rate, and optical depth for both models. The corresponding numerical values are also provided in Table~\ref{tab:selected_info} for clarity.} 
\label{fig:2x2}
\end{center}
\end{figure}

\texttt{SEDust} constructs an extensive grid of radiative transfer models in which several key quantities are treated as free parameters: the stellar effective temperature, the dust condensation temperature, and the optical depth of the circumstellar envelope. The grid spans stellar effective temperatures from 2000 to 4000~K in steps of 100~K. Dust condensation temperatures are sampled from 100 to 1500~K with the same spacing. The optical depth is defined at the $K$ band ($2.2\,\mu$m) and ranges from $\tau_{K}=0.01$ to 2, with steps of 0.01 up to $\tau_{K}=0.5$ and 0.05 at higher values. This sampling was chosen to provide fine resolution over the range occupied by the observed sources. Nearly all stars in our sample have best-fitting optical depths below $\tau_{K}\simeq1$, and extending the grid to $\tau_{K}=2$ was found to be sufficient, with no impact on the derived physical parameters. Figure~\ref{fig:2x2} illustrates representative \texttt{SEDust} fits and lists the parameters corresponding to the best-fitting SEDs.

Dust chemistry within the grid is tailored to the inferred atmospheric type. For carbon stars, we adopt a dust mixture of 85\% amorphous carbon and 15\% silicon carbide following \citet{Groenewegen06}, using the optical constants of \citet{Hanner88} and \citet{Pegourie88}, respectively. Oxygen-rich AGB stars are modeled using astronomical silicates \citep{Draine84}. We therefore compute separate grids for O-rich and C-rich sources, each containing 25200 models in the final library used for fitting.

The circumstellar envelopes are assumed to be spherically symmetric. Rather than prescribing a fixed density profile or wind velocity, we employ the self-consistent radiatively driven wind solution implemented in \texttt{DUSTY} \citep{Ivezic97}. In this approach, the density structure and terminal outflow velocity are computed by \texttt{DUSTY} from the adopted stellar and dust properties. The resulting terminal velocity, which varies across the model grid, is then used in the \texttt{DUSTY} scaling relations to derive the corresponding mass-loss rate.

The final catalog of this study contains 1610 sources with photometric measurements in all eight bands used in the main SED fitting, namely $J$, $H$, $K$, 3.6, 4.5, 5.8, 8.0, and 24~$\mu$m. The \texttt{SEDust} pipeline can also accommodate additional photometric measurements at other wavelengths. For each available observation, the corresponding wavelength is identified in the model SED and the model flux is evaluated at that wavelength. The observed and model fluxes are then compared using the reduced $\chi^2$, calculated using the available photometric measurements. When a supplementary band is unavailable for a given source, that wavelength is simply omitted from the $\chi^2$ sum for that star, rather than treated as an upper limit; no sources in the primary eight-band fit required such treatment. The model with the minimum reduced $\chi^2$ is adopted as the best-fitting model.

\subsection{Uncertainties}
\label{sec:uncertainties}

The uncertainties associated with the mass-loss rates derived using \texttt{SEDust} originate from two main sources: systematic uncertainties related to the underlying radiative transfer modeling, and statistical uncertainties driven by photometric measurement errors.

Systematic uncertainties arise from the assumptions inherent to the \texttt{DUSTY} radiative-transfer framework, including the adopted dust properties, shell geometry, and the limited sensitivity of the SED to certain physical parameters. In particular, \citet{Ivezic97} estimate an inherent uncertainty of approximately 30\% in the mass-loss rate and terminal outflow velocity derived from the radiatively driven wind solution, associated with the gravitational correction: varying this correction by up to 50\% has little effect on the modeled spectrum. We therefore adopt a 30\% systematic uncertainty for the \texttt{DUSTY}-derived mass-loss rates.

Statistical uncertainties were estimated using a Monte Carlo approach. For a representative star with complete photometric coverage across all available filters, the observed magnitudes were randomly perturbed according to their reported uncertainties, assuming Gaussian error distributions. A total of 50 perturbed realizations were generated and independently fitted using \texttt{SEDust}, considering both carbon-rich and oxygen-rich dust chemistries. The distributions of the resulting fitted parameters were found to be approximately Gaussian, and their standard deviations were therefore adopted as estimates of the statistical uncertainties.

For the carbon-rich model, the statistical uncertainty in the mass-loss rate is found to be $8.5\%$, corresponding to $\pm 2.50 \times 10^{-8}\ \msun \mathrm{yr}^{-1}$ for a reference value of $\dot{M} = 2.96 \times 10^{-7}\ \msun \mathrm{yr}^{-1}$. The associated luminosity uncertainty is $4.3\%$. For the oxygen-rich model, larger statistical uncertainties are obtained, amounting to $28\%$ in the mass-loss rate and $4.1\%$ in luminosity, reflecting the increased degeneracy of oxygen-rich SEDs.

As discussed above, the systematic uncertainty is therefore taken to be
\[
\sigma_{\rm sys}=0.3\,\dot{M},
\]
and the total uncertainty is obtained by combining the statistical and systematic contributions in quadrature,

\begin{equation}
\sigma_{\mathrm{tot}} = \sqrt{\sigma_{\mathrm{stat}}^2 + \left(0.3\ \dot{M}\right)^2}
\end{equation}

%For carbon-rich models, this results in a typical total uncertainty of $\sim31\%$, while for oxygen-rich models the total uncertainty increases to $\sim41\%$. These values are adopted as representative uncertainties for the \texttt{SEDust} mass-loss estimates throughout this work. In systems with comparable photometric quality and observational setups, the same statistical uncertainties are assumed to apply.

For carbon-rich models, this results in a typical total uncertainty of $\sim31\%$, while for oxygen-rich models the total uncertainty increases to $\sim41\%$. These values are adopted as representative uncertainties for the \texttt{SEDust} mass-loss estimates throughout this work.
 
We note that this Monte Carlo test was performed on a single star with complete, high-quality photometric coverage, and the resulting fractional uncertainties were then applied uniformly to the full sample. Photometric uncertainties in the \citet{Hirschauer20} catalog span a wide range (Sect.~\ref{sec:data}; e.g., 0.02--0.30~mag at 3.6 and 4.5~$\mu$m and up to 0.85~mag in $J$), so this uniform assumption likely underestimates the true statistical uncertainty for sources with sparser photometric coverage or lower signal-to-noise, particularly at the faint end of the sample. We therefore regard the quoted 31\% (C-rich) and 41\% (O-rich) total uncertainties as representative, best-case values applicable to well-detected sources with photometry across all available bands, rather than as a rigorous per-star error budget.

\section{Results}
\label{sec:result}

The spectral energy distributions of the 1610 evolved stars in our sample were modeled using the SEDust pipeline described in Sect.~\ref{sec:model}. The fitting procedure provided estimates of the stellar luminosity, dust optical depth, and total (gas+dust) mass-loss rate for each source, together with their associated uncertainties. Representative uncertainties for these quantities also were estimated as described in Sect. \ref{sec:uncertainties}. In this section, these derived parameters are presented and compared with theoretical expectations and previous observational studies of evolved stars.

\subsection{Mass-loss rates}
\label{sec:mdot}

Figure~\ref{fig:L_MDOT} presents the relation between stellar luminosity and mass-loss rate for the evolved stars in NGC~6822. The derived total mass-loss rates of the evolved stars in this galaxy range from $(1.0 \pm 0.31)\times10^{-7}$ to $(7.3 \pm 2.99)\times10^{-5} \msun \mathrm{yr}^{-1}$ (Table~\ref{tab:Max-min}). Carbon stars in the Milky Way reach mass-loss rates of up to $\sim10^{-4} \msun \mathrm{yr}^{-1}$ \citep{Whitelock06}, whereas carbon stars in the Large Magellanic Cloud and M33 typically exhibit mass-loss rates of the order of $10^{-5} \msun \mathrm{yr}^{-1}$ \citep{Gullieuszik12,Javadi13_III}. The values derived for the carbon stars in NGC~6822 are therefore consistent with those observed in other Local Group galaxies. Likewise, M-type AGB stars in the solar neighborhood commonly exhibit mass-loss rates of a few $10^{-5} \msun \mathrm{yr}^{-1}$ \citep{Jura89}, in agreement with the upper range derived for the M-type stars in our sample.

Using the \citet{Hirschauer20} catalog, the integrated mass-loss rate of the 560 carbon-rich AGB stars is estimated to be $(9.8 \pm 2.7)\times10^{-4} \msun \mathrm{yr}^{-1}$, while the corresponding value for the 1050 oxygen-rich stars is $(2.7 \pm 1.1)\times10^{-3} \msun \mathrm{yr}^{-1}$. The resulting integrated mass-loss rate of the AGB population in NGC~6822 is therefore $(3.7 \pm 1.1)\times10^{-3} \msun \mathrm{yr}^{-1}$.

For the LPV sample of \citet{Whitelock13}, which was cross-matched with the AGB catalog to obtain homogeneous multi-wavelength photometry, the integrated mass-loss rates are $(1.3 \pm 0.4)\times10^{-4} \msun \mathrm{yr}^{-1}$ for the 50 carbon-rich stars and $(3.7 \pm 1.5)\times10^{-5} \msun \mathrm{yr}^{-1}$ for the nine oxygen-rich stars, yielding a total LPV mass-loss rate of $(1.7 \pm 0.4)\times10^{-4} \msun \mathrm{yr}^{-1}$. Although LPVs constitute only 3.6\% of the AGB population, they account for 4.6\% of the total integrated mass-loss. This close correspondence indicates that the LPV population provides a representative measure of the total AGB mass-loss in NGC~6822, consistent with the results previously reported for NGC~147 and NGC~185 by \citet{Mahani25}.  This direct \texttt{SEDust} fit to the LPV subsample is distinct from the results of the \emph{LPV method} used in Sect.~\ref{Sec:LPV_Method} to estimate the mass-loss rate of the full 1610-star catalog.

\subsection{Luminosity}
\label{sec:Luminosity}

\begin{figure}
\begin{center}
\includegraphics[width=0.45\textwidth]{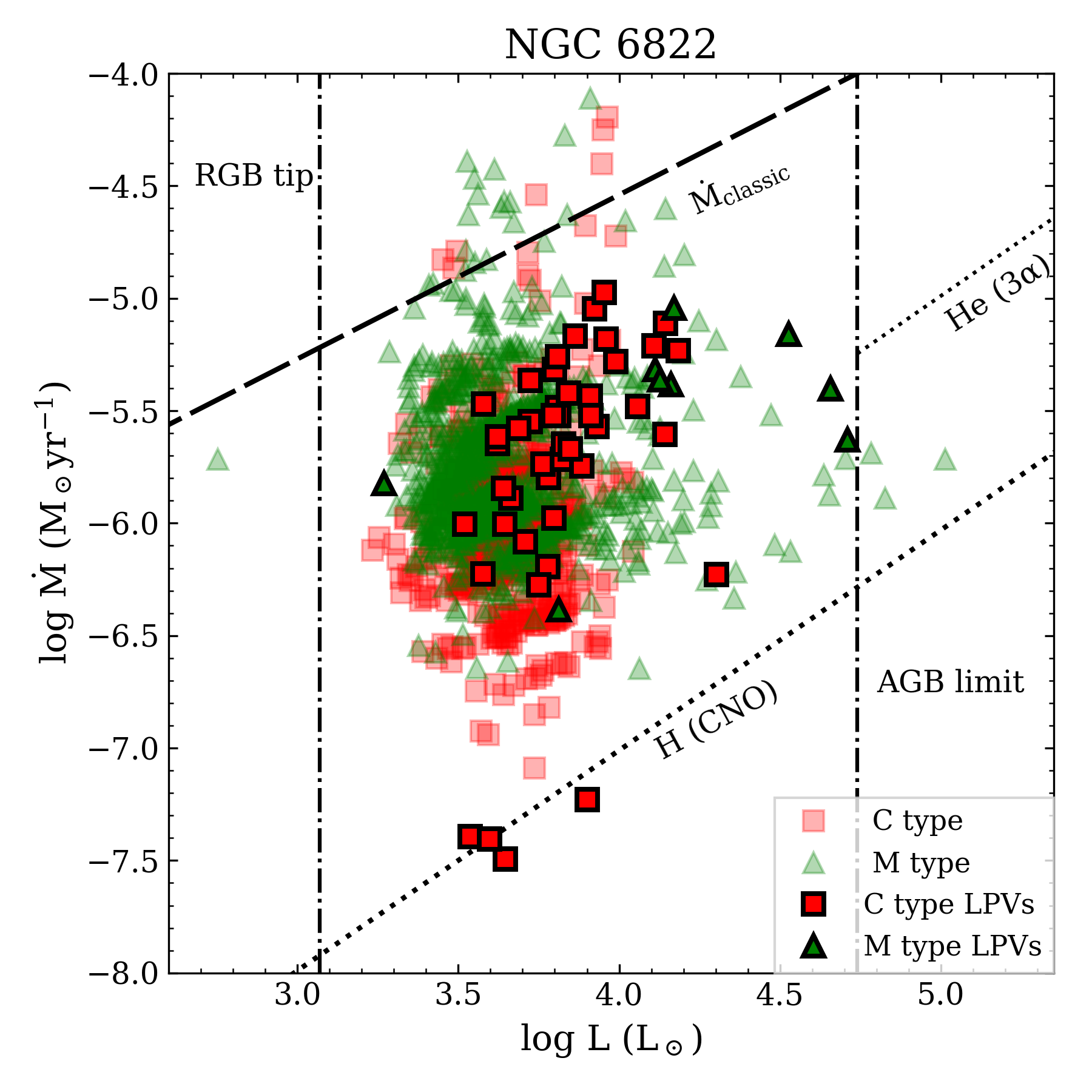}
\caption{ Mass-loss rate as a function of stellar luminosity for evolved stars in NGC~6822. Oxygen-rich (M-type) stars are shown as green triangles, while carbon-rich (C-type) stars are shown as red squares. The full sample is plotted using semi-transparent symbols. LPVs are highlighted with the same symbols outlined in black. The dashed and dotted curves indicate empirical and theoretical mass-loss prescriptions, including the classical mass-loss relation \citep{vanLoon99}, hydrogen-shell burning (H(CNO)), and helium-shell burning (He($3\alpha$)). Vertical dash-dotted lines mark the approximate locations of the RGB tip and the upper AGB luminosity limit. } 
\label{fig:L_MDOT}
\end{center}
\end{figure}

The stellar luminosities derived with \texttt{SEDust} span a broad range, from $5\times10^{2}$ to $1.0\times10^{5} \lsun$ for M-type stars and from $1.7\times10^{3}$ to $1.3\times10^{4} \lsun$ for C-type stars (Table~\ref{tab:Max-min}). Figure~\ref{fig:L_MDOT} shows the relation between stellar luminosity and mass-loss rate for the evolved stars in NGC~6822. A general increase in mass-loss rate with luminosity is evident, although considerable scatter is present.

Higher stellar luminosity is expected to increase the radiation pressure acting on circumstellar dust grains, thereby strengthening stellar winds and enhancing mass-loss rates in AGB stars \citep{Habing13,Hofner18}. This positive correlation between luminosity and mass-loss is a well-established prediction of theoretical models, in which the mass-loss process is regulated by the balance between outward radiation pressure on dust and the inward gravitational force of the star. As luminosity increases, the photon flux rises, leading to stronger radiation pressure that facilitates the levitation and subsequent ejection of material from the outer stellar atmosphere. This provides a physical explanation for the observed dependence of the mass-loss rate on stellar luminosity \citep{vanLoon99,Srinivasan09}.

In addition to this positive correlation, a segregation between C- and M-type stars is expected. Previous studies have demonstrated that stars with initial masses between $1.1 \msun$ and $4 \msun$ follow distinct evolutionary paths that yield different surface chemistries \citep{Karakas14, Karakas14_II}. Carbon is produced in stars during helium burning through the triple-alpha process \citep{Kippenhahn90}. The amount of carbon that reaches the stellar surface depends on the stellar mass and the efficiency of internal mixing processes. In low-mass stars, strong mixing occurs, but the convective envelope does not extend deep enough to bring carbon from the core to the surface. As a result, carbon does not accumulate in the atmosphere. In contrast, in stars with masses greater than $4\msun$, the carbon produced in the core is consumed through subsequent nuclear fusion processes before it can be dredged up to the surface \citep{Mowlavi99}. Only stars within the intermediate mass range of approximately $1.1\msun$ to $4\msun$ can transport sufficient carbon to the surface through the third dredge-up process. This makes them the primary progenitors of carbon stars, as their atmospheres become enriched in carbon during their evolution \citep{Marini21}.

Carbon stars are drawn from the narrower intermediate initial-mass range identified above ($\sim1.1$--$4\,\msun$), whereas M-type AGB stars span a broader range of initial masses. Because AGB luminosity tracks initial mass, this difference should show up mainly as a luminosity segregation between the two chemical types, not as a difference in mass-loss rate. This luminosity-based segregation between O- and C-rich stars, driven by initial mass, is illustrated in Fig.~27 of \citet{Abdollahi23}. The clearer segregation seen in that study may in part reflect the nature of the host galaxy, Andromeda~IX, a quenched dwarf galaxy in which the evolved stellar population is expected to lie predominantly near the terminal stages of AGB evolution.

NGC~6822, by contrast, has an extended star-formation history, so its evolved stars span a wider range of ages and evolutionary phases along the AGB. Mass-loss rate, moreover, does not track stellar mass alone: it also depends on effective temperature, pulsation properties, and wind-driving efficiency, all of which evolve over the AGB lifetime \citep{vanLoon99,vanLoon05}. A clean separation between C- and M-type stars in the mass-loss--luminosity plane (Fig.~\ref{fig:L_MDOT}) is therefore not necessarily expected, even where the underlying mass-driven luminosity segregation is present. The considerable scatter in Fig.~\ref{fig:L_MDOT} is consistent with this picture: stars of both chemical types span a range of evolutionary stages along the AGB, evolving toward higher luminosities and larger mass-loss rates as they age \citep{vanLoon99,vanLoon05}. This age-driven spread blurs any sharp chemical-type boundary in mass-loss rate, even though the segregation is present in luminosity.

%For improved readability, a clearer and more simplified representation of the mass-loss–luminosity relation is shown in Fig. \ref{fig:L_MDOT2x2}, in which different methods and their combined sample are presented separately (see Appendix \ref{app:Comparison}).

\begin{table*}
\caption{Minimum and maximum values calculated in different models of evolved stars. }
\begin{tabular}{lllllll}
\hline
\hline
\multirow{2}{*}{}                  & \multicolumn{2}{c|}{SEDust}                               & \multicolumn{2}{c|}{LPVs}                                 & \multicolumn{2}{c}{DESK}                                 \\ \cline{2-7} 
                                   & \multicolumn{1}{c|}{M-type} & \multicolumn{1}{c|}{C-type} & \multicolumn{1}{c|}{M-type} & \multicolumn{1}{c|}{C-type} & \multicolumn{1}{c|}{M-type} & \multicolumn{1}{c}{C-type} \\ \hline
\multicolumn{1}{l|}{Luminosity [$10^{3} \lsun]$}    & 0.5 - 102.8                       & 1.7 - 12.8                       & 2.1 - 17.1                       & 2.8 - 13.4                       & 2.0 - 89.9                       & 2.8 - 53.9                      \\ \hline
\multicolumn{1}{l|}{Mass-loss [$10^{-6} \msun \mathrm{yr}^{-1}$]}     & 0.2 - 73.2                       & 0.1 - 63.7                       & 1.2 - 10.6                       & 0.4 - 18.3                       & 0.4 - 34.8                       & 1.8 - 40.7                      \\ \hline
\multicolumn{1}{l|}{Optical depth} & 0.01 - 0.30                       & 0.01 - 0.49                       & 0.01 - 0.22                       & 0.02 - 0.95                       & 0.1 - 0.7    & 0.1 - 0.1                      \\ \hline
\hline
\end{tabular}
\label{tab:Max-min}
\end{table*}

\subsection{Optical depth}
\label{Sec:Opticaldepth}

Stellar luminosity provides information on a star’s mass, radius, and evolutionary stage. In AGB stars, however, the presence of a dense circumstellar envelope can reduce the observed luminosity. This reduction is caused by self-extinction, in which dust absorbs stellar radiation and re-emits it at longer wavelengths. The effect is strongest at short wavelengths and depends on the envelope density, the gas-to-dust ratio, and the optical depth \citep{Whitelock06}.

Mass-loss in AGB stars is driven by the coupling between dust opacity and the stellar radiation field. Dust grains absorb photons and are accelerated outward by radiation pressure. The efficiency of this acceleration increases with optical depth \citep{Bladh12}. Stellar pulsations generate shock waves that lift material to regions where dust can form. Together, radiation pressure on dust and pulsation-induced shocks regulate the mass-loss rate.

\begin{figure}
\begin{center}
\includegraphics[width=0.45\textwidth]{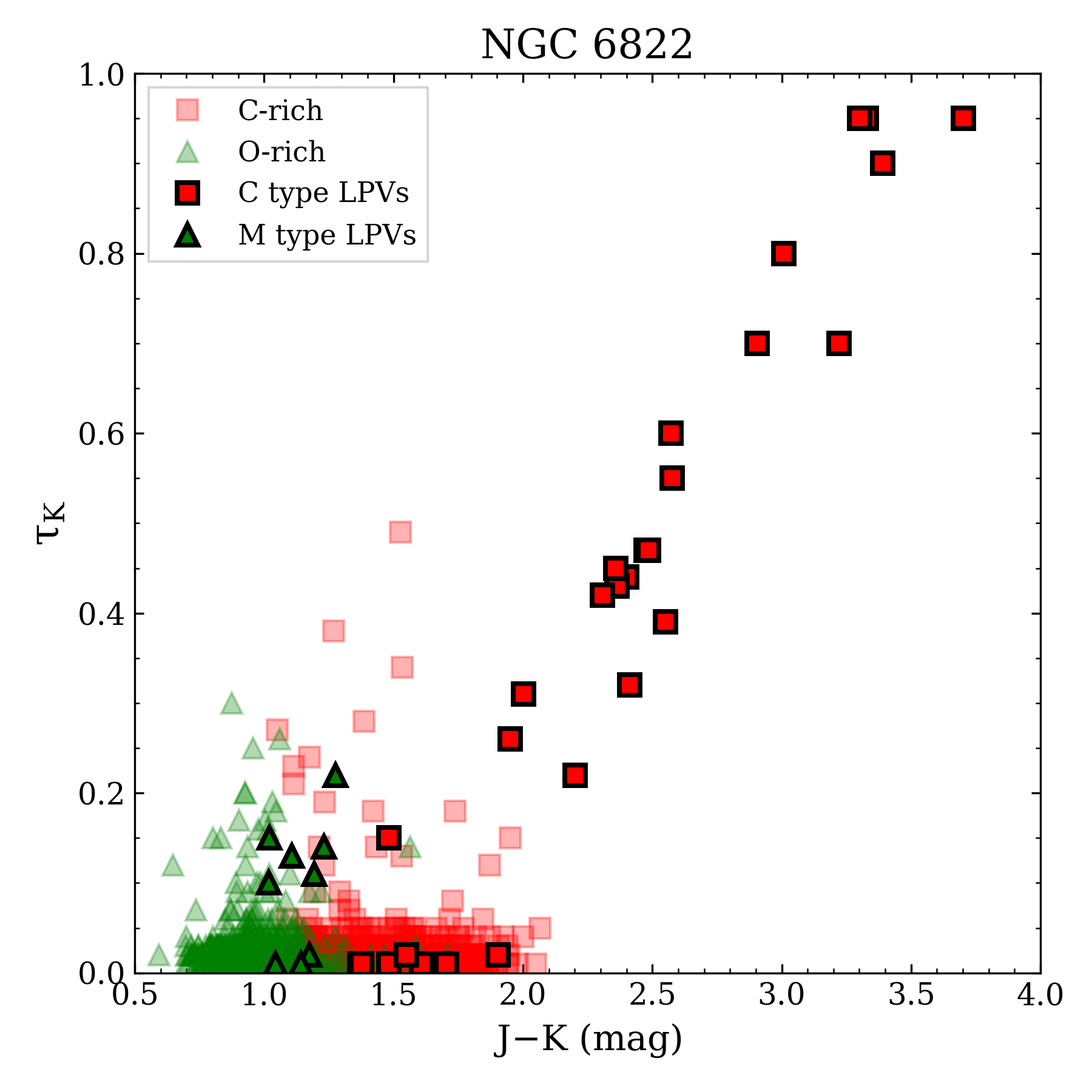}
\caption{Optical depth at the $K$ band, $\tau_{K}$, as a function of near-infrared color $J-K$ for evolved stars in NGC~6822. Carbon-rich stars are shown as red squares, while oxygen-rich stars are shown as green triangles. The full sample is plotted using semi-transparent symbols. LPVs are highlighted with the same symbols outlined in black.} 
\label{fig:optical_depth}
\end{center}
\end{figure}

Figure \ref{fig:optical_depth} presents the relation between optical depth and near-infrared color. For LPVs, redder $J-K$ colors correspond to higher optical depths. The AGB catalog reaches a maximum observed color of $J-K=2.07$~mag, the reddest source recovered by \citet{Hirschauer20}; as a result, the increase in optical depth with reddening is not clearly visible across the full \citet{Hirschauer20} sample. Some scatter is present, which is expected given the variable nature of AGB stars. Despite this limitation, oxygen-rich and carbon-rich stars are clearly separated \citep{Guglielmo93, Zijlstra96, vanLoon97, vanLoon98}. This separation has also been reported by \citet{Javadi13_III} and \citet{Mahani25} and reflects differences in dust opacity at $2.2\,\mu$m, the wavelength at which the optical depth $\tau_{K}$ is defined: carbon-rich dust is more opaque than oxygen-rich dust at this wavelength \citep{Suh99,Suh2000}.

%Figure \ref{fig:optical_depth} presents the relation between optical depth and mid-infrared color (see Fig. \ref{fig:optical_depth_2x2} for the individual methods). For LPVs, redder colors correspond to higher optical depths. The AGB catalog is limited to a maximum color of 2.07 mag. As a result, the increase in optical depth with reddening is not clearly visible in the \citet{Hirschauer20} data. Some scatter is present, which is expected given the variable nature of AGB stars. Despite this limitation, oxygen-rich and carbon-rich stars form distinct sequences \citep{Guglielmo93, Zijlstra96, vanLoon97, vanLoon98}.

%This separation has also been reported by \citet{Javadi13_III, Mahani25} and reflects differences in dust opacity. Carbon-rich dust is more opaque than oxygen-rich dust. Consequently, at a given optical depth, carbon stars appear redder than oxygen-rich stars. 

\subsection{Color}
\label{Sec:color}

\begin{figure}
\begin{center}
\includegraphics[width=0.45\textwidth]{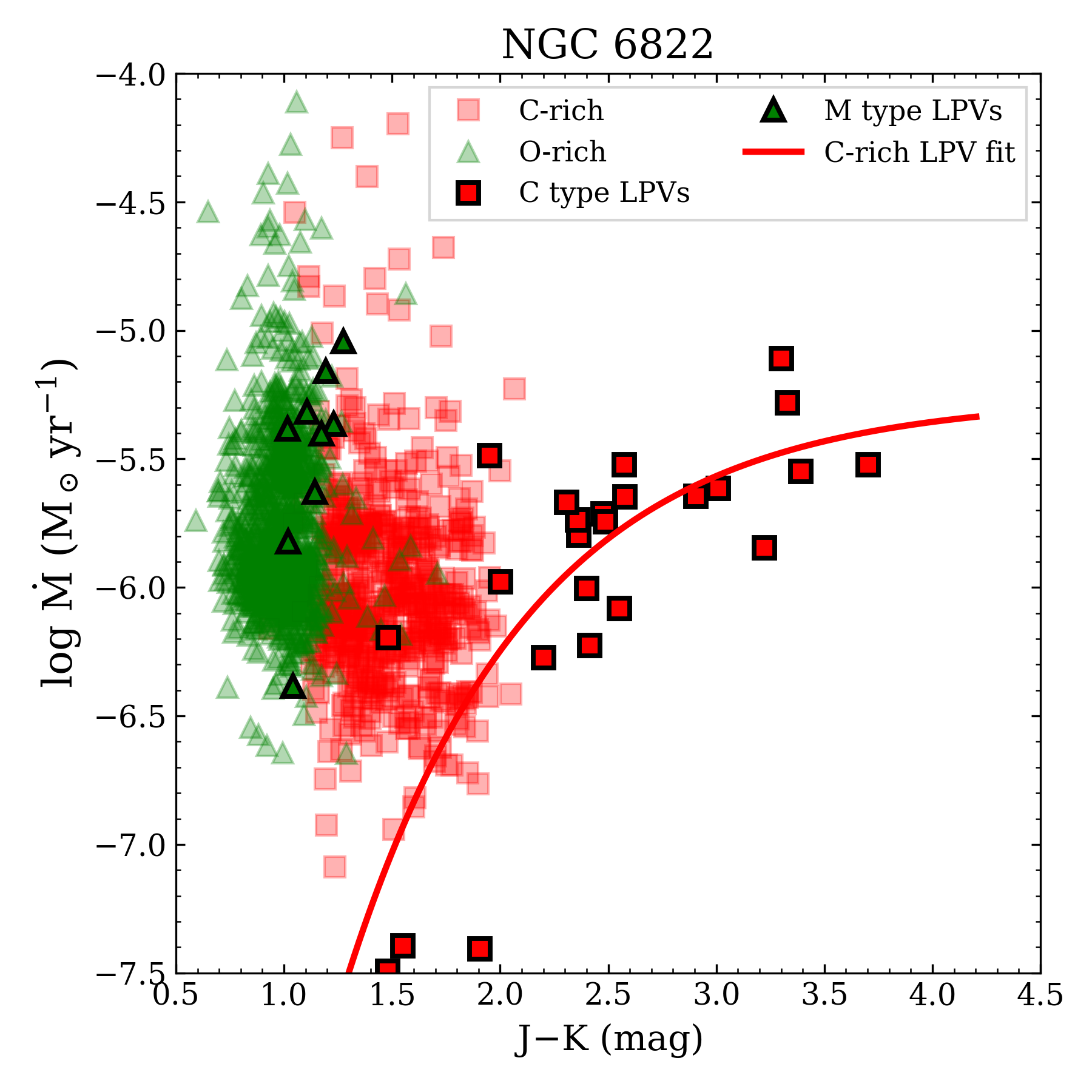}
\caption{Mass-loss rate as a function of near-infrared color $J-K$ for evolved stars in NGC~6822. Carbon-rich stars are shown as red squares, while oxygen-rich stars are shown as green triangles. The full catalog is plotted using semi-transparent symbols. LPVs are highlighted with identical symbols outlined in black. The solid red curve shows the best-fit model for the carbon-rich LPVs, illustrating a rapid increase in mass-loss rate with increasing $J-K$ color, followed by a plateau at the reddest colors. The fit is calibrated on Mira-type LPVs specifically and is not expected to describe the broader, non-Mira AGB population.} 
\label{fig:Mdot_color}
\end{center}
\end{figure}

Figure \ref{fig:Mdot_color} shows the relation between the mass-loss rate and the J$-$K color. This color is commonly used as a proxy for both the stellar effective temperature and the dust content of the circumstellar envelope \citep{Javadi13_III}. As LPV stars evolve along the asymptotic giant branch, they cool down, and their J$-$K colors become redder. This change reflects the influence of molecular absorption and circumstellar dust on the observed fluxes \citep{Prager22}. The associated reddening is closely linked to higher mass-loss rates, since dust formation follows the ejection of material from the stellar atmosphere \citep{Whitelock06, vanLoon08, Matsuura09}.

Red J$-$K colors indicate cool and extended stellar atmospheres that favor dust formation. As discussed in Section~\ref{Sec:Opticaldepth}, radiation pressure acting on newly formed dust grains, together with stellar pulsations, drives the mass-loss process. Consequently, stars with higher mass-loss rates generally exhibit redder J$-$K colors due to the combined effects of lower effective temperatures and enhanced circumstellar extinction.

Figure \ref{fig:Mdot_color} shows a clear increase in mass-loss rate with increasing $J-K$ color up to $J-K\approx2$~mag. Carbon-rich LPVs populate the reddest regions of the diagram, as expected. However, for $J-K$ values greater than about 2~mag, the relation flattens and the derived mass-loss rate plateaus (Fig.~\ref{fig:Mdot_color}). This flattening is consistent with a saturation of near-infrared colors at high circumstellar optical depth: once the envelope becomes sufficiently optically thick, the observed $J$- and $K$-band flux is increasingly dominated by scattered and reprocessed light rather than direct stellar photons, so further increases in $\tau$ no longer translate into a proportional increase in $J-K$ \citep{Riebel12,Eriksson14}. As a result, $J-K$ becomes an increasingly insensitive tracer of mass-loss rate at the reddest colors, and the reddest stars are poorly constrained by this color alone. Variations in dust geometry and distribution may also contribute scatter to the observed colors at fixed mass-loss rate \citep{Brunner18}.

Although LPVs are themselves AGB stars, within our catalog they specifically comprise Mira-type variables \citep{Whitelock13}, that is, large-amplitude, long-period, quasi-regular pulsators characteristic of the most advanced stages of AGB evolution. The tight relation traced by the solid curve is therefore specific to this Mira subsample, whose regular, large-amplitude pulsations efficiently and predictably drive dust formation and, in turn, mass-loss. The broader AGB population also includes stars with weaker or irregular pulsation, or that have not yet reached the Mira stage; for these sources the coupling between near-infrared color and mass-loss rate is less direct, and they do not necessarily follow the same trend as the LPV/Mira subsample.

\section{Discussion}
\label{sec:Discussion}

\subsection{Integrated AGB Mass-loss Rates}
\label{Sec:Total_mass_loss}

The integrated mass-loss rate of the AGB population derived from the \texttt{SEDust} analysis is presented in Sect.~\ref{sec:mdot}. To assess the robustness of this result, we compare it with integrated mass-loss rates obtained using two independent methods: an estimate based on LPVs and that derived using the Dusty-Evolved-Star-Kit (\texttt{DESK}). The results of these comparisons, including the differences between the methods and their possible sources, are discussed in Sect.~\ref{Sec:ThreeMethodComparison}.

\subsubsection{LPV Method}
\label{Sec:LPV_Method}

Previous studies by \citet{Javadi13_III} and \citet{Mahani25} have demonstrated that LPVs can be used to obtain reliable estimates of the total mass-loss rate of the AGB population.  This method is particularly useful when the homogeneous multi-wavelength photometry required for direct \texttt{SEDust} fitting is unavailable. For NGC~6822, LPVs identified by \citet{Whitelock13} provide an independent means of estimating the mass-loss properties of the evolved stellar population. 

\begin{figure}
\begin{center}
\includegraphics[width=0.45\textwidth]{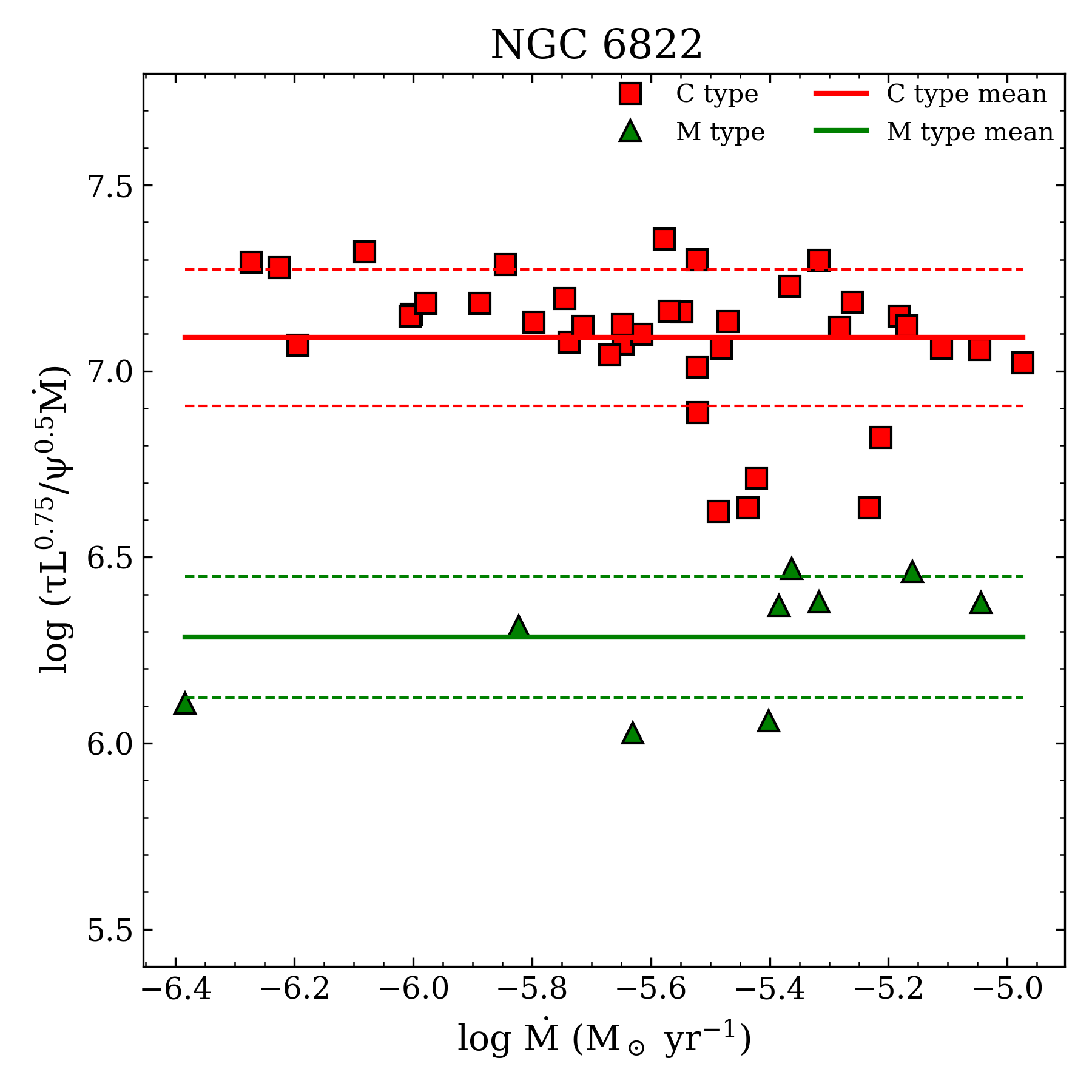}
\caption{Comparison between the derived mass-loss rates and the parameter constructed from optical depth, stellar luminosity, and the gas-to-dust mass ratio, as defined in Eq.~\ref{eq:mdot_const}. Carbon-rich stars are indicated by red symbols, and oxygen-rich stars by green symbols. The dashed lines mark the $1\sigma$ dispersion about the mean trend.} 
\label{fig:C_cnst}
\end{center}
\end{figure}

The method relies on the fact that a specific combination of stellar parameters remains nearly constant, within a small dispersion. As shown in Fig.~\ref{fig:C_cnst}, the scaling relations implemented in \texttt{DUSTY} \citep{Ivezic97} imply that

\begin{equation}
C = \frac{\tau L^{3/4}}{\psi^{1/2}\dot{M}}
\label{eq:mdot_const}
\end{equation}

The parameter $C$ is not a universal constant; rather, its value depends on the adopted dust chemistry and the corresponding dust properties. As shown in Fig. \ref{fig:C_cnst}, the mean values differ by approximately 0.81 dex between the C-rich and O-rich populations, with (log$ C \simeq7.09$) and  (log$ C \simeq6.28$), respectively. We therefore adopt separate values of $C$ for the two dust chemistries when deriving the mass-loss rates. In the Eq.\ref{eq:mdot_const}, $\tau$ is the optical depth, $L$ is the stellar luminosity, $\dot{M}$ is the stellar mass-loss rate, and $\psi=200$ is the gas-to-dust mass ratio \citep{Ivezic97}.

\begin{figure}
\begin{center}
\includegraphics[width=0.45\textwidth]{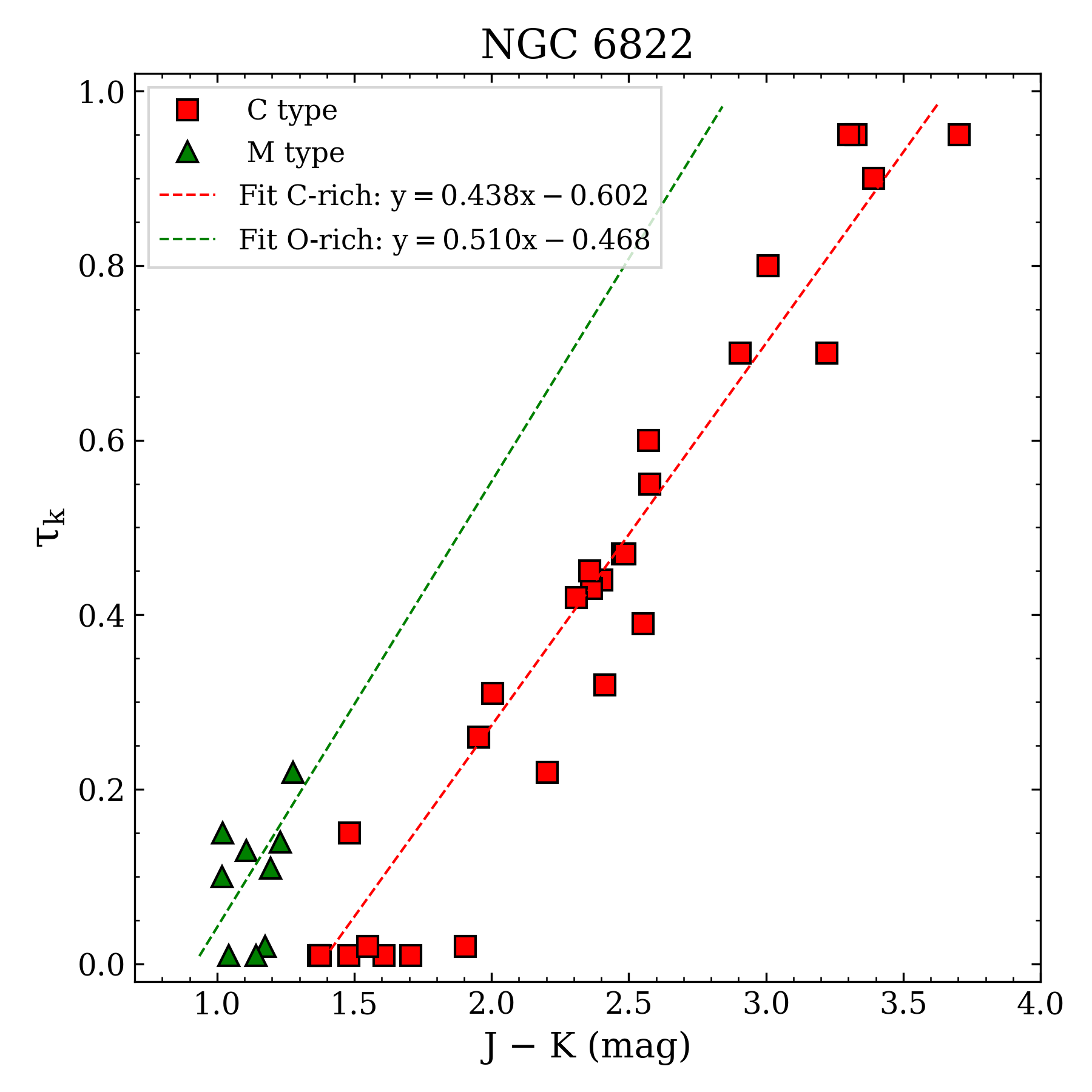}
\caption{Optical depth at the $K$ band, $\tau_{K}$, as a function of near-infrared color $J-K$ for LPVs in NGC~6822. Carbon-rich LPVs are shown as red squares, while oxygen-rich LPVs are shown as green triangles. Dashed lines indicate linear least-squares fits for each chemical type.} 
\label{fig:Tau_JK_LPVs_FIT}
\end{center}
\end{figure}

\begin{figure}
\begin{center}
\includegraphics[width=0.45\textwidth]{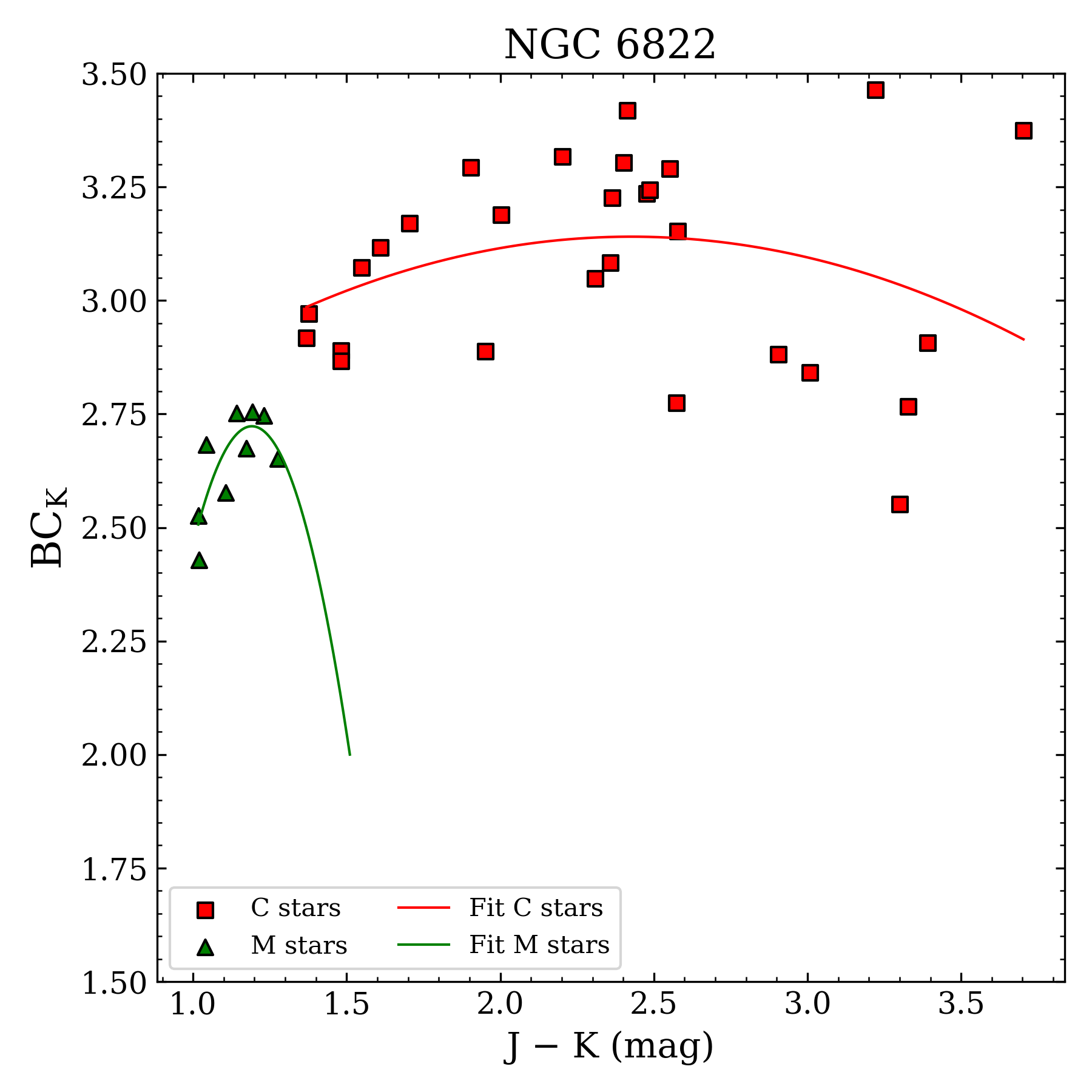}
\caption{Bolometric correction in the $K$ band as a function of near-infrared color $J-K$ for long-period variables in NGC~6822. Carbon-rich stars are shown as red squares, while oxygen-rich stars are shown as green triangles. Solid curves represent best fits for each chemical type. The corresponding fit parameters are listed in Table~\ref{tab:Fits_LPVs}.} 
\label{fig:BC_LPVs}
\end{center}
\end{figure}

The derivation of the integrated AGB mass-loss rate from the LPV population involves three steps. First, the optical depth of each AGB is estimated from empirical relations between optical depth and near-infrared color, calibrated using LPVs with homogeneous multi-wavelength photometry (Fig.~\ref{fig:Tau_JK_LPVs_FIT}; Table~\ref{tab:Fits_LPVs}). Second, the stellar luminosity is determined using the bolometric correction relations shown in Fig.~\ref{fig:BC_LPVs}. Finally, the individual mass-loss rates are calculated from Eq.~\ref{eq:mdot_const} and summed to obtain the integrated mass-loss rate of the AGB population; this is referred to hereafter as the "LPV method," reflecting the origin of the calibration rather than a restriction to the LPV subsample alone. This differs from the \texttt{SEDust}-fitted mass-loss rate of the 59-star LPV subsample itself reported in Sect.~\ref{sec:mdot} ($1.7\times10^{-4}\ \msun\ \mathrm{yr}^{-1}$): here the LPV-calibrated $\tau_K$--$(J-K)$ and BC$_K$--$(J-K)$ relations are applied to the near-infrared colors of the full 1610-star catalog, not to the LPV subsample alone. We note that the M-type BC$_{K}$ relation is calibrated on only nine O-rich LPVs, and the individual polynomial coefficients (Table~\ref{tab:Fits_LPVs}) carry fractional uncertainties of 50–100\%. However, because these coefficients are strongly covariant, the propagated uncertainty on the predicted luminosities and mass-loss rates is much smaller than this would naively suggest; we quantify this explicitly in Appendix~\ref{app:Propagation}.

\begin{table}
\begin{center}
\caption{A list of the parameters of the fitted curves in Figures \ref{fig:Tau_JK_LPVs_FIT} and \ref{fig:BC_LPVs}. }
\begin{tabular}{cccc}
\hline
\hline
\multicolumn{4}{c}{$\tau_{k} = a \times (J-K) + b$}     \\ \hline
            &    & a          & b                      \\ \hline
M-type      &    &   0.51 $\pm$ 0.01         &   $-$0.47 $\pm$ 0.04                    \\ \hline
C-type      &    &   0.44 $\pm$ 0.05         &   $-$0.60 $\pm$ 0.11                     \\ \hline
\hline
\multicolumn{4}{c}{{$BC_{K} = A \times (J-K)^{2} + B \times (J-K)+ C$}} \\ \hline
                 & A          & B          & C          \\ \hline
M-type           &     $-$7.11 $\pm$ 4.03       &   16.93 $\pm$ 9.15         &    $-$7.37 $\pm$ 5.17        \\ \hline
C-type           &    $-$0.14 $\pm$ 0.01        &   0.67 $\pm$ 0.47         &    2.32 $\pm$ 0.55        \\ \hline
\hline
\end{tabular}
\label{tab:Fits_LPVs}
\end{center}
\end{table}

Figure~\ref{fig:Tau_JK_LPVs_FIT} shows that Mira-type LPVs generally have higher optical depths and redder colors than typical AGB stars. This behavior reflects the fact that LPVs represent the final evolutionary stages of the AGB phase. As AGB stars evolve toward the tip of the AGB, they are expected to show similar increases in color and optical depth. Although optical depths derived from LPVs may overestimate those of the broader AGB population, this effect is partly mitigated because the \citet{Hirschauer20} catalogue contains relatively few extremely red sources. This likely reflects incompleteness at the reddest colours, where heavily obscured AGB stars become difficult to detect at the distance of NGC~6822 \citep{vanLoon97}. The total AGB mass-loss rate inferred from the LPV method is $(1.6 \pm 0.6) \times 10^{-3} \msun \mathrm{yr}^{-1}$. This uncertainty already incorporates, in quadrature, the calibration uncertainty propagated from the BC$_{K}$ fit covariance (
$\sigma \approx 7.5 \times 10^{-7} \msun \mathrm{yr}^{-1}$; Appendix~\ref{app:Propagation}), which is negligible compared to the photometric and systematic contributions and does not change the quoted value.

\subsubsection{Comparison with DESK}
\label{Sec:DESK_comparison}

We also estimated the total AGB mass-loss rate using the \texttt{DESK} code, a Python-based SED fitting package designed for dusty evolved stars \citep{Goldman20}. \texttt{DESK} fits multi-wavelength photometry or spectra to grids of radiative transfer models, enabling a quantitative comparison of different stellar samples and model assumptions. In this work, we adopted the \texttt{silicates} grid for oxygen-rich stars and the \texttt{amorphous-carbon} grid for carbon-rich stars. We note that the  \texttt{DESK} output used in this study does not provide a reduced $\chi^2$ statistic; therefore, a direct comparison of the fitting quality based on  $\chi^2$ is not possible.

As shown in Fig.~\ref{fig:L_MDOT2x2}, the \texttt{DESK} results exhibit reduced dispersion and occupy a narrower region of the luminosity–mass-loss plane (see also Fig.~\ref{fig:optical_depth_2x2} for the relation between optical depth and near-infrared color). This behavior arises from the lower resolution of the \texttt{DESK} models compared to our grids. Consequently, multiple stars with similar properties are assigned identical mass-loss rates. The total mass-loss rates derived from the three different methods are summarized in Table~\ref{tab:mass-loss_vals}.

For oxygen-rich stars, the mass-loss rates derived with \texttt{DESK} are generally consistent with those obtained from \texttt{SEDust}. In contrast, for the C-rich sources, \texttt{DESK} models generally reproduce the observed photometry less accurately and tend to predict higher luminosities than the corresponding \texttt{SEDust} fits (Fig.~\ref{fig:DESK2x2}). This difference is likely related to the substantially coarser parameter grid adopted by \texttt{DESK}. For the present observations, all best-fitting C-rich \texttt{DESK} models converge to the $\tau_K=0.1$, although $\tau_K=0.1$ is not imposed as a fixed value for all C-rich models in the \texttt{DESK} grid. At this grid resolution, the limited sampling of optical depth reduces the flexibility of the best-fitting models to reproduce the observed SEDs. Consequently, the fitting procedure can compensate for the limited optical-depth sampling by selecting a higher stellar luminosity to improve the overall agreement with the observed photometry. This provides a plausible explanation for the systematically higher luminosities obtained for the C-rich sources with \texttt{DESK}.

The source-by-source comparison demonstrates that this difference is not restricted to a small number of extreme sources. The minimum mass-loss rate among the \texttt{DESK} C-rich sources is $1.8\times10^{-6} \msun{\rm yr}^{-1}$, while 83.8\% of the matched C-rich sources have lower mass-loss rates in the \texttt{SEDust} analysis than this minimum \texttt{DESK} value. We also verified that the relevant IRAC photometry is not affected by saturation and that the chemical classifications used for the \texttt{SEDust} and \texttt{DESK} analyses are identical. Therefore, the systematic difference in the C-rich mass-loss rates is not attributable to photometric saturation or differences in chemical classification, but is primarily associated with the different model grids and parameter sampling adopted by the two methods.

\begin{table}
\caption{Total mass-loss rates derived using three different methods (SEDust, LPVs, and DESK), separated into M-type and C-type sources. Values are given in units of  $10^{-4}[\msun \mathrm{yr}^{-1}]$.}
\begin{tabular}{llll}
\hline
\hline
                            & \multicolumn{1}{l}{ SEDust} & \multicolumn{1}{l}{LPVs} & DESK \\ \hline
\multicolumn{1}{l|}{M-type }  & 27.1 $\pm$ 11.3                  & 14.6 $\pm$ 6.1                         & 12.4 $\pm$ 5.2    \\ \hline
\multicolumn{1}{l|}{C-type }  & 9.8 $\pm$  2.7                    & 1.6 $\pm$ 0.4                        & 76.3 $\pm$  21.1   \\ \hline
\multicolumn{1}{l|}{Total }   & 37.0 $\pm$ 11.6                    & 16.2 $\pm$ 6.1                       & 88.7 $\pm$ 21.7    \\ \hline
\hline
\end{tabular}
\label{tab:mass-loss_vals}
\end{table}

\subsubsection{Comparison of the Three Methods}
\label{Sec:ThreeMethodComparison}

The total mass-loss rates derived from \texttt{SEDust}, the LPV method, and \texttt{DESK} are $(3.70\pm1.16)\times10^{-3}$, $(1.62\pm0.61)\times10^{-3}$, and $(8.87\pm2.17)\times10^{-3}\ M_\odot\ {\rm yr}^{-1}$, respectively (Table~\ref{tab:mass-loss_vals}). Thus, the LPV and \texttt{DESK} estimates differ from \texttt{SEDust} by factors of approximately 2.3 and 2.4, respectively. These differences are dominated by the C-type population, for which the integrated rates are $1.6\times10^{-4}$, $9.8\times10^{-4}$, and $7.63\times10^{-3}\,M_\odot\,{\rm yr}^{-1}$ for the LPV, \texttt{SEDust}, and \texttt{DESK} methods, respectively, a difference of nearly an order of magnitude between \texttt{DESK} and \texttt{SEDust}. For the oxygen-rich population, by contrast, the integrated rates are $2.71$, $1.46$, and $1.24\times10^{-3}\,M_\odot\,{\rm yr}^{-1}$ for \texttt{SEDust}, the LPV method, and \texttt{DESK}, respectively, showing much closer agreement.

We do not interpret these differences as being encompassed by the formal statistical uncertainties of the \texttt{SEDust} fits; instead, we regard them as evidence for significant method-dependent systematic uncertainties. The \texttt{DESK} fits to C-type stars tend to yield higher luminosities and reproduce the observed photometry less accurately than the corresponding \texttt{SEDust} fits (Sect.~\ref{Sec:DESK_comparison}), whereas the LPV method relies on an empirical bolometric-correction relation calibrated using a relatively small sample (Sect.~\ref{Sec:LPV_Method}; Appendix~\ref{app:Propagation}). These methodological differences can produce substantially different integrated mass-loss rates even when applied to broadly the same evolved-star population. We retain \texttt{SEDust} as our primary estimate because it provides a homogeneous analysis of the full sample using the same radiative-transfer framework throughout, while the LPV and \texttt{DESK} results are used below to characterize the sensitivity of the integrated mass-loss budget to alternative methodologies.

\subsection{Contribution of Red Supergiants}
\label{Sec:RSG}

The present analysis is restricted to AGB and LPV stars and does not include red supergiants, which can also drive substantial mass-loss and contribute dust to the ISM \citep{vanLoon25, Christodoulou25, deWit25}. Although \citet{Hirschauer20} report 1292 RSG candidates in NGC~6822, this sample has an estimated contamination rate of $52.93\%$ as reported by the authors. In the present work, we specifically focus on the mass-loss and dust production associated with AGB and LPV stars, and therefore do not include RSGs in the analysis. Their potential contribution to the overall mass and dust budget of NGC~6822 is considered here to assess the extent to which their exclusion may affect our results.

To estimate the plausible scale of the RSG contribution, we compare with \citet{Antoniadis25}, who used \texttt{DUSTY} to derive individual mass-loss rates for 396 RSG candidates in NGC~6822. They find a mean mass-loss rate of $\dot{M}\approx1.5\times10^{-7} \msun \mathrm{yr}^{-1}$ per source, corresponding to an integrated RSG mass-loss rate of $\sim 5.9 \times10^{-5} \msun \mathrm{yr}^{-1}$ for their NGC~6822 sample. For comparison, scaling this mean rate to the full 1292 RSG candidates reported by \citet{Hirschauer20} gives $\sim 1.9 \times10^{-4} \msun \mathrm{yr}^{-1}$. This latter value should be regarded only as a scaling estimate because the \citet{Hirschauer20} sample has substantial contamination and the individual mass-loss rates of these candidates are not available.

Both estimates are substantially below the total AGB mass-loss rate derived in this work, $(3.7\pm1.1)\times10^{-3} \msun \mathrm{yr}^{-1}$ (Sect.~\ref{sec:mdot}). The integrated RSG mass-loss rate from \citet{Antoniadis25} corresponds to approximately $1.6\%$ of the AGB rate, while the scaled value based on the full \citet{Hirschauer20} candidate sample corresponds to approximately $5.1\%$. We therefore conclude that RSGs are a subdominant, though not entirely negligible, contributor to the present-day mass and dust budget of NGC~6822. Their exclusion does not materially affect the conclusions drawn in Sect.~\ref{Sec:Dust} and \ref{Sec:Specific_mass_return_rates} regarding the AGB-only mass-return and dust-injection rates. A dedicated extension of \texttt{SEDust}, or a comparable RSG-specific method, applied to a vetted RSG catalogue of NGC~6822 would nevertheless provide a useful complement to the present AGB/LPV census.

\subsection{Dust}
\label{Sec:Dust}

As described in Sect.~\ref{sec:mdot}, the mass-loss rates of individual AGB stars were derived by fitting their spectral energy distributions with the \texttt{DUSTY} radiative transfer code. For the adopted \texttt{DUSTY} wind solution, the corresponding dust-injection rate is obtained by dividing the model mass-loss rate by the adopted gas-to-dust  mass ratio. We therefore retain $\psi=200$, consistently with the \texttt{DUSTY} models used in this work. This yields an integrated dust-injection rate of $\approx1.9\times10^{-5}\ \msun\ \mathrm{yr}^{-1}$ (see Sect.~\ref{Sec:Specific_mass_return_rates} for the corresponding total, gas-inclusive mass-loss rate and its dependence on the assumed gas-to-dust ratio).

For comparison, \citet{Galametz10} estimated a total dust mass of $2.7\times10^{5}\ \msun$ in the central region of NGC~6822 from \textit{Herschel} observations. If the present-day dust-injection rate had remained constant over the past 10~Gyr, the cumulative dust mass returned by AGB stars would be approximately $1.9\times10^{5}\ \msun$.

This estimate assumes a constant AGB dust-injection rate, equal to its present-day value, combined with a formally infinite dust-survival (destruction) timescale. In general, the AGB dust-injection rate at any epoch $t$ follows the convolution:
\begin{equation}
\dot{M}_{\rm d}^{\rm AGB}(t) = \int_{M_{\rm min}}^{M_{\rm max}}
\mathrm{SFR}\!\left[t-\tau(M)\right]\,\xi(M)\,y_{\rm d}(M,Z)\,dM,
\end{equation}
where $\mathrm{SFR}(t)$ is the star-formation rate, $\tau(M)$ is the mass-dependent evolutionary delay to the AGB, $\xi(M)$ is the stellar initial mass function, and $y_{\rm d}(M,Z)$ is the dust yield of an AGB star of initial mass $M$ and metallicity $Z$ \citep{Zhukovska13}. For a Kroupa IMF \citep{Kroupa01} and the approximate main-sequence lifetime relation $\tau_{\rm MS}\simeq10\,\mathrm{Gyr}\,(M/M_\odot)^{-2.5}$, progenitors reaching the AGB today correspond to $M\simeq1.3\,M_\odot$, while progenitors responding to enhanced star formation $\sim$2.6--3~Gyr ago correspond to $M\simeq1.6$--$1.7\,M_\odot$ \citep{Gallart96_III}. Because NGC~6822's star-formation rate is not constant, the AGB population, and hence the true dust-injection rate, must have varied over the galaxy's history; however, the magnitude and timing of this variation cannot be established from the present analysis without adopting a specific SFH, stellar-lifetime prescription, and mass- and metallicity-dependent dust-yield model.

Neglect of grain destruction via supernova shocks \citep[e.g.,][]{Zhukovska08,Nanni20} is a separate, independent simplification: it means our estimate of the dust mass surviving to the present day is, if anything, an overestimate of the true present-day AGB-sourced dust reservoir. Since neither effect can be reliably quantified with the present pipeline, we retain the $1.9\times10^5\ M_\odot$ value only as an illustrative order-of-magnitude comparison with the observed dust mass, not as a rigorous cumulative production history or a prediction of the surviving dust mass. A fully consistent calculation would require an SFH-weighted, time-dependent chemical-evolution model incorporating the IMF, stellar lifetimes, metallicity evolution, mass-dependent AGB dust yields, and dust destruction, which is beyond the scope of the present pipeline-focused study and is left to future work.

%This value is lower than the observed dust mass. Therefore, the present dust reservoir of NGC~6822 cannot be explained by AGB stars alone, consistent with similar conclusions reached for the evolved-star dust budgets of the Magellanic Clouds \citep{Srinivasan16}.

This value is approximately 30\% lower than the observed dust mass. However, this difference is comparable to the representative uncertainties of 31\%--41\% adopted for the \texttt{SEDust} mass-loss estimates (Sect.~\ref{sec:uncertainties}) and is well within the spread among the integrated mass-loss rates obtained using the three independent methods (Sect.~\ref{Sec:ThreeMethodComparison}). Therefore, this difference does not establish a significant deficit in the AGB contribution to the present dust reservoir. Nevertheless, other dust production mechanisms may also contribute to the observed dust mass.

In particular, additional dust growth within the interstellar medium, especially in dense molecular clouds, is likely required to explain the observed dust mass \citep{Zhukovska08}. However, the efficiency of this process remains uncertain and may depend on the adopted grain-growth timescales and the physical conditions of the ISM \citep{Nanni20}. Supernovae have also been proposed as potential dust producers. However, their contribution in dwarf galaxies remains uncertain, since a significant fraction of the dust and metals they eject may escape the shallow gravitational potential of these systems rather than being retained in the ISM. Wolf--Rayet stars may also contribute to the dust budget, although their overall contribution is expected to be small.

\subsection{Specific mass return rates}
\label{Sec:Specific_mass_return_rates}

The dust-injection rate derived in Sect.~\ref{Sec:Dust} corresponds to a total (gas+dust) AGB mass-loss rate of $\dot{M}_{\rm total}=\psi\,\dot{M}_{\rm d}$. Adopting $\psi=200$, consistent with the \texttt{DUSTY} model assumptions, gives $\dot{M}_{\rm total}\approx 3.7\times10^{-3}\ \msun\ \mathrm{yr}^{-1}$. \citet{Galametz10} independently derived $\psi=186$ for the diffuse ISM of NGC~6822, which would give $\dot{M}_{\rm total}\approx3.5\times10^{-3}\ \msun\,\mathrm{yr}^{-1}$; since this describes the general ISM rather than AGB circumstellar envelopes, we retain $\psi=200$ for consistency with the \texttt{DUSTY} models below. The gas-to-dust ratio is also known to depend on metallicity: \citet{RemyRuyer14} found a significantly steeper increase toward lower metallicity below $12+\log({\rm O/H})\sim8.0$--8.1, and applying their relation at NGC~6822's oxygen abundance of $8.14\pm0.08$ \citep{DuartePuertas24} gives $\psi\sim10^3$, which would instead raise the total rate to $\dot{M}_{\rm total}\approx1.9\times10^{-2}\ \msun\,\mathrm{yr}^{-1}$. However, the \citet{RemyRuyer14} relation describes the global ISM gas-to-dust ratio of galaxies and does not necessarily apply to the circumstellar envelopes of individual AGB stars, so we adopt $\psi=200$ as our fiducial value below and note this ISM-based estimate as an upper-range alternative.

\citet{Gallart96_III} estimated the total stellar mass of NGC~6822 to be $5.0\times10^{7}\ \msun$. Using $\dot{M}_{\rm total}=3.7\times10^{-3}\ \msun\,\mathrm{yr}^{-1}$, we derive a specific mass-return rate of $7.4\times10^{-11}\ \mathrm{yr}^{-1}$, defined as the total mass-return rate divided by the total stellar mass. The inverse of this quantity corresponds to a characteristic timescale of $1.4\times10^{10}\ \mathrm{yr}$. This timescale exceeds the epoch of the first major star-formation peak in NGC~6822, approximately $5.0\times10^{9}\ \mathrm{yr}$ ago \citep{Galametz10}.

Since the present-day mass-return rate is not expected to represent the long-term average, as it depends on the star-formation history of the galaxy, a more direct comparison is with the current star-formation rate. \citet{Efremova11} derived a recent star-formation rate of $\mathrm{SFR}\approx1.0\times10^{-2}\ \msun\ \mathrm{yr}^{-1}$ for NGC~6822 from ultraviolet observations. The current AGB mass-return rate therefore corresponds to $\dot{M}_{\rm AGB}/\mathrm{SFR}\approx0.37$, indicating that AGB stars presently return gas at only about one-third of the current star-formation rate. This suggests that mass returned by AGB stars alone is insufficient to sustain the current level of star formation. Additional gas must therefore be supplied through other channels, such as mass return from massive stars or external gas accretion. We note that under the higher, metallicity-based estimate of $\psi\sim10^3$, this ratio would instead approach $\dot{M}_{\rm AGB}/\mathrm{SFR}\approx1.9$, i.e. AGB mass return could rival or exceed the current SFR; resolving this depends on which gas-to-dust ratio is appropriate for AGB envelopes in a low-metallicity dwarf like NGC~6822.

\subsection{Comparison with JWST Results}
\label{Sec:JWST_comparison}

The JWST photometry used in the present work was selected from the catalog of \citet{Nally24} following the quality-control procedure described in Appendix~\ref{app:JWST}, and the selected measurements were incorporated into our final photometric catalog. In contrast, the comparison presented here is with the independent analysis of \citet{Nally26}, which is based on a new PSF-matched \textit{JWST} NIRCam/MIRI catalog. Thus, the JWST photometry used in our analysis and that used by \citet{Nally26} are drawn from different catalogs and processed independently. Both our study and \citet{Nally26} derive dust-production rates through radiative-transfer modeling, but the two studies differ substantially in their photometric data sets, spatial coverage, and fitting methodologies.

The most important difference lies in the stellar sample. The analysis of \citet{Nally26} is based on the \textit{JWST} footprint covering approximately $29~\mathrm{arcmin}^{2}$ in the central bar of NGC~6822, where 610 AGB stars were identified. The derived dust-production rate was subsequently extrapolated to the full galaxy using the larger catalog of \citet{Hirschauer20}. In contrast, our analysis was performed directly on the full \citet{Hirschauer20} catalog, covering approximately $3~\mathrm{deg}^{2}$ and containing 1610 AGB stars (1050 O-rich and 560 C-rich), together with the LPV catalog of \citet{Whitelock13}. Consequently, our global mass-loss estimate does not rely on scaling from a limited spatial region. If the dust-production efficiency varies with galactocentric radius, as suggested by the different spatial distributions of evolved stars and star-forming regions, the two approaches do not need to yield identical global dust-production rates.

A second difference concerns the adopted radiative-transfer models. \citet{Nally26} employed the GRAMS model grid \citep{Sargent11,Srinivasan11}, whereas the present work is based on the \texttt{DUSTY} radiative-transfer code \citep{Ivezic97}. Although both approaches solve the radiative transfer through dusty circumstellar envelopes, they differ in the adopted dust optical properties, model grids, and parameter sampling. Such differences directly affect the conversion between the observed infrared SED, optical depth, and the inferred mass-loss rate, and systematic offsets between the two approaches are therefore expected even when identical photometry is analyzed.

The two studies also differ in their fitting methodology. \citet{Nally26} assign each source a probability of belonging to the carbon-rich or oxygen-rich population by considering the likelihoods of the full ensemble of GRAMS models, whereas \texttt{SEDust} adopts the best-fitting model from independently constructed O-rich and C-rich grids through a minimum-$\chi^{2}$ procedure. The probabilistic approach naturally accounts for cases in which the two chemical classifications produce comparably good fits, while the minimum-$\chi^{2}$ solution assigns a unique chemical type. This difference may contribute to the relative fractions of carbon-rich and oxygen-rich stars recovered by the two analyses.

Quantitatively, \citet{Nally26} derived a global dust-production rate of $1.9\times10^{-6}\ \msun\ \mathrm{yr}^{-1}$, whereas the present work finds a dust-injection rate of $1.9\times10^{-5}\ \msun\ \mathrm{yr}^{-1}$, approximately one order of magnitude higher. Given the substantial differences in spatial coverage, stellar samples, radiative-transfer models, and fitting methodology, such a discrepancy is not unexpected, and it is presently difficult to identify the dominant contributor to the offset.

\citet{Nally26} report that $N=458$ of the 1610 sources in the \citet{Hirschauer20} AGB catalog fall within their combined NIRCam/MIRI footprint ($\approx28\%$ sample coverage), which they use to scale their field-limited AGB dust-production rate ($5.3\times10^{-7}\ \msun\,\mathrm{yr}^{-1}$) to the global estimate of $1.9\times10^{-6}\ \msun\,\mathrm{yr}^{-1}$ quoted above. A direct, source-by-source comparison of individual dust-production rates for the overlapping sample -- once Tables 4 and 6 of \citet{Nally26} become available via VizieR -- would help determine whether the remaining order-of-magnitude offset reflects a systematic effect (e.g., dust opacity or model-grid differences) or is driven by a subset of discrepant sources.

Despite this quantitative difference, the two studies reach a similar qualitative conclusion. In the present work, the integrated mass-loss rate is dominated by M-type stars, which contribute approximately $73\%$ of the total, while C-type stars contribute the remaining $27\%$ (Table~\ref{tab:mass-loss_vals}). Likewise, \citet{Nally26} find that oxygen-rich stars dominate the present-day dust production in NGC~6822. 
%The agreement between two independent analyses, based on different photometric catalogs and radiative-transfer implementations, strengthens the conclusion that oxygen-rich stars currently dominate the dust return in NGC~6822, despite its relatively low metallicity.
Although the relative contribution of each chemical type is influenced by the underlying stellar population, the agreement between two independent analyses, based on different photometric catalogs and radiative-transfer implementations, supports the conclusion that oxygen-rich stars currently dominate the integrated dust return in NGC~6822.

\subsection{Spatial Distribution}
\label{Sec:Spatial_Distribution}

\begin{figure}
\begin{center}
\includegraphics[width=0.45\textwidth]{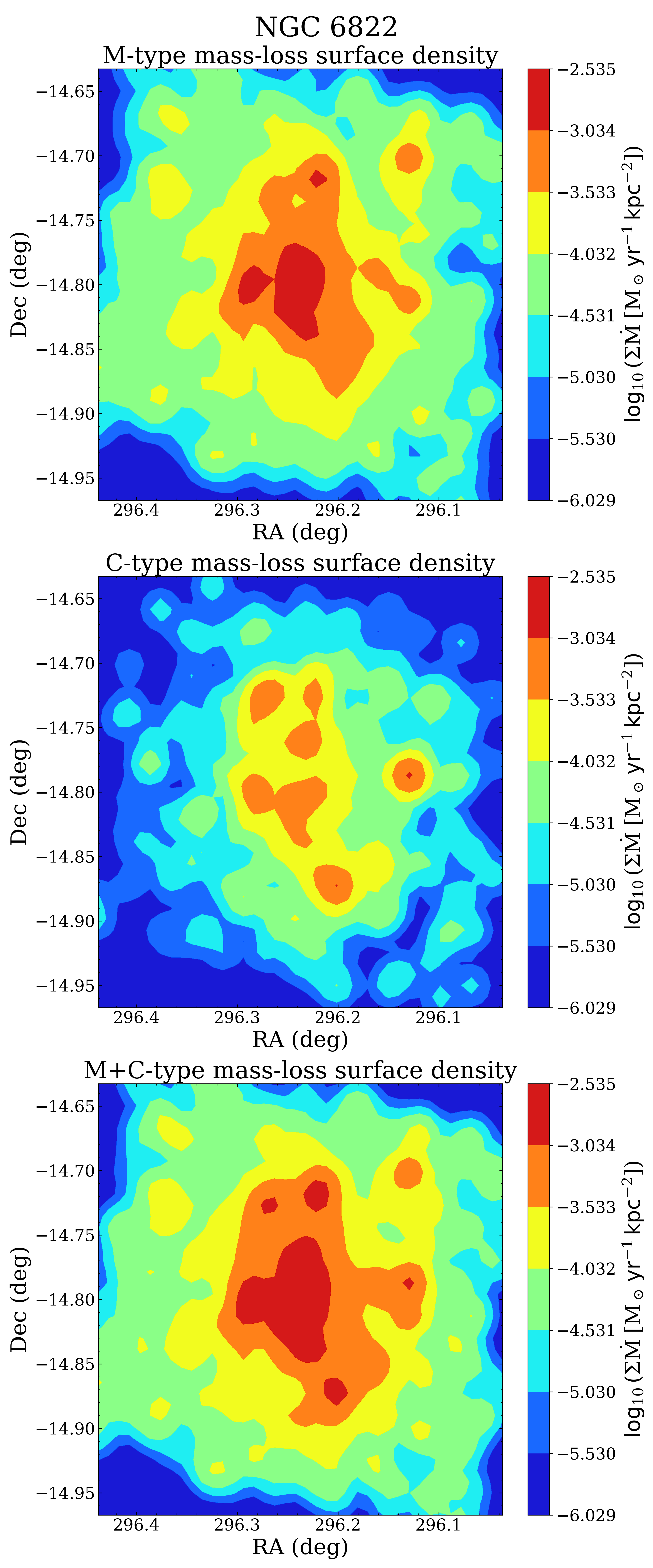}
\caption{Surface density maps of the mass-loss rate for carbon-rich (C-type), oxygen-rich (M-type), and combined (M+C-type) stars in NGC~6822.} 
\label{fig:MassLoss_Density}
\end{center}
\end{figure}

The spatial distribution of AGB stars in NGC~6822 is shown in Fig.~\ref{fig:MassLoss_Density}. Two-dimensional maps of the mass-loss surface density are presented for M-type stars, C-type stars, and the combined population. These maps were constructed by binning stellar positions into a regular grid in right ascension and declination and weighting each bin by the stellar mass-loss rate. The centrally concentrated AGB mass-loss distribution is broadly consistent with the concentration of recent star formation reported by \citet{Kinson21}.
% However, the distribution extends beyond the central bar and declines gradually with galactocentric radius, reflecting the smoother, approximately spheroidal distribution of AGB stars reported by \citet{Kinson21,Letarte02,Hirschauer20}.

The mass-loss from both chemical types is concentrated toward the central regions of the galaxy. The C-type population appears to be slightly more centrally concentrated than the M-type population, with median galactocentric radii of 0.63 and 0.68~kpc, respectively. A two-sample Kolmogorov--Smirnov test indicates that the two radial distributions differ statistically (KS = 0.100, $p = 0.0012$), although the effect is modest.

\begin{figure}
\begin{center}
\includegraphics[width=0.45\textwidth]{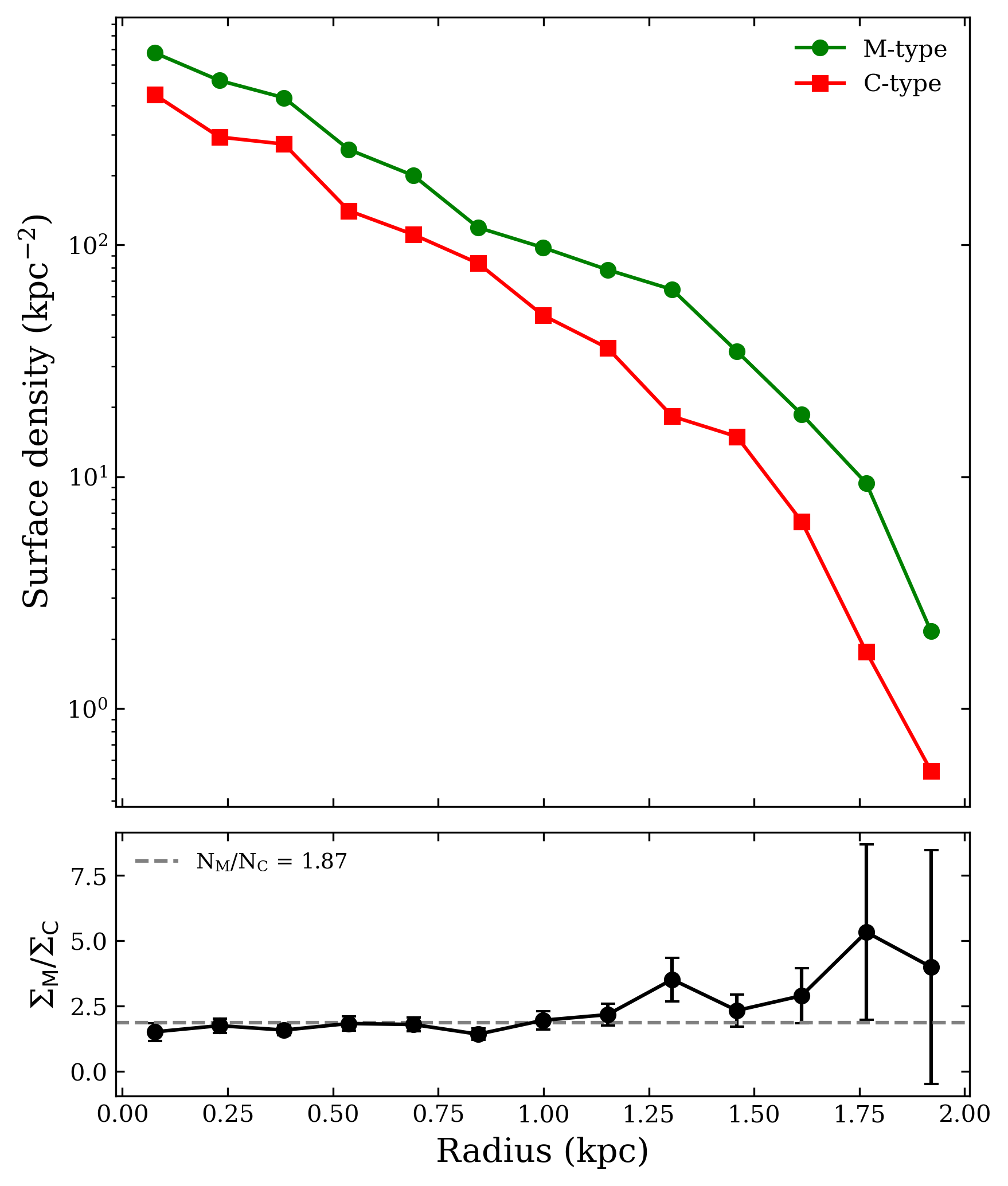}
\caption{Radial surface-density profiles of M-type (green circles) and C-type (red squares) evolved stars in NGC~6822, together with their ratio (lower panel). The surface density declines smoothly with radius for both populations. The dashed line in the lower panel marks the galaxy-wide ratio $N_{\rm M}/N_{\rm C} \approx 1.87$ expected if the two populations shared an identical radial distribution; the rise of $\Sigma_{\rm M}/\Sigma_{\rm C}$ above this line at $r \gtrsim 1$~kpc reflects the modestly more compact spatial distribution of C-type stars relative to M-type stars (Sect.~\ref{app:spatial_metrics}).}
\label{fig:radial_profiles_BL}
\end{center}
\end{figure}

To further quantify the spatial distributions, radial surface-density profiles were constructed following the procedure described in Appendix~\ref{app:spatial_metrics}. As shown in the upper panel of Fig.~\ref{fig:radial_profiles_BL}, the surface densities of both populations decline smoothly with increasing radius and show no evidence for significant spatial substructure. The lower panel presents the ratio of the M-type to C-type surface densities. This ratio remains approximately constant within the central $\sim1$~kpc, but increases at larger radii, indicating a relative excess of M-type stars in the outer regions. This behavior is consistent with the C-type population being more centrally concentrated than the M-type population. However, this result should be interpreted with caution. The C-type sample (560 stars) is substantially smaller than the M-type sample (1050 stars), resulting in larger statistical uncertainties at large galactocentric radii. In addition, the AGB catalog is likely incomplete at the reddest colors, where the most heavily dust-enshrouded carbon stars are expected to reside \citep{vanLoon97}. Such selection effects may influence the inferred radial distributions and the observed surface-density ratio.

\section{Conclusions}
\label{sec:Conclusions}

We introduced \texttt{SEDust}, a dedicated and automated spectral energy distribution fitting pipeline based on radiative transfer modeling, designed to derive stellar and circumstellar properties of evolved stars. The pipeline was applied to AGB stars in the nearby dwarf irregular galaxy NGC~6822, enabling a homogeneous determination of stellar luminosities, mass-loss rates, and dust production properties.

Using \texttt{SEDust}, we estimated a total stellar mass-loss rate of $(3.7 \pm 1.1) \times 10^{-3}\msun \mathrm{yr}^{-1}$ from the AGB population in NGC~6822. This result provides a quantitative assessment of the contribution of evolved stars to the recycling of gas and dust in a low-metallicity environment. Comparisons between the derived dust injection rates and the dust masses inferred from \textit{Herschel} observations indicate that the present-day AGB dust-injection rate is insufficient to account for the observed dust reservoir of the galaxy. This suggests that additional sources, such as dust growth in the interstellar medium and external gas accretion, are likely to play an important role in the evolution of the interstellar medium of NGC~6822.

An analysis of the spatial distribution of AGB stars reveals no evidence for fragmentation or asymmetry in the associated gas and dust structures. The smooth spatial distribution of the AGB population does not reveal an obvious signature of recent tidal disturbance. Together, these results highlight the importance of multiple stellar feedback channels in shaping the interstellar medium of dwarf galaxies and demonstrate the utility of \texttt{SEDust} for studying mass-loss and dust production in evolved stellar populations.

\section*{Acknowledgments}
H.M. acknowledges the High-Performance Computing (HPC) and Turin Cloud Services of the Institute for Research in Fundamental Sciences (IPM) for providing the computational resources used to execute the codes in this work.

%@@@@@@@@@@@@@@@@@@@@@@@@@@@@@@@@@@@@@@@@@@@@@@@@@@@@@@@@@@@@@@@@@@@@@@@@@@@@@@@@@@@@@@@@@@

\bibliography{NGC6822}{}

@book{Habing13,
  title={Asymptotic giant branch stars},
  author={Habing, Harm J and Olofsson, Hans},
  year={2013},
  publisher={Springer Science \& Business Media}
}

@ARTICLE{Hofner18,
       author = {{H{\"o}fner}, Susanne and {Olofsson}, Hans},
        title = "{Mass loss of stars on the asymptotic giant branch. Mechanisms, models and measurements}",
      journal = {\aapr},
         year = 2018,
        month = jan,
       volume = {26},
       number = {1},
          eid = {1},
        pages = {1},
          doi = {10.1007/s00159-017-0106-5},
       adsurl = {https://ui.adsabs.harvard.edu/abs/2018A&ARv..26....1H}
}

@article{lee1993tip,
  title={The tip of the red giant branch as a distance indicator for resolved galaxies},
  author={Lee, Myung Gyoon and Freedman, Wendy L and Madore, Barry F},
  journal={Astrophysical Journal v. 417, p. 553},
  volume={417},
  pages={553},
  year={1993}
}

@article{hodge1991cosmos,
  title={A COSMOS study of the structure and content of NGC 6822},
  author={Hodge, Paul and Smith, Toby and Eskridge, Paul and MacGillivray, Harvey and Beard, Steven},
  journal={Astrophysical Journal, Part 1 (ISSN 0004-637X), vol. 379, Oct. 1, 1991, p. 621-630.},
  volume={379},
  pages={621--630},
  year={1991}
}

@article{de2000evidence,
  title={Evidence for tidal interaction and a supergiant hi shell in the local group dwarf galaxy ngc 6822},
  author={De Blok, WJG and Walter, F},
  journal={The Astrophysical Journal},
  volume={537},
  number={2},
  pages={L95},
  year={2000},
  publisher={IOP Publishing}
}

@article{davidge2003metallicity,
  title={The Metallicity of the Red Giant Branch in the Disk of NGC 6822},
  author={Davidge, TJ},
  journal={Publications of the Astronomical Society of the Pacific},
  volume={115},
  number={808},
  pages={635},
  year={2003},
  publisher={IOP Publishing}
}

@article{gallart1996local,
  title={The Local Group dwarf irregular galaxy NGC 6822. I. The stellar content},
  author={Gallart, C and Aparicio, A and Vilchez, JM},
  journal={The Astronomical Journal},
  volume={112},
  pages={1928},
  year={1996}
}

@article{wyder2001star,
  title={The star formation history of NGC 6822},
  author={Wyder, Ted K},
  journal={The Astronomical Journal},
  volume={122},
  number={5},
  pages={2490},
  year={2001},
  publisher={IOP Publishing}
}

@ARTICLE{Javadi13_III,
       author = {{Javadi}, Atefeh and {van Loon}, Jacco Th. and {Khosroshahi}, Habib and {Mirtorabi}, Mohammad Taghi},
        title = "{The UK Infrared Telescope M33 monitoring project - III. Feedback from dusty stellar winds in the central square kiloparsec}",
      journal = {\mnras},
         year = 2013,
        month = jul,
       volume = {432},
       number = {4},
        pages = {2824-2836},
          doi = {10.1093/mnras/stt640},
archivePrefix = {arXiv},
       eprint = {1304.3782},
 primaryClass = {astro-ph.SR},
       adsurl = {https://ui.adsabs.harvard.edu/abs/2013MNRAS.432.2824J}
}

@ARTICLE{Hirschauer20,
       author = {{Hirschauer}, Alec S. and {Gray}, Laurin and {Meixner}, Margaret and {Jones}, Olivia C. and {Srinivasan}, Sundar and {Boyer}, Martha L. and {Sargent}, B.~A.},
        title = "{Dusty Stellar Birth and Death in the Metal-poor Galaxy NGC 6822}",
      journal = {\apj},
         year = 2020,
        month = apr,
       volume = {892},
       number = {2},
          eid = {91},
        pages = {91},
          doi = {10.3847/1538-4357/ab7b60},
archivePrefix = {arXiv},
       eprint = {2002.12437},
 primaryClass = {astro-ph.GA},
       adsurl = {https://ui.adsabs.harvard.edu/abs/2020ApJ...892...91H}
}

@ARTICLE{Nally24,
       author = {{Nally}, Conor and {Jones}, Olivia C. and {Lenki{\'c}}, Laura and {Habel}, Nolan and {Hirschauer}, Alec S. and {Meixner}, Margaret and {Kavanagh}, P.~J. and {Boyer}, Martha L. and {Ferguson}, Annette M.~N. and {Sargent}, B.~A. and {Nayak}, Omnarayani and {Temim}, Tea},
        title = "{JWST MIRI and NIRCam unveil previously unseen infrared stellar populations in NGC 6822}",
      journal = {\mnras},
         year = 2024,
        month = jun,
       volume = {531},
       number = {1},
        pages = {183-198},
          doi = {10.1093/mnras/stae1163},
archivePrefix = {arXiv},
       eprint = {2309.13521},
 primaryClass = {astro-ph.GA},
       adsurl = {https://ui.adsabs.harvard.edu/abs/2024MNRAS.531..183N}
}

@ARTICLE{Mowlavi99,
       author = {{Mowlavi}, N.},
        title = "{On the third dredge-up phenomenon in asymptotic giant branch stars}",
      journal = {\aap},
         year = 1999,
        month = apr,
       volume = {344},
        pages = {617-631},
          doi = {10.48550/arXiv.astro-ph/9903473},
archivePrefix = {arXiv},
       eprint = {astro-ph/9903473},
 primaryClass = {astro-ph},
       adsurl = {https://ui.adsabs.harvard.edu/abs/1999A&A...344..617M}
}

@ARTICLE{Uttenthaler19,
       author = {{Uttenthaler}, S. and {McDonald}, I. and {Bernhard}, K. and {Cristallo}, S. and {Gobrecht}, D.},
        title = "{Interplay between pulsation, mass loss, and third dredge-up: More about Miras with and without technetium}",
      journal = {\aap},
         year = 2019,
        month = feb,
       volume = {622},
          eid = {A120},
        pages = {A120},
          doi = {10.1051/0004-6361/201833794},
archivePrefix = {arXiv},
       eprint = {1812.07434},
 primaryClass = {astro-ph.SR},
       adsurl = {https://ui.adsabs.harvard.edu/abs/2019A&A...622A.120U}
}

@ARTICLE{Nowotny03,
       author = {{Nowotny}, W. and {Kerschbaum}, F. and {Olofsson}, H. and {Schwarz}, H.~E.},
        title = "{A census of AGB stars in Local Group galaxies.  II. NGC 185 and NGC 147}",
      journal = {\aap},
         year = 2003,
        month = may,
       volume = {403},
        pages = {93-103},
          doi = {10.1051/0004-6361:20030282},
       adsurl = {https://ui.adsabs.harvard.edu/abs/2003A&A...403...93N}
}

@ARTICLE{Ferrarotti06,
       author = {{Ferrarotti}, A.~S. and {Gail}, H. -P.},
        title = "{Composition and quantities of dust produced by AGB-stars and returned to the interstellar medium}",
      journal = {\aap},
         year = 2006,
        month = feb,
       volume = {447},
       number = {2},
        pages = {553-576},
          doi = {10.1051/0004-6361:20041198},
       adsurl = {https://ui.adsabs.harvard.edu/abs/2006A&A...447..553F}
}

@ARTICLE{Ivezic97,
       author = {{Ivezic}, Zeljko and {Elitzur}, Moshe},
        title = "{Self-similarity and scaling behaviour of infrared emission from radiatively heated dust - I. Theory}",
      journal = {\mnras},
         year = 1997,
        month = jun,
       volume = {287},
       number = {4},
        pages = {799-811},
          doi = {10.1093/mnras/287.4.799},
archivePrefix = {arXiv},
       eprint = {astro-ph/9612164},
 primaryClass = {astro-ph},
       adsurl = {https://ui.adsabs.harvard.edu/abs/1997MNRAS.287..799I}
}

@MISC{Hanner88,
       author = {{Hanner}, Martha},
        title = "{Grain optical properties.}",
 howpublished = {In NASA, Washington, Infrared Observations of Comets Halley and Wilson and Properties of the Grains p 22-49 (SEE N89-13330 04-89)},
         year = 1988,
        month = sep,
        pages = {22-49},
       adsurl = {https://ui.adsabs.harvard.edu/abs/1988ioch.rept...22H}
}

@ARTICLE{Pegourie88,
       author = {{Pegourie}, B.},
        title = "{Optical properties of alpha silicon carbide.}",
      journal = {\aap},
         year = 1988,
        month = apr,
       volume = {194},
        pages = {335-339},
       adsurl = {https://ui.adsabs.harvard.edu/abs/1988A&A...194..335P}
}

@ARTICLE{Draine84,
       author = {{Draine}, B.~T. and {Lee}, H.~M.},
        title = "{Optical Properties of Interstellar Graphite and Silicate Grains}",
      journal = {\apj},
         year = 1984,
        month = oct,
       volume = {285},
        pages = {89},
          doi = {10.1086/162480},
       adsurl = {https://ui.adsabs.harvard.edu/abs/1984ApJ...285...89D}
}

@ARTICLE{Letarte02,
       author = {{Letarte}, Bruno and {Demers}, Serge and {Battinelli}, Paolo and {Kunkel}, W.~E.},
        title = "{The Extent of NGC 6822 Revealed by Its C Star Population}",
      journal = {\aj},
         year = 2002,
        month = feb,
       volume = {123},
       number = {2},
        pages = {832-839},
          doi = {10.1086/338319},
archivePrefix = {arXiv},
       eprint = {astro-ph/0110026},
 primaryClass = {astro-ph},
       adsurl = {https://ui.adsabs.harvard.edu/abs/2002AJ....123..832L}
}

@ARTICLE{vanLoon99,
       author = {{van Loon}, J. Th. and {Groenewegen}, M.~A.~T. and {de Koter}, A. and {Trams}, N.~R. and {Waters}, L.~B.~F.~M. and {Zijlstra}, A.~A. and {Whitelock}, P.~A. and {Loup}, C.},
        title = "{Mass-loss rates and luminosity functions of dust-enshrouded AGB stars and red supergiants in the LMC}",
      journal = {\aap},
         year = 1999,
        month = nov,
       volume = {351},
        pages = {559-572},
archivePrefix = {arXiv},
       eprint = {astro-ph/9909416},
 primaryClass = {astro-ph},
       adsurl = {https://ui.adsabs.harvard.edu/abs/1999A&A...351..559V}
}

@ARTICLE{Abdollahi23,
       author = {{Abdollahi}, Hedieh and {Javadi}, Atefeh and {Mirtorabi}, Mohammad Taghi and {Saremi}, Elham and {van Loon}, Jacco Th. and {Khosroshahi}, Habib G. and {McDonald}, Iain and {Khalouei}, Elahe and {Mahani}, Hamidreza and {Aghdam}, Sima Taefi and {Saberi}, Maryam and {Torki}, Maryam},
        title = "{The Isaac Newton Telescope Monitoring Survey of Local Group Dwarf Galaxies. VI. The Star Formation History and Dust Production in Andromeda IX}",
      journal = {\apj},
         year = 2023,
        month = may,
       volume = {948},
       number = {1},
          eid = {63},
        pages = {63},
          doi = {10.3847/1538-4357/acbbc9},
archivePrefix = {arXiv},
       eprint = {2302.07674},
 primaryClass = {astro-ph.GA},
       adsurl = {https://ui.adsabs.harvard.edu/abs/2023ApJ...948...63A}
}

@ARTICLE{Mahani25,
       author = {{Mahani}, Hamidreza and {Javadi}, Atefeh and {van Loon}, Jacco Th. and {Kemper}, Francisca and {Golshan}, Roya Hamedani and {McDonald}, Iain and {Khosroshahi}, Habib G. and {Abdollahi}, Hedieh and {Mahdizadeh}, Sajjad},
        title = "{Long-period Variable Stars in NGC 147 and NGC 185. II. Their Dust Production}",
      journal = {\apj},
         year = 2025,
        month = oct,
       volume = {992},
       number = {1},
          eid = {94},
        pages = {94},
          doi = {10.3847/1538-4357/adfa2b},
archivePrefix = {arXiv},
       eprint = {2508.05596},
 primaryClass = {astro-ph.GA},
       adsurl = {https://ui.adsabs.harvard.edu/abs/2025ApJ...992...94M}
}

@ARTICLE{Srinivasan09,
       author = {{Srinivasan}, Sundar and {Meixner}, Margaret and {Leitherer}, Claus and {Vijh}, Uma and {Volk}, Kevin and {Blum}, Robert D. and {Babler}, Brian L. and {Block}, Miwa and {Bracker}, Steve and {Cohen}, Martin and {Engelbracht}, Charles W. and {For}, Bi-Qing and {Gordon}, Karl D. and {Harris}, Jason and {Hora}, Joseph L. and {Indebetouw}, Remy and {Markwick-Kemper}, Francisca and {Meade}, Marilyn and {Misselt}, Karl A. and {Sewilo}, Marta and {Whitney}, Barbara},
        title = "{The Mass Loss Return from Evolved Stars to the Large Magellanic Cloud: Empirical Relations for Excess Emission at 8 and 24 {\ensuremath{\mu}}m}",
      journal = {\aj},
         year = 2009,
        month = jun,
       volume = {137},
       number = {6},
        pages = {4810-4823},
          doi = {10.1088/0004-6256/137/6/4810},
archivePrefix = {arXiv},
       eprint = {0903.1661},
 primaryClass = {astro-ph.CO},
       adsurl = {https://ui.adsabs.harvard.edu/abs/2009AJ....137.4810S}
}

@ARTICLE{Boyer11,
       author = {{Boyer}, Martha L. and {Srinivasan}, Sundar and {van Loon}, Jacco Th. and {McDonald}, Iain and {Meixner}, Margaret and {Zaritsky}, Dennis and {Gordon}, Karl D. and {Kemper}, F. and {Babler}, Brian and {Block}, Miwa and {Bracker}, Steve and {Engelbracht}, Charles W. and {Hora}, Joe and {Indebetouw}, Remy and {Meade}, Marilyn and {Misselt}, Karl and {Robitaille}, Thomas and {Sewi{\l}o}, Marta and {Shiao}, Bernie and {Whitney}, Barbara},
        title = "{Surveying the Agents of Galaxy Evolution in the Tidally Stripped, Low Metallicity Small Magellanic Cloud (SAGE-SMC). II. Cool Evolved Stars}",
      journal = {\aj},
         year = 2011,
        month = oct,
       volume = {142},
       number = {4},
          eid = {103},
        pages = {103},
          doi = {10.1088/0004-6256/142/4/103},
archivePrefix = {arXiv},
       eprint = {1106.5026},
 primaryClass = {astro-ph.SR},
       adsurl = {https://ui.adsabs.harvard.edu/abs/2011AJ....142..103B}
}

@ARTICLE{Boyer17,
       author = {{Boyer}, M.~L. and {McQuinn}, K.~B.~W. and {Groenewegen}, M.~A.~T. and {Zijlstra}, A.~A. and {Whitelock}, P.~A. and {van Loon}, J. Th. and {Sonneborn}, G. and {Sloan}, G.~C. and {Skillman}, E.~D. and {Meixner}, M. and {McDonald}, I. and {Jones}, O.~C. and {Javadi}, A. and {Gehrz}, R.~D. and {Britavskiy}, N. and {Bonanos}, A.~Z.},
        title = "{An Infrared Census of DUST in Nearby Galaxies with Spitzer (DUSTiNGS). IV. Discovery of High-redshift AGB Analogs}",
      journal = {\apj},
         year = 2017,
        month = dec,
       volume = {851},
       number = {2},
          eid = {152},
        pages = {152},
          doi = {10.3847/1538-4357/aa9892},
archivePrefix = {arXiv},
       eprint = {1711.02129},
 primaryClass = {astro-ph.SR},
       adsurl = {https://ui.adsabs.harvard.edu/abs/2017ApJ...851..152B}
}

@ARTICLE{Prager22,
       author = {{Prager}, Henry A. and {Willson}, Lee Anne and {Marengo}, Massimo and {Creech-Eakman}, Michelle J.},
        title = "{Relation of Observable Stellar Parameters to Mass-loss Rate of AGB Stars in the LMC}",
      journal = {\apj},
         year = 2022,
        month = dec,
       volume = {941},
       number = {1},
          eid = {44},
        pages = {44},
          doi = {10.3847/1538-4357/ac9e57},
archivePrefix = {arXiv},
       eprint = {2210.15039},
 primaryClass = {astro-ph.SR},
       adsurl = {https://ui.adsabs.harvard.edu/abs/2022ApJ...941...44P}
}

@ARTICLE{Karakas14,
       author = {{Karakas}, Amanda I. and {Lattanzio}, John C.},
        title = "{The Dawes Review 2: Nucleosynthesis and Stellar Yields of Low- and Intermediate-Mass Single Stars}",
      journal = {\pasa},
         year = 2014,
        month = jul,
       volume = {31},
          eid = {e030},
        pages = {e030},
          doi = {10.1017/pasa.2014.21},
archivePrefix = {arXiv},
       eprint = {1405.0062},
 primaryClass = {astro-ph.SR},
       adsurl = {https://ui.adsabs.harvard.edu/abs/2014PASA...31...30K}
}

@ARTICLE{Karakas14_II,
       author = {{Karakas}, Amanda I.},
        title = "{Helium enrichment and carbon-star production in metal-rich populations}",
      journal = {\mnras},
         year = 2014,
        month = nov,
       volume = {445},
       number = {1},
        pages = {347-358},
          doi = {10.1093/mnras/stu1727},
archivePrefix = {arXiv},
       eprint = {1408.5936},
 primaryClass = {astro-ph.SR},
       adsurl = {https://ui.adsabs.harvard.edu/abs/2014MNRAS.445..347K}
}

@BOOK{Kippenhahn90,
       author = {{Kippenhahn}, Rudolf and {Weigert}, Alfred},
        title = "{Stellar Structure and Evolution}",
         year = 1990,
       adsurl = {https://ui.adsabs.harvard.edu/abs/1990sse..book.....K},
      publisher = {Springer-Verlag}
}

@ARTICLE{Marini21,
       author = {{Marini}, E. and {Dell'Agli}, F. and {Groenewegen}, M.~A.~T. and {Garc{\'\i}a-Hern{\'a}ndez}, D.~A. and {Mattsson}, L. and {Kamath}, D. and {Ventura}, P. and {D'Antona}, F. and {Tailo}, M.},
        title = "{Understanding the evolution and dust formation of carbon stars in the Large Magellanic Cloud via the JWST}",
      journal = {\aap},
         year = 2021,
        month = mar,
       volume = {647},
          eid = {A69},
        pages = {A69},
          doi = {10.1051/0004-6361/202039613},
archivePrefix = {arXiv},
       eprint = {2012.12289},
 primaryClass = {astro-ph.SR},
       adsurl = {https://ui.adsabs.harvard.edu/abs/2021A&A...647A..69M}
}

@ARTICLE{Whitelock06,
       author = {{Whitelock}, Patricia A. and {Feast}, Michael W. and {Marang}, Freddy and {Groenewegen}, M.~A.~T.},
        title = "{Near-infrared photometry of carbon stars}",
      journal = {\mnras},
         year = 2006,
        month = jun,
       volume = {369},
       number = {2},
        pages = {751-782},
          doi = {10.1111/j.1365-2966.2006.10322.x},
archivePrefix = {arXiv},
       eprint = {astro-ph/0603504},
 primaryClass = {astro-ph},
       adsurl = {https://ui.adsabs.harvard.edu/abs/2006MNRAS.369..751W}
}

@ARTICLE{Gullieuszik12,
       author = {{Gullieuszik}, M. and {Groenewegen}, M.~A.~T. and {Cioni}, M. -R.~L. and {de Grijs}, R. and {van Loon}, J. Th. and {Girardi}, L. and {Ivanov}, V.~D. and {Oliveira}, J.~M. and {Emerson}, J.~P. and {Guandalini}, R.},
        title = "{The VMC survey. III. Mass-loss rates and luminosities of LMC AGB stars}",
      journal = {\aap},
         year = 2012,
        month = jan,
       volume = {537},
          eid = {A105},
        pages = {A105},
          doi = {10.1051/0004-6361/201117493},
archivePrefix = {arXiv},
       eprint = {1110.4497},
 primaryClass = {astro-ph.CO},
       adsurl = {https://ui.adsabs.harvard.edu/abs/2012A&A...537A.105G}
}

@ARTICLE{Jura89,
       author = {{Jura}, M. and {Kleinmann}, S.~G.},
        title = "{Dust-enshrouded Asymptotic Giant Branch Stars in the Solar Neighborhood}",
      journal = {\apj},
         year = 1989,
        month = jun,
       volume = {341},
        pages = {359},
          doi = {10.1086/167499},
       adsurl = {https://ui.adsabs.harvard.edu/abs/1989ApJ...341..359J}
}

@ARTICLE{Jura90,
       author = {{Jura}, M. and {Kleinmann}, S.~G.},
        title = "{Very Dusty Carbon-rich Asymptotic Giant Branch Stars between approximately 1 and approximately 2.5 Kiloparsecs from the Sun}",
      journal = {\apj},
         year = 1990,
        month = dec,
       volume = {364},
        pages = {663},
          doi = {10.1086/169449},
       adsurl = {https://ui.adsabs.harvard.edu/abs/1990ApJ...364..663J}
}

@ARTICLE{Bladh12,
       author = {{Bladh}, S. and {H{\"o}fner}, S.},
        title = "{Exploring wind-driving dust species in cool luminous giants. I. Basic criteria and dynamical models of M-type AGB stars}",
      journal = {\aap},
         year = 2012,
        month = oct,
       volume = {546},
          eid = {A76},
        pages = {A76},
          doi = {10.1051/0004-6361/201219138},
archivePrefix = {arXiv},
       eprint = {1209.2112},
 primaryClass = {astro-ph.SR},
       adsurl = {https://ui.adsabs.harvard.edu/abs/2012A&A...546A..76B}
}

@article{matsuura09,
  title={The global gas and dust budget of the Large Magellanic Cloud: AGB stars and supernovae, and the impact on the ISM evolution},
  author={Matsuura, M and Barlow, MJ and Zijlstra, AA and Whitelock, PA and Cioni, M-RL and Groenewegen, MAT and Volk, K and Kemper, F and Kodama, T and Lagadec, E and others},
  journal={Monthly Notices of the Royal Astronomical Society},
  volume={396},
  number={2},
  pages={918--934},
  year={2009},
  publisher={Blackwell Publishing Ltd Oxford, UK}
}

@ARTICLE{Riebel12,
       author = {{Riebel}, D. and {Srinivasan}, S. and {Sargent}, B. and {Meixner}, M.},
        title = "{The Mass-loss Return from Evolved Stars to the Large Magellanic Cloud. VI. Luminosities and Mass-loss Rates on Population Scales}",
      journal = {\apj},
         year = 2012,
        month = jul,
       volume = {753},
       number = {1},
          eid = {71},
        pages = {71},
          doi = {10.1088/0004-637X/753/1/71},
archivePrefix = {arXiv},
       eprint = {1205.0280},
 primaryClass = {astro-ph.SR},
       adsurl = {https://ui.adsabs.harvard.edu/abs/2012ApJ...753...71R}
}

@ARTICLE{Eriksson14,
       author = {{Eriksson}, K. and {Nowotny}, W. and {H{\"o}fner}, S. and {Aringer}, B. and {Wachter}, A.},
        title = "{Synthetic photometry for carbon-rich giants. IV. An extensive grid of dynamic atmosphere and wind models}",
      journal = {\aap},
         year = 2014,
        month = jun,
       volume = {566},
          eid = {A95},
        pages = {A95},
          doi = {10.1051/0004-6361/201323241},
archivePrefix = {arXiv},
       eprint = {1404.7515},
 primaryClass = {astro-ph.SR},
       adsurl = {https://ui.adsabs.harvard.edu/abs/2014A&A...566A..95E}
}

@ARTICLE{Gallart96_III,
       author = {{Gallart}, C. and {Aparicio}, A. and {Bertelli}, G. and {Chiosi}, C.},
        title = "{The Local Group Dwarf Irregular Galaxy NGC 6822.III. The Recent Star Formation History}",
      journal = {\aj},
         year = 1996,
        month = dec,
       volume = {112},
        pages = {2596},
          doi = {10.1086/118205},
       adsurl = {https://ui.adsabs.harvard.edu/abs/1996AJ....112.2596G}
}

@ARTICLE{Galametz10,
       author = {{Galametz}, M. and {Madden}, S.~C. and {Galliano}, F. and {Hony}, S. and {Sauvage}, M. and {Pohlen}, M. and {Bendo}, G.~J. and {Auld}, R. and {Baes}, M. and {Barlow}, M.~J. and {Bock}, J.~J. and {Boselli}, A. and {Bradford}, M. and {Buat}, V. and {Castro-Rodr{\'\i}guez}, N. and {Chanial}, P. and {Charlot}, S. and {Ciesla}, L. and {Clements}, D.~L. and {Cooray}, A. and {Cormier}, D. and {Cortese}, L. and {Davies}, J.~I. and {Dwek}, E. and {Eales}, S.~A. and {Elbaz}, D. and {Gear}, W.~K. and {Glenn}, J. and {Gomez}, H.~L. and {Griffin}, M. and {Isaak}, K.~G. and {Levenson}, L.~R. and {Lu}, N. and {O'Halloran}, B. and {Okumura}, K. and {Oliver}, S. and {Page}, M.~J. and {Panuzzo}, P. and {Papageorgiou}, A. and {Parkin}, T.~J. and {P{\'e}rez-Fournon}, I. and {Rangwala}, N. and {Rigby}, E.~E. and {Roussel}, H. and {Rykala}, A. and {Sacchi}, N. and {Schulz}, B. and {Schirm}, M.~R.~P. and {Smith}, M.~W.~L. and {Spinoglio}, L. and {Stevens}, J.~A. and {Sundar}, S. and {Symeonidis}, M. and {Trichas}, M. and {Vaccari}, M. and {Vigroux}, L. and {Wilson}, C.~D. and {Wozniak}, H. and {Wright}, G.~S. and {Zeilinger}, W.~W.},
        title = "{Herschel photometric observations of the nearby low metallicity irregular galaxy NGC 6822}",
      journal = {\aap},
         year = 2010,
        month = jul,
       volume = {518},
          eid = {L55},
        pages = {L55},
          doi = {10.1051/0004-6361/201014604},
archivePrefix = {arXiv},
       eprint = {1005.1730},
 primaryClass = {astro-ph.CO},
       adsurl = {https://ui.adsabs.harvard.edu/abs/2010A&A...518L..55G}
}

@ARTICLE{Whitelock13,
       author = {{Whitelock}, Patricia A. and {Menzies}, John W. and {Feast}, Michael W. and {Nsengiyumva}, Francois and {Matsunaga}, Noriyuki},
        title = "{The Local Group galaxy NGC 6822 and its asymptotic giant branch stars}",
      journal = {\mnras},
         year = 2013,
        month = jan,
       volume = {428},
       number = {3},
        pages = {2216-2231},
          doi = {10.1093/mnras/sts188},
archivePrefix = {arXiv},
       eprint = {1210.3695},
 primaryClass = {astro-ph.GA},
       adsurl = {https://ui.adsabs.harvard.edu/abs/2013MNRAS.428.2216W}
}

@ARTICLE{Brunner18,
       author = {{Brunner}, M. and {Maercker}, M. and {Mecina}, M. and {Khouri}, T. and {Kerschbaum}, F.},
        title = "{Modelling the carbon AGB star R Sculptoris. Constraining the dust properties in the detached shell based on far-infrared and sub-millimeter observations}",
      journal = {\aap},
         year = 2018,
        month = jun,
       volume = {614},
          eid = {A17},
        pages = {A17},
          doi = {10.1051/0004-6361/201732065},
       adsurl = {https://ui.adsabs.harvard.edu/abs/2018A&A...614A..17B}
}

@article{Goldman20,
  doi = {10.21105/joss.02554},
  url = {https://joss.theoj.org/papers/10.21105/joss.02554},
  year = {2020},
  publisher = {The Open Journal},
  volume = {5},
  number = {54},
  pages = {2554},
  author = {Steven R. Goldman},
  title = {The Dusty Evolved Star Kit (DESK): A Python package for fitting the Spectral Energy Distribution of Evolved Stars},
  journal = {Journal of Open Source Software}
}

@ARTICLE{Astropy2013,
       author = {{Astropy Collaboration} and {Robitaille}, Thomas P. and {Tollerud}, Erik J. and {Greenfield}, Perry and {Droettboom}, Michael and {Bray}, Erik and {Aldcroft}, Tom and {Davis}, Matt and {Ginsburg}, Adam and {Price-Whelan}, Adrian M. and {Kerzendorf}, Wolfgang E. and {Conley}, Alexander and {Crighton}, Neil and {Barbary}, Kyle and {Muna}, Demitri and {Ferguson}, Henry and {Grollier}, Fr{\'e}d{\'e}ric and {Parikh}, Madhura M. and {Nair}, Prasanth H. and {Unther}, Hans M. and {Deil}, Christoph and {Woillez}, Julien and {Conseil}, Simon and {Kramer}, Roban and {Turner}, James E.~H. and {Singer}, Leo and {Fox}, Ryan and {Weaver}, Benjamin A. and {Zabalza}, Victor and {Edwards}, Zachary I. and {Azalee Bostroem}, K. and {Burke}, D.~J. and {Casey}, Andrew R. and {Crawford}, Steven M. and {Dencheva}, Nadia and {Ely}, Justin and {Jenness}, Tim and {Labrie}, Kathleen and {Lim}, Pey Lian and {Pierfederici}, Francesco and {Pontzen}, Andrew and {Ptak}, Andy and {Refsdal}, Brian and {Servillat}, Mathieu and {Streicher}, Ole},
        title = "{Astropy: A community Python package for astronomy}",
      journal = {\aap},
         year = 2013,
        month = oct,
       volume = {558},
          eid = {A33},
        pages = {A33},
          doi = {10.1051/0004-6361/201322068},
archivePrefix = {arXiv},
       eprint = {1307.6212},
 primaryClass = {astro-ph.IM},
       adsurl = {https://ui.adsabs.harvard.edu/abs/2013A&A...558A..33A}
}

@ARTICLE{Gholami25,
       author = {{Gholami}, Mahtab and {Javadi}, Atefeh and {Mahani}, Hamidreza and {van Loon}, Jacco Th. and {Khosroshahi}, Habib and {Saremi}, Elham and {McDonald}, Iain and {Eftekhari}, Samaneh and {Ren}, Yi and {Altafi}, Hamed},
        title = "{The Isaac Newton Telescope Monitoring Survey of Local Group Dwarf Galaxies. VII. Long-period Variable Stars in the Nearest Starburst Dwarf Galaxy, IC 10}",
      journal = {\aj},
         year = 2025,
        month = jul,
       volume = {170},
       number = {1},
          eid = {54},
        pages = {54},
          doi = {10.3847/1538-3881/adda2b},
archivePrefix = {arXiv},
       eprint = {2505.03972},
 primaryClass = {astro-ph.GA},
       adsurl = {https://ui.adsabs.harvard.edu/abs/2025AJ....170...54G}
}

@INPROCEEDINGS{Mahani23,
       author = {{Mahani}, Hamidreza and {Javadi}, Atefeh and {van Loon}, Jacco Th. and {Khosroshahi}, Habib and {Saremi}, Elham and {Hamedani Golshan}, Roya and {Navabi}, Mahdieh and {Hashemi}, Seyed Azim and {Gholami}, Mahtab and {Aghdam}, Sima Taefi},
        title = "{From evolved stars to the formation and evolution of galaxies}",
    booktitle = {Resolving the Rise and Fall of Star Formation in Galaxies},
         year = 2023,
       editor = {{Wong}, Tony and {Kim}, Woong-Tae},
       series = {IAU Symposium},
       volume = {373},
        month = jan,
        pages = {264-267},
          doi = {10.1017/S1743921322004951},
       adsurl = {https://ui.adsabs.harvard.edu/abs/2023IAUS..373..264M}
}

@ARTICLE{Mateo98,
       author = {{Mateo}, Mario L.},
        title = "{Dwarf Galaxies of the Local Group}",
      journal = {\araa},
         year = 1998,
        month = jan,
       volume = {36},
        pages = {435-506},
          doi = {10.1146/annurev.astro.36.1.435},
archivePrefix = {arXiv},
       eprint = {astro-ph/9810070},
 primaryClass = {astro-ph},
       adsurl = {https://ui.adsabs.harvard.edu/abs/1998ARA&A..36..435M}
}

@ARTICLE{Tolstoy09,
       author = {{Tolstoy}, Eline and {Hill}, Vanessa and {Tosi}, Monica},
        title = "{Star-Formation Histories, Abundances, and Kinematics of Dwarf Galaxies in the Local Group}",
      journal = {\araa},
         year = 2009,
        month = sep,
       volume = {47},
       number = {1},
        pages = {371-425},
          doi = {10.1146/annurev-astro-082708-101650},
archivePrefix = {arXiv},
       eprint = {0904.4505},
 primaryClass = {astro-ph.CO},
       adsurl = {https://ui.adsabs.harvard.edu/abs/2009ARA&A..47..371T}
}

@ARTICLE{Abdollahi26,
       author = {{Abdollahi}, Hedieh and {Javadi}, Atefeh and {van Loon}, Jacco Th. and {McDonald}, Iain and {Abdollahi}, Mahdi and {Saremi}, Elham and {Khosroshahi}, Habib G. and {Moln{\'a}r}, L{\'a}szl{\'o} and {Mahani}, Hamidreza},
        title = "{The Isaac Newton Telescope Monitoring Survey of Local Group Dwarf Galaxies. VIII. A Census of Long-period Variable Stars across the Andromeda Dwarf Satellite System}",
      journal = {\apj},
         year = 2026,
        month = mar,
       volume = {1000},
       number = {1},
          eid = {69},
        pages = {69},
          doi = {10.3847/1538-4357/ae40f3},
archivePrefix = {arXiv},
       eprint = {2601.06924},
 primaryClass = {astro-ph.GA},
       adsurl = {https://ui.adsabs.harvard.edu/abs/2026ApJ..1000...69A}
}

@ARTICLE{vanLoon25,
       author = {{van Loon}, Jacco Th.},
        title = "{Red Supergiant Mass Loss and Mass-Loss Rates}",
      journal = {Galaxies},
         year = 2025,
        month = jun,
       volume = {13},
       number = {4},
          eid = {72},
        pages = {72},
          doi = {10.3390/galaxies13040072},
archivePrefix = {arXiv},
       eprint = {2507.15971},
 primaryClass = {astro-ph.SR},
       adsurl = {https://ui.adsabs.harvard.edu/abs/2025Galax..13...72V}
}

@ARTICLE{Jones15,
       author = {{Jones}, Olivia C. and {Meixner}, Margaret and {Sargent}, Benjamin A. and {Boyer}, Martha L. and {Sewi{\l}o}, Marta and {Hony}, Sacha and {Roman-Duval}, Julia},
        title = "{The Dustiest Post-Main Sequence Stars in the Magellanic Clouds}",
      journal = {\apj},
         year = 2015,
        month = oct,
       volume = {811},
       number = {2},
          eid = {145},
        pages = {145},
          doi = {10.1088/0004-637X/811/2/145},
archivePrefix = {arXiv},
       eprint = {1508.06999},
 primaryClass = {astro-ph.SR},
       adsurl = {https://ui.adsabs.harvard.edu/abs/2015ApJ...811..145J}
}

@ARTICLE{Kinson21,
       author = {{Kinson}, David A. and {Oliveira}, Joana M. and {van Loon}, Jacco Th},
        title = "{Massive young stellar objects in the Local Group irregular galaxy NGC 6822 identified using machine learning}",
      journal = {\mnras},
         year = 2021,
        month = nov,
       volume = {507},
       number = {4},
        pages = {5106-5131},
          doi = {10.1093/mnras/stab2386},
archivePrefix = {arXiv},
       eprint = {2108.07105},
 primaryClass = {astro-ph.GA},
       adsurl = {https://ui.adsabs.harvard.edu/abs/2021MNRAS.507.5106K}
}

@ARTICLE{vanLoon05,
       author = {{van Loon}, J. Th. and {Marshall}, J.~R. and {Zijlstra}, A.~A.},
        title = "{Dust-enshrouded giants in clusters in the Magellanic Clouds}",
      journal = {\aap},
         year = 2005,
        month = nov,
       volume = {442},
       number = {2},
        pages = {597-613},
          doi = {10.1051/0004-6361:20053528},
archivePrefix = {arXiv},
       eprint = {astro-ph/0507571},
 primaryClass = {astro-ph},
       adsurl = {https://ui.adsabs.harvard.edu/abs/2005A&A...442..597V}
}

@ARTICLE{Guglielmo93,
       author = {{Guglielmo}, F. and {Epchtein}, N. and {Le Bertre}, T. and {Fouque}, P. and {Hron}, J. and {Kerschbaum}, F. and {Lepine}, J.~R.~D.},
        title = "{Identification of 106 new infrared carbon stars in the IRAS Point Source Catalog : Near-infrared photometry and their space distribution in the galaxy.}",
      journal = {\aaps},
         year = 1993,
        month = may,
       volume = {99},
        pages = {31-69},
       adsurl = {https://ui.adsabs.harvard.edu/abs/1993A&AS...99...31G}
}

@ARTICLE{Zijlstra96,
       author = {{Zijlstra}, Albert A. and {Loup}, Cecile and {Waters}, L.~B.~F.~M. and {Whitelock}, P.~A. and {van Loon}, Jacco Th. and {Guglielmo}, F.},
        title = "{Obscured asymptotic giant branch stars in the Magellanic Clouds - II. Near-infrared and mid-infrared counterparts}",
      journal = {\mnras},
         year = 1996,
        month = mar,
       volume = {279},
       number = {1},
        pages = {32-62},
          doi = {10.1093/mnras/279.1.32},
archivePrefix = {arXiv},
       eprint = {astro-ph/9510069},
 primaryClass = {astro-ph},
       adsurl = {https://ui.adsabs.harvard.edu/abs/1996MNRAS.279...32Z}
}

@ARTICLE{vanLoon97,
       author = {{van Loon}, J.~T. and {Zijlstra}, A.~A. and {Whitelock}, P.~A. and {Waters}, L.~B.~F.~M. and {Loup}, C. and {Trams}, N.~R.},
        title = "{Obscured Asymptotic Giant Branch stars in the Magellanic Clouds. III. New IRAS counterparts.}",
      journal = {\aap},
         year = 1997,
        month = sep,
       volume = {325},
        pages = {585-600},
          doi = {10.48550/arXiv.astro-ph/9704201},
archivePrefix = {arXiv},
       eprint = {astro-ph/9704201},
 primaryClass = {astro-ph},
       adsurl = {https://ui.adsabs.harvard.edu/abs/1997A&A...325..585V}
}

@ARTICLE{vanLoon98,
       author = {{van Loon}, Jacco Th. and {Zijlstra}, Albert A. and {Whitelock}, Patricia A. and {te Lintel Hekkert}, Peter and {Chapman}, Jessica M. and {Loup}, Cecile and {Groenewegen}, M.~A.~T. and {Waters}, L.~B.~F.~M. and {Trams}, Norman R.},
        title = "{Obscured Asymptotic Giant Branch stars in the Magellanic Clouds. IV. Carbon stars and OH/IR stars}",
      journal = {\aap},
         year = 1998,
        month = jan,
       volume = {329},
        pages = {169-185},
          doi = {10.48550/arXiv.astro-ph/9709119},
archivePrefix = {arXiv},
       eprint = {astro-ph/9709119},
 primaryClass = {astro-ph},
       adsurl = {https://ui.adsabs.harvard.edu/abs/1998A&A...329..169V}
}

@ARTICLE{vanLoon08,
       author = {{van Loon}, Jacco Th. and {Boyer}, Martha L. and {McDonald}, Iain},
        title = "{Spitzer Space Telescope Evidence in NGC 6791: No Super Mass Loss at Supersolar Metallicity to Explain Helium White Dwarfs?}",
      journal = {\apjl},
         year = 2008,
        month = jun,
       volume = {680},
       number = {1},
        pages = {L49},
          doi = {10.1086/589711},
archivePrefix = {arXiv},
       eprint = {0804.4559},
 primaryClass = {astro-ph},
       adsurl = {https://ui.adsabs.harvard.edu/abs/2008ApJ...680L..49V}
}

@ARTICLE{Zhukovska08,
       author = {{Zhukovska}, S. and {Gail}, H.-P. and {Trieloff}, M.},
        title = "{Evolution of interstellar dust and stardust in the solar neighbourhood}",
      journal = {\aap},
         year = 2008,
        month = feb,
       volume = {479},
       number = {2},
        pages = {453-480},
          doi = {10.1051/0004-6361:20077789},
archivePrefix = {arXiv},
       eprint = {0706.1155},
 primaryClass = {astro-ph},
       adsurl = {https://ui.adsabs.harvard.edu/abs/2008A&A...479..453Z}
}

@ARTICLE{Nanni20,
       author = {{Nanni}, A. and {Burgarella}, D. and {Theul{\'e}}, P. and {C{\^o}t{\'e}}, B. and {Hirashita}, H.},
        title = "{The gas, metal, and dust evolution in low-metallicity local and high-redshift galaxies}",
      journal = {\aap},
         year = 2020,
        month = sep,
       volume = {641},
          eid = {A168},
        pages = {A168},
          doi = {10.1051/0004-6361/202037833},
archivePrefix = {arXiv},
       eprint = {2006.15146},
 primaryClass = {astro-ph.GA},
       adsurl = {https://ui.adsabs.harvard.edu/abs/2020A&A...641A.168N}
}

@ARTICLE{Efremova11,
       author = {{Efremova}, Boryana V. and {Bianchi}, Luciana and {Thilker}, David A. and {Neill}, James D. and {Burgarella}, Denis and {Wyder}, Ted K. and {Madore}, Barry F. and {Rey}, Soo-Chang and {Barlow}, Tom A. and {Conrow}, Tim and {Forster}, Karl and {Friedman}, Peter G. and {Martin}, D. Christopher and {Morrissey}, Patrick and {Neff}, Susan G. and {Schiminovich}, David and {Seibert}, Mark and {Small}, Todd},
        title = "{The Recent Star Formation in NGC 6822: An Ultraviolet Study}",
      journal = {\apj},
         year = 2011,
        month = apr,
       volume = {730},
       number = {2},
          eid = {88},
        pages = {88},
          doi = {10.1088/0004-637X/730/2/88},
archivePrefix = {arXiv},
       eprint = {1101.6051},
 primaryClass = {astro-ph.GA},
       adsurl = {https://ui.adsabs.harvard.edu/abs/2011ApJ...730...88E}
}

@ARTICLE{Nally26,
       author = {{Nally}, Conor and {Jones}, Olivia C. and {Lenki{\'c}}, Laura and {Ferguson}, Annette M.~N. and {Habel}, Nolan and {Hirschauer}, Alec S. and {Meixner}, Margaret and {Kavanagh}, P.~J. and {Boyer}, Martha L. and {Nayak}, Omnarayani and {Sargent}, B. and {Scicluna}, P.},
        title = "{JWST NIRCam and MIRI Reveal the Dust-Producing AGB Population of NGC 6822}",
      journal = {arXiv e-prints},
         year = 2026,
        month = jul,
          eid = {arXiv:2607.11682},
        pages = {arXiv:2607.11682},
          doi = {10.48550/arXiv.2607.11682},
archivePrefix = {arXiv},
       eprint = {2607.11682},
 primaryClass = {astro-ph.GA},
       adsurl = {https://ui.adsabs.harvard.edu/abs/2026arXiv260711682N}
}

@ARTICLE{Sargent11,
       author = {{Sargent}, Benjamin A. and {Srinivasan}, S. and {Meixner}, M.},
        title = "{The Mass-loss Return from Evolved Stars to the Large Magellanic Cloud. IV. Construction and Validation of a Grid of Models for Oxygen-rich AGB Stars, Red Supergiants, and Extreme AGB Stars}",
      journal = {\apj},
         year = 2011,
        month = feb,
       volume = {728},
       number = {2},
          eid = {93},
        pages = {93},
          doi = {10.1088/0004-637X/728/2/93},
archivePrefix = {arXiv},
       eprint = {1407.8452},
 primaryClass = {astro-ph.SR},
       adsurl = {https://ui.adsabs.harvard.edu/abs/2011ApJ...728...93S}
}

@ARTICLE{Srinivasan11,
       author = {{Srinivasan}, S. and {Sargent}, B.~A. and {Meixner}, M.},
        title = "{The mass-loss return from evolved stars to the Large Magellanic Cloud. V. The GRAMS carbon-star model grid}",
      journal = {\aap},
         year = 2011,
        month = aug,
       volume = {532},
          eid = {A54},
        pages = {A54},
          doi = {10.1051/0004-6361/201117033},
archivePrefix = {arXiv},
       eprint = {1106.3256},
 primaryClass = {astro-ph.SR},
       adsurl = {https://ui.adsabs.harvard.edu/abs/2011A&A...532A..54S}
}

@ARTICLE{Hewett06,
       author = {{Hewett}, P.~C. and {Warren}, S.~J. and {Leggett}, S.~K. and {Hodgkin}, S.~T.},
        title = "{The UKIRT Infrared Deep Sky Survey ZY JHK photometric system: passbands and synthetic colours}",
      journal = {\mnras},
         year = 2006,
        month = apr,
       volume = {367},
       number = {2},
        pages = {454-468},
          doi = {10.1111/j.1365-2966.2005.09969.x},
archivePrefix = {arXiv},
       eprint = {astro-ph/0601592},
 primaryClass = {astro-ph},
       adsurl = {https://ui.adsabs.harvard.edu/abs/2006MNRAS.367..454H}
}

@ARTICLE{Fazio04,
       author = {{Fazio}, G.~G. and {Hora}, J.~L. and {Allen}, L.~E. and {Ashby}, M.~L.~N. and {Barmby}, P. and {Deutsch}, L.~K. and {Huang}, J.-S. and {Kleiner}, S. and {Marengo}, M. and {Megeath}, S.~T. and {Melnick}, G.~J. and {Pahre}, M.~A. and {Patten}, B.~M. and {Polizotti}, J. and {Smith}, H.~A. and {Taylor}, R.~S. and {Wang}, Z. and {Willner}, S.~P. and {Hoffmann}, W.~F. and {Pipher}, J.~L. and {Forrest}, W.~J. and {McMurty}, C.~W. and {McCreight}, C.~R. and {McKelvey}, M.~E. and {McMurray}, R.~E. and {Koch}, D.~G. and {Moseley}, S.~H. and {Arendt}, R.~G. and {Mentzell}, J.~E. and {Marx}, C.~T. and {Losch}, P. and {Mayman}, P. and {Eichhorn}, W. and {Krebs}, D. and {Jhabvala}, M. and {Gezari}, D.~Y. and {Fixsen}, D.~J. and {Flores}, J. and {Shakoorzadeh}, K. and {Jungo}, R. and {Hakun}, C. and {Workman}, L. and {Karpati}, G. and {Kichak}, R. and {Whitley}, R. and {Mann}, S. and {Tollestrup}, E.~V. and {Eisenhardt}, P. and {Stern}, D. and {Gorjian}, V. and {Bhattacharya}, B. and {Carey}, S. and {Nelson}, B.~O. and {Glaccum}, W.~J. and {Lacy}, M. and {Lowrance}, P.~J. and {Laine}, S. and {Reach}, W.~T. and {Stauffer}, J.~A. and {Surace}, J.~A. and {Wilson}, G. and {Wright}, E.~L. and {Hoffman}, A. and {Domingo}, G. and {Cohen}, M.},
        title = "{The Infrared Array Camera (IRAC) for the Spitzer Space Telescope}",
      journal = {\apjs},
         year = 2004,
        month = sep,
       volume = {154},
       number = {1},
        pages = {10-17},
          doi = {10.1086/422843},
archivePrefix = {arXiv},
       eprint = {astro-ph/0405616},
 primaryClass = {astro-ph},
       adsurl = {https://ui.adsabs.harvard.edu/abs/2004ApJS..154...10F}
}

@ARTICLE{Rieke04,
       author = {{Rieke}, G.~H. and {Young}, E.~T. and {Engelbracht}, C.~W. and {Kelly}, D.~M. and {Low}, F.~J. and {Haller}, E.~E. and {Beeman}, J.~W. and {Gordon}, K.~D. and {Stansberry}, J.~A. and {Misselt}, K.~A. and {Cadien}, J. and {Morrison}, J.~E. and {Rivlis}, G. and {Latter}, W.~B. and {Noriega-Crespo}, A. and {Padgett}, D.~L. and {Stapelfeldt}, K.~R. and {Hines}, D.~C. and {Egami}, E. and {Muzerolle}, J. and {Alonso-Herrero}, A. and {Blaylock}, M. and {Dole}, H. and {Hinz}, J.~L. and {Le Floc'h}, E. and {Papovich}, C. and {P{\'e}rez-Gonz{\'a}lez}, P.~G. and {Smith}, P.~S. and {Su}, K.~Y.~L. and {Bennett}, L. and {Frayer}, D.~T. and {Henderson}, D. and {Lu}, N. and {Masci}, F. and {Pesenson}, M. and {Rebull}, L. and {Rho}, J. and {Keene}, J. and {Stolovy}, S. and {Wachter}, S. and {Wheaton}, W. and {Werner}, M.~W. and {Richards}, P.~L.},
        title = "{The Multiband Imaging Photometer for Spitzer (MIPS)}",
      journal = {\apjs},
         year = 2004,
        month = sep,
       volume = {154},
       number = {1},
        pages = {25-29},
          doi = {10.1086/422717},
       adsurl = {https://ui.adsabs.harvard.edu/abs/2004ApJS..154...25R}
}

@INPROCEEDINGS{Nagashima99,
       author = {{Nagashima}, C. and {Nagayama}, T. and {Nakajima}, Y. and {Tamura}, M. and {Sugitani}, K. and {Nagata}, T. and {Hirao}, T. and {Nakaya}, H. and {Yanagisawa}, K. and {Sato}, S.},
        title = "{Development of SIRIUS --- A Simultaneous-Color InfraRed Imager for Unbiased Survey}",
    booktitle = {Star Formation 1999},
         year = 1999,
       editor = {{Nakamoto}, T.},
        month = dec,
        pages = {397-398},
       adsurl = {https://ui.adsabs.harvard.edu/abs/1999sf99.proc..397N}
}

@INPROCEEDINGS{Cuillandre01,
       author = {{Cuillandre}, Jean-Charles and {Luppino}, Gerry and {Starr}, Barry and {Isani}, Sidik},
        title = "{CFH12K: the 12,288 by 8,192 pixel CCD mosaic camera for the CFHT prime focus}",
    booktitle = {SF2A-2001: Semaine de l'Astrophysique Francaise},
         year = 2001,
       editor = {{Combes}, F. and {Barret}, D. and {Th{\'e}venin}, F.},
        month = may,
        pages = {605},
       adsurl = {https://ui.adsabs.harvard.edu/abs/2001sf2a.conf..605C}
}

@ARTICLE{Rieke23,
       author = {{Rieke}, Marcia J. and {Kelly}, Douglas M. and {Misselt}, Karl and {Stansberry}, John and {Boyer}, Martha and {Beatty}, Thomas and {Egami}, Eiichi and {Florian}, Michael and {Greene}, Thomas P. and {Hainline}, Kevin and {Leisenring}, Jarron and {Roellig}, Thomas and {Schlawin}, Everett and {Sun}, Fengwu and {Tinnin}, Lee and {Williams}, Christina C. and {Willmer}, Christopher N.~A. and {Wilson}, Debra and {Clark}, Charles R. and {Rohrbach}, Scott and {Brooks}, Brian and {Canipe}, Alicia and {Correnti}, Matteo and {DiFelice}, Audrey and {Gennaro}, Mario and {Girard}, Julien H. and {Hartig}, George and {Hilbert}, Bryan and {Koekemoer}, Anton M. and {Nikolov}, Nikolay K. and {Pirzkal}, Norbert and {Rest}, Armin and {Robberto}, Massimo and {Sunnquist}, Ben and {Telfer}, Randal and {Wu}, Chi Rai and {Ferry}, Malcolm and {Lewis}, Dan and {Baum}, Stefi and {Beichman}, Charles and {Doyon}, Ren{\'e} and {Dressler}, Alan and {Eisenstein}, Daniel J. and {Ferrarese}, Laura and {Hodapp}, Klaus and {Horner}, Scott and {Jaffe}, Daniel T. and {Johnstone}, Doug and {Krist}, John and {Martin}, Peter and {McCarthy}, Donald W. and {Meyer}, Michael and {Rieke}, George H. and {Trauger}, John and {Young}, Erick T.},
        title = "{Performance of NIRCam on JWST in Flight}",
      journal = {\pasp},
         year = 2023,
        month = feb,
       volume = {135},
       number = {1044},
          eid = {028001},
        pages = {028001},
          doi = {10.1088/1538-3873/acac53},
archivePrefix = {arXiv},
       eprint = {2212.12069},
 primaryClass = {astro-ph.IM},
       adsurl = {https://ui.adsabs.harvard.edu/abs/2023PASP..135b8001R}
}

@ARTICLE{Dicken24,
       author = {{Dicken}, Dan and {Mar{\'\i}n}, Macarena Garc{\'\i}a and {Shivaei}, Irene and {Guillard}, Pierre and {Libralato}, Mattia and {Glasse}, Alistair and {Gordon}, Karl D. and {Cossou}, Christophe and {Kavanagh}, Patrick and {Temim}, Tea and {Flagey}, Nicolas and {Klaassen}, Pamela and {Rieke}, George H. and {Wright}, Gillian and {Alberts}, Stacey and {Azzollini}, Ruyman and {{\'A}lvarez-M{\'a}rquez}, Javier and {Bouchet}, Patrice and {Bright}, Stacey and {Cracraft}, Misty and {Coulais}, Alain and {Detre}, Ors Hunor and {Engesser}, Mike and {Fox}, Ori D. and {Gaspar}, Andras and {Gastaud}, Ren{\'e} and {Glauser}, Adrian M. and {Hines}, Dean C. and {Kendrew}, Sarah and {Labiano}, Alvaro and {Lagage}, Pierre-Oliver and {Lee}, David and {Law}, David R. and {Morrison}, Jane E. and {Noriega-Crespo}, Alberto and {Jones}, Olivia and {Patapis}, Polychronis and {Scheithauer}, Silvia and {Sloan}, G.~C. and {Tamas}, Laszlo},
        title = "{JWST MIRI flight performance: Imaging}",
      journal = {\aap},
         year = 2024,
        month = sep,
       volume = {689},
          eid = {A5},
        pages = {A5},
          doi = {10.1051/0004-6361/202449451},
archivePrefix = {arXiv},
       eprint = {2403.16686},
 primaryClass = {astro-ph.IM},
       adsurl = {https://ui.adsabs.harvard.edu/abs/2024A&A...689A...5D}
}

@ARTICLE{Zhukovska13,
       author = {{Zhukovska}, S. and {Henning}, T.},
        title = "{Dust input from AGB stars in the Large Magellanic Cloud}",
      journal = {\aap},
         year = 2013,
        month = jul,
       volume = {555},
          eid = {A99},
        pages = {A99},
          doi = {10.1051/0004-6361/201321368},
archivePrefix = {arXiv},
       eprint = {1305.3521},
 primaryClass = {astro-ph.GA},
       adsurl = {https://ui.adsabs.harvard.edu/abs/2013A&A...555A..99Z}
}

@article{Kroupa01,
  title={On the variation of the initial mass function},
  author={Kroupa, Pavel},
  journal={Monthly Notices of the Royal Astronomical Society},
  volume={322},
  number={2},
  pages={231--246},
  year={2001},
  publisher={Blackwell Science Ltd Oxford, UK}
}

@ARTICLE{Antoniadis25,
       author = {{Antoniadis}, K. and {Zapartas}, E. and {Bonanos}, A.~Z. and {Maravelias}, G. and {Vlassis}, S. and {Mu{\~n}oz-Sanchez}, G. and {Nally}, C. and {Meixner}, M. and {Jones}, O.~C. and {Lenki{\'c}}, L. and {Kavanagh}, P.~J.},
        title = "{Investigating the metallicity dependence of the mass-loss rate relation of red supergiants}",
      journal = {\aap},
         year = 2025,
        month = oct,
       volume = {702},
          eid = {A178},
        pages = {A178},
          doi = {10.1051/0004-6361/202554416},
archivePrefix = {arXiv},
       eprint = {2503.05876},
 primaryClass = {astro-ph.SR},
       adsurl = {https://ui.adsabs.harvard.edu/abs/2025A&A...702A.178A}
}

@ARTICLE{RemyRuyer14,
       author = {{R{\'e}my-Ruyer}, A. and {Madden}, S.~C. and {Galliano}, F. and {Galametz}, M. and {Takeuchi}, T.~T. and {Asano}, R.~S. and {Zhukovska}, S. and {Lebouteiller}, V. and {Cormier}, D. and {Jones}, A. and {Bocchio}, M. and {Baes}, M. and {Bendo}, G.~J. and {Boquien}, M. and {Boselli}, A. and {DeLooze}, I. and {Doublier-Pritchard}, V. and {Hughes}, T. and {Karczewski}, O. {\L}. and {Spinoglio}, L.},
        title = "{Gas-to-dust mass ratios in local galaxies over a 2 dex metallicity range}",
      journal = {\aap},
         year = 2014,
        month = mar,
       volume = {563},
          eid = {A31},
        pages = {A31},
          doi = {10.1051/0004-6361/201322803},
archivePrefix = {arXiv},
       eprint = {1312.3442},
 primaryClass = {astro-ph.GA},
       adsurl = {https://ui.adsabs.harvard.edu/abs/2014A&A...563A..31R}
}

@ARTICLE{DuartePuertas24,
       author = {{Duarte Puertas}, Salvador and {Drissen}, Laurent and {Robert}, Carmelle and {Rousseau-Nepton}, Laurie and {Martin}, Ren{\'e} Pierre and {Amram}, Philippe and {Martin}, Thomas},
        title = "{Properties of supernova remnants in SIGNALS galaxies - I. NGC 6822 and M33}",
      journal = {\mnras},
         year = 2024,
        month = sep,
       volume = {533},
       number = {3},
        pages = {2677-2704},
          doi = {10.1093/mnras/stae1641},
archivePrefix = {arXiv},
       eprint = {2407.01153},
 primaryClass = {astro-ph.GA},
       adsurl = {https://ui.adsabs.harvard.edu/abs/2024MNRAS.533.2677D}
}

@ARTICLE{Schlegel98,
       author = {{Schlegel}, David J. and {Finkbeiner}, Douglas P. and {Davis}, Marc},
        title = "{Maps of Dust Infrared Emission for Use in Estimation of Reddening and Cosmic Microwave Background Radiation Foregrounds}",
      journal = {\apj},
         year = 1998,
        month = jun,
       volume = {500},
       number = {2},
        pages = {525-553},
          doi = {10.1086/305772},
archivePrefix = {arXiv},
       eprint = {astro-ph/9710327},
 primaryClass = {astro-ph},
       adsurl = {https://ui.adsabs.harvard.edu/abs/1998ApJ...500..525S}
}

@ARTICLE{Groenewegen06,
       author = {{Groenewegen}, M.~A.~T.},
        title = "{The mid- and far-infrared colours of AGB and post-AGB stars}",
      journal = {\aap},
         year = 2006,
        month = mar,
       volume = {448},
       number = {1},
        pages = {181-187},
          doi = {10.1051/0004-6361:20054163},
archivePrefix = {arXiv},
       eprint = {astro-ph/0511475},
 primaryClass = {astro-ph},
       adsurl = {https://ui.adsabs.harvard.edu/abs/2006A&A...448..181G}
}

@ARTICLE{Suh99,
       author = {{Suh}, Kyung-Won},
        title = "{Optical properties of the silicate dust grains in the envelopes around asymptotic giant branch stars}",
      journal = {\mnras},
         year = 1999,
        month = apr,
       volume = {304},
       number = {2},
        pages = {389-405},
          doi = {10.1046/j.1365-8711.1999.02317.x},
       adsurl = {https://ui.adsabs.harvard.edu/abs/1999MNRAS.304..389S}
}

@ARTICLE{Suh2000,
       author = {{Suh}, Kyung-Won},
        title = "{Optical properties of the carbon dust grains in the envelopes around asymptotic giant branch stars}",
      journal = {\mnras},
         year = 2000,
        month = jul,
       volume = {315},
       number = {4},
        pages = {740-750},
          doi = {10.1046/j.1365-8711.2000.03482.x},
       adsurl = {https://ui.adsabs.harvard.edu/abs/2000MNRAS.315..740S}
}

@ARTICLE{deWit25,
       author = {{de Wit}, S. and {Mu{\~n}oz-Sanchez}, G. and {Maravelias}, G. and {Bonanos}, A.~Z. and {Antoniadis}, K. and {Garc{\'\i}a-{\'A}lvarez}, D. and {Britavskiy}, N. and {Ruiz}, A. and {Philippopoulou}, A.},
        title = "{Investigating episodic mass loss in evolved massive stars: III. Spectroscopy of dusty massive stars in three northern galaxies}",
      journal = {\aap},
         year = 2025,
        month = jun,
       volume = {698},
          eid = {A279},
        pages = {A279},
          doi = {10.1051/0004-6361/202453571},
archivePrefix = {arXiv},
       eprint = {2505.01498},
 primaryClass = {astro-ph.SR},
       adsurl = {https://ui.adsabs.harvard.edu/abs/2025A&A...698A.279D}
}

@ARTICLE{Christodoulou25,
       author = {{Christodoulou}, E. and {de Wit}, S. and {Bonanos}, A.~Z. and {Mu{\~n}oz-Sanchez}, G. and {Maravelias}, G. and {Ruiz}, A. and {Antoniadis}, K. and {Garc{\'\i}a-{\'A}lvarez}, D. and {Rubio D{\'\i}ez}, M.~M.},
        title = "{Investigating episodic mass loss in evolved massive stars: IV. Comprehensive analysis of dusty red supergiants in NGC 6822, IC 10, and WLM}",
      journal = {\aap},
         year = 2025,
        month = dec,
       volume = {704},
          eid = {A96},
        pages = {A96},
          doi = {10.1051/0004-6361/202554071},
archivePrefix = {arXiv},
       eprint = {2510.26674},
 primaryClass = {astro-ph.SR},
       adsurl = {https://ui.adsabs.harvard.edu/abs/2025A&A...704A..96C}
}
\bibliographystyle{aasjournal}

%@@@@@@@@@@@@@@@@@@@@@@@@@@@@@@@@@@@@@@@@@@@@@@@@@@@@@@@@@@@@@@@@@@@@@@@@@@@@@@@@@@@@@@@@@@

\appendix
\renewcommand{\thefigure}{A\arabic{figure}}
\renewcommand{\thetable}{A\arabic{table}}
\setcounter{figure}{0} % Reset the figure counter
\setcounter{table}{0}  % Reset the table counter

\section{JWST selection rules}
\label{app:JWST}

We utilized publicly available JWST source catalog data from the Mikulski Archive for Space Telescopes (MAST). All JWST catalog data used in this paper are available in MAST: \dataset[https://doi.org/10.17909/7ehm-dn38]{https://doi.org/10.17909/7ehm-dn38}. These data are associated with JWST Program ID 1234 and include source catalogs based on NIRCam (F115W, F200W, F356W, F444W) and MIRI (F770W, F1000W, F1500W, F2100W) imaging observations \citep{Nally24}. All photometric measurements were extracted from the standard Level-3 source catalogs generated by the JWST calibration pipeline (CALJWST, version 1.20.2).

Despite the high quality of the pipeline products, source confusion, imperfect cross-band matching, and detector-specific artifacts can still affect JWST photometry in deep and crowded fields \citep[e.g.,][]{Nally24}.

To mitigate these effects, we cross-matched the JWST catalog with the archival UKIRT and \textit{Spitzer} photometric catalogs presented by \citet{Hirschauer20}. Specifically, the JWST F115W, F200W, F356W, F444W, and F770W bands were compared with the corresponding J, K, 3.6\,$\mu$m, 4.5\,$\mu$m, and 8.0\,$\mu$m bands. For each filter pair, we calculated the magnitude residuals, 
\begin{equation}
    \Delta m = m_{\text{JWST}} - m_{\text{ref}},
\end{equation}
where $m_{\text{ref}}$ denotes the magnitude from the reference catalog. 

\begin{figure*}
\begin{center}
\includegraphics[width=\textwidth]{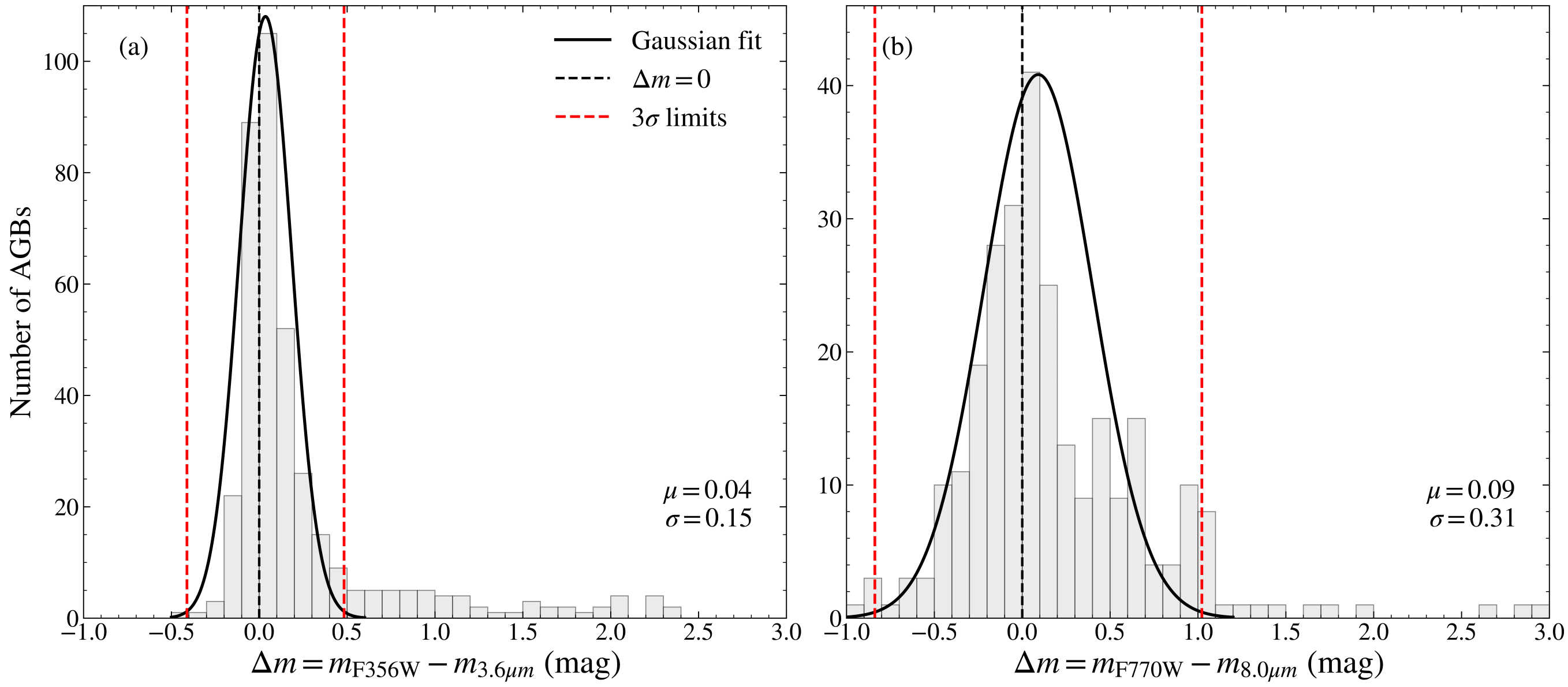}
\caption{ Histograms of the photometric residuals, $\Delta m = m_{\rm JWST}-m_{\rm ref}$, for the F356W–3.6$\mu m$ (left) and F770W–8.0$\mu m$ (right) filter pairs. The black curves show Gaussian fits to the central distributions, and the red dashed lines indicate the adopted $\pm3\sigma$ clipping limits. Sources outside these limits were excluded from the final sample. For clarity, only the range $\Delta m \leq 3$ mag is shown.
} 
\label{fig:JWST_histograms}
\end{center}
\end{figure*}

Histograms of the $\Delta m $ distributions show a clear peak near $\Delta m \approx 0$, with a small fraction of sources distributed far from the central peak. Examples for the F356W and F770W filters are shown in Fig.~\ref{fig:JWST_histograms}. We therefore fitted a Gaussian function to the central peak and applied a conservative 3$\sigma$ clipping threshold to reject extreme outliers, following standard sigma-clipping procedures commonly used in astronomical data analysis \citep{Astropy2013}. The 3$\sigma$ threshold was chosen to retain the bulk of reliable measurements while rejecting sources with anomalous photometric residuals. Sources satisfying $|\Delta m - \mu| > 3\sigma$ were excluded from the final sample, where $\mu$ and $\sigma$ denote the mean and standard deviation of the fitted Gaussian distribution, respectively. This quality-control procedure removed measurements with anomalous photometric behavior while preserving the dominant astrophysical population, resulting in the rejection of approximately 10--35\% of the initially matched sources, depending on the filter.

The resulting quality-controlled subset of JWST photometry from \citet{Nally24} was incorporated into our final photometric catalog and used in the SED fitting described in Sect.~\ref{sec:model}.

\section{Method-to-Method Comparison}
\label{app:Comparison}

This appendix provides a direct comparison between the results obtained with \texttt{SEDust}, the LPV-based method described in Section~\ref{Sec:Total_mass_loss}, and the \texttt{DESK} code. The purpose of this comparison is to illustrate the level of agreement among the three approaches and to highlight systematic differences arising from their underlying assumptions and model resolutions.

\begin{figure*}
\begin{center}
\includegraphics[width=\textwidth]{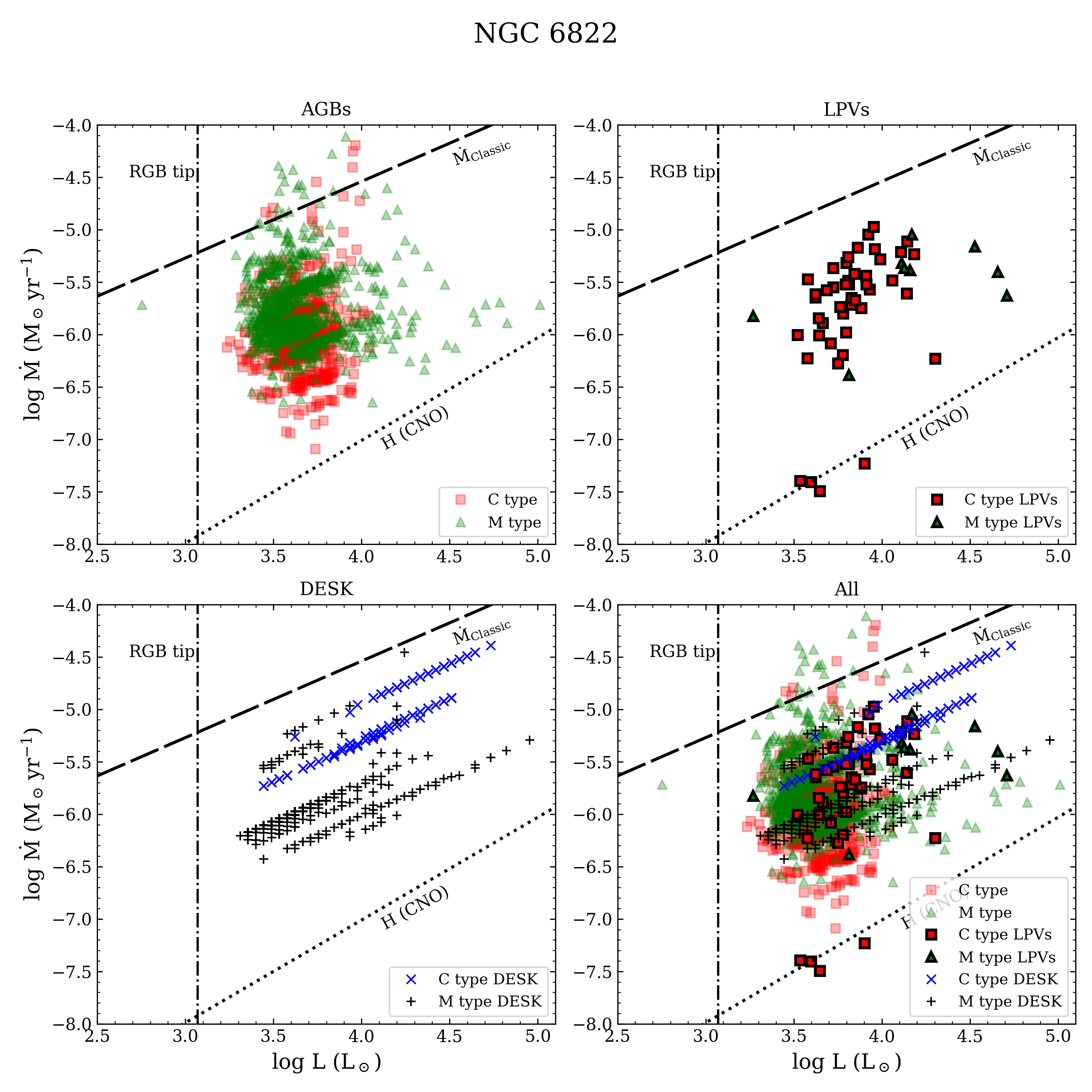}
\caption{ Mass-loss rate as a function of luminosity for evolved stars in NGC 6822.
The four panels show (top left) AGB stars, (top right) LPVs, (bottom left) DESK results, and (bottom right) the combined sample. Carbon-rich and oxygen-rich stars are indicated by red squares and green triangles, respectively; LPVs are highlighted with black-edged symbols, while DESK results are shown with crosses and plus symbols. The dashed and dotted black lines indicate the classical mass-loss prescription and the H-burning (CNO) limit, respectively. The vertical dash-dotted line marks the RGB tip. All quantities are shown in logarithmic units.} 
\label{fig:L_MDOT2x2}
\end{center}
\end{figure*}

Fig.~\ref{fig:L_MDOT2x2} presents the mass-loss rate--luminosity relation for the individual methods and for the combined sample. Although the overall trends are similar, the \texttt{DESK} results occupy a more restricted region of parameter space and exhibit noticeably less dispersion than the \texttt{SEDust} and LPV-based estimates. This behavior reflects the coarser parameter sampling adopted by \texttt{DESK}, which can assign identical physical properties to multiple stars with similar photometric characteristics.

\begin{figure*}
\begin{center}
\includegraphics[width=\textwidth]{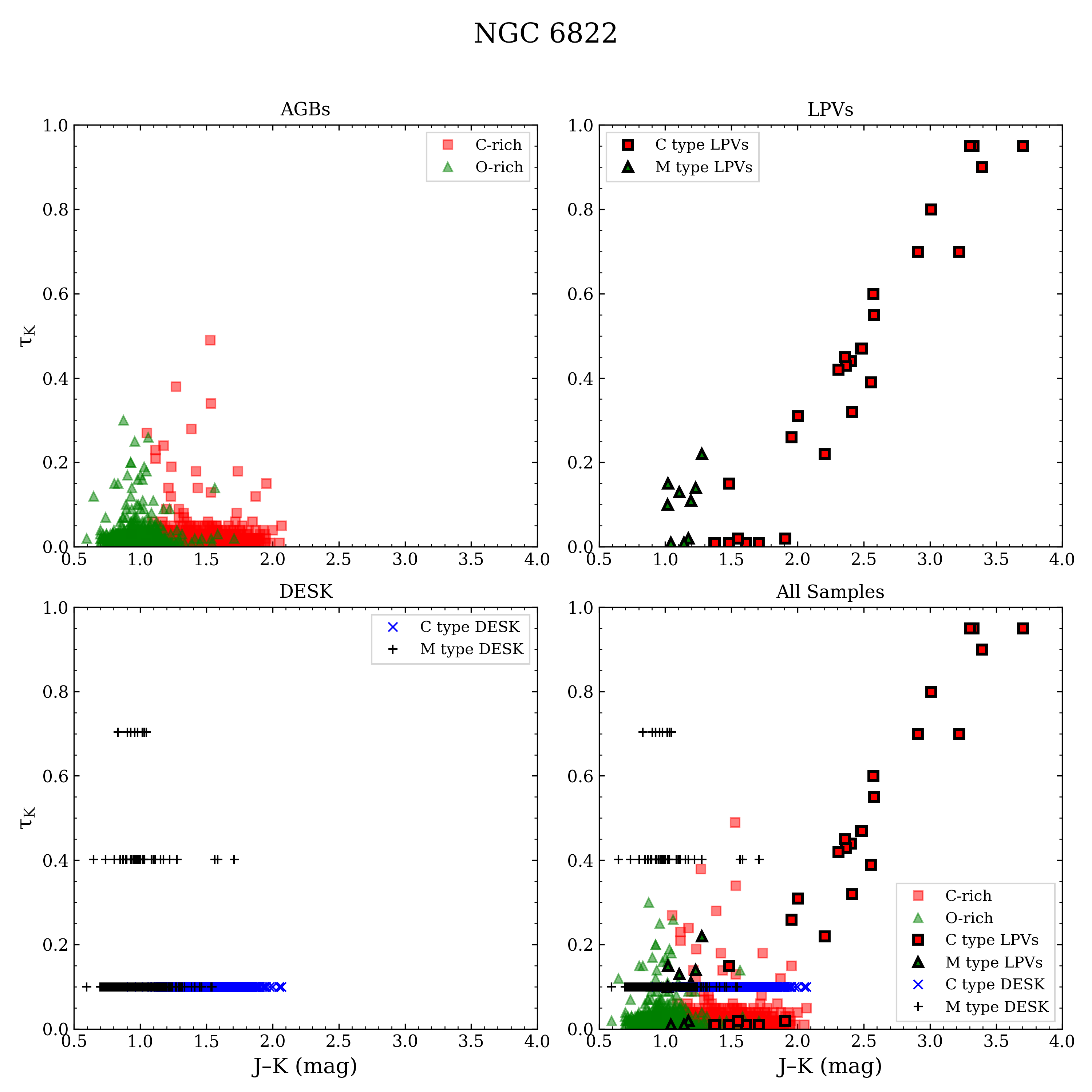}
\caption{ Optical depth in the K band, $\tau_{K}$, as a function of near-infrared color $(J-K)$ for evolved stars in NGC 6822. Carbon-rich and oxygen-rich sources are shown as red squares and green triangles, respectively. LPVs are highlighted with black-edged symbols, while DESK sources are indicated by crosses and plus symbols. For both chemical types, the optical depths derived from the \texttt{DESK} code show limited resolution. In particular, the code reports a single optical depth value of 0.1 for carbon stars. For oxygen-rich stars, the available optical depth values are limited to 0.1, 0.4, and 0.7.
} 
\label{fig:optical_depth_2x2}
\end{center}
\end{figure*}

Fig.~\ref{fig:optical_depth_2x2} compares the relation between optical depth and near-infrared color. The optical depths derived with \texttt{SEDust} span a continuous range of values, whereas the \texttt{DESK} results are concentrated at a small number of discrete optical depths. In particular, all carbon-rich stars are assigned $\tau = 0.1$, while oxygen-rich stars are restricted to $\tau = 0.1$, 0.4, or 0.7. As a result, the optical depth distribution derived with \texttt{DESK} exhibits substantially lower resolution than that obtained with \texttt{SEDust}.

\begin{figure}
\begin{center}
\includegraphics[width=0.45 \textwidth]{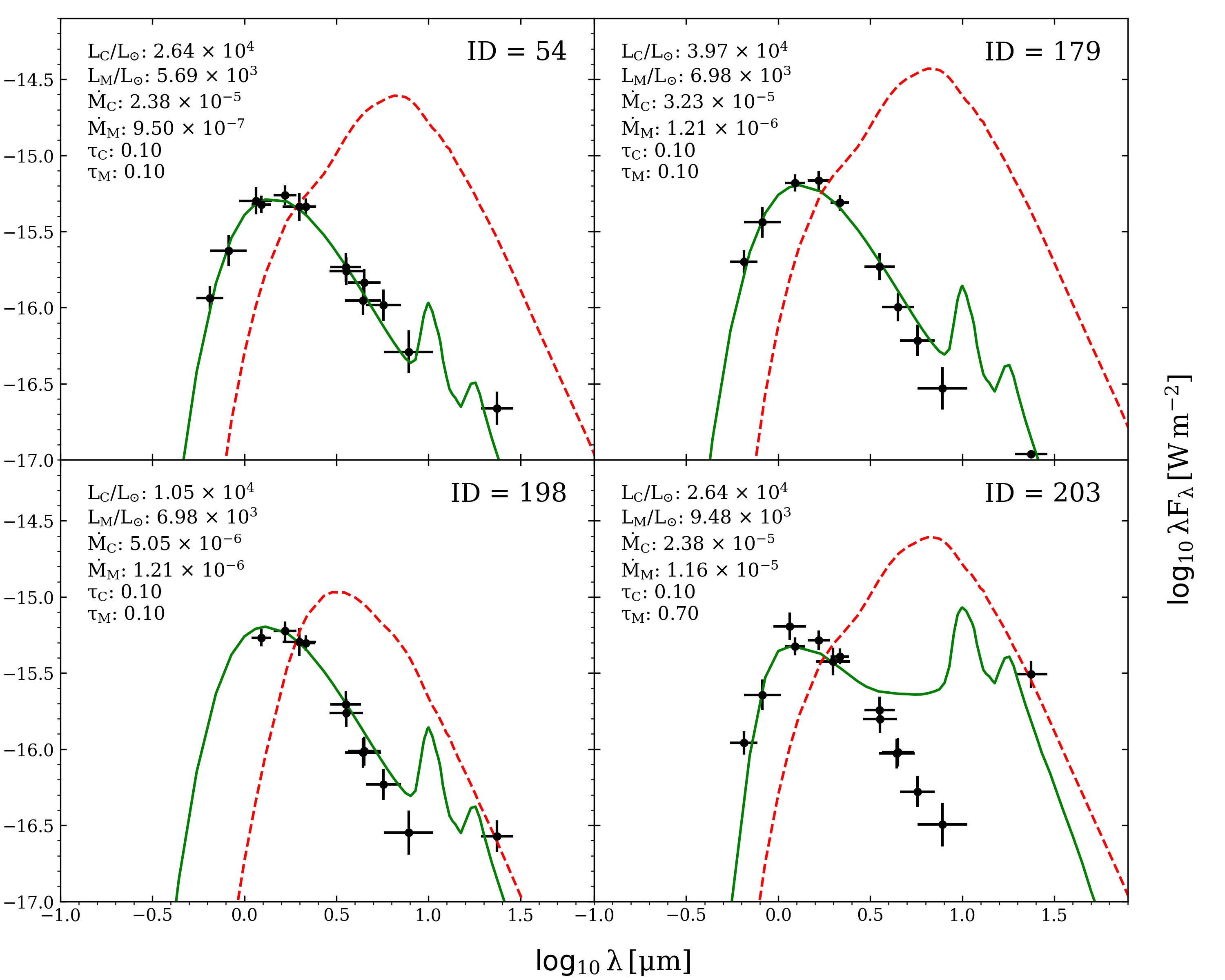}
\caption{Spectral energy distributions for the same stars shown in Fig.~\ref{fig:2x2}, fitted using the \texttt{DESK} code (For clarity, see Table~\ref{tab:selected_info}).}
\label{fig:DESK2x2}
\end{center}
\end{figure}

Additional examples of the fitted SEDs are shown in Fig.~\ref{fig:DESK2x2}, and the corresponding stellar and circumstellar parameters derived with both methods are summarized in Table~\ref{tab:selected_info}.

\begin{table*}
\begin{center}
\caption{Comparison of the stellar and circumstellar parameters derived with \texttt{SEDust} and \texttt{DESK} for the representative AGBs shown in Figures~\ref{fig:2x2} and \ref{fig:DESK2x2}. For each star, the best-fitting optical depth ($\tau$), luminosity ($L$), and mass-loss rate ($\dot{M}$) are listed for both carbon-rich and oxygen-rich models.}
\label{tab:selected_info}
\begin{tabular}{cccccccc}
\hline
\hline
ID &
Method &
$\tau_{\rm C}$ &
$\tau_{\rm M}$ &
$L_{\rm C}$ &
$L_{\rm M}$ &
$\dot{M}_{\rm C}$ &
$\dot{M}_{\rm M}$ \\
&
&
&
&
[$L_\odot$] &
[$L_\odot$] &
[$M_\odot\,{\rm yr}^{-1}$] &
[$M_\odot\,{\rm yr}^{-1}$] \\
\hline
54 &
SEDust &
0.05 $\pm$ 0.05 &
0.06 $\pm$ 0.05 &
(5.8 $\pm$ 0.2) $\times 10^{3}$ &
(6.1 $\pm$ 0.2) $\times 10^{3}$ &
(1.1 $\pm$ 0.3) $\times 10^{-6}$ &
(2.3 $\pm$ 0.9) $\times 10^{-6}$ \\
&
DESK &
0.1 $\pm$ 0.1 &
0.1 $\pm$ 0.1 &
(2.6 $\pm$ 0.1) $\times 10^{4}$ &
(5.7 $\pm$ 0.2) $\times 10^{3}$ &
(2.3 $\pm$ 0.7) $\times 10^{-5}$ &
(9.5 $\pm$ 3.8) $\times 10^{-7}$ \\
\hline
179 &
SEDust &
0.01 $^{0.05}_{0.01} $ &
0.01 $^{0.05}_{0.01} $ &
(6.7 $\pm$ 0.2) $\times 10^{3}$ &
(6.7 $\pm$ 0.2) $\times 10^{3}$ &
(1.2 $\pm$ 0.3) $\times 10^{-6}$ &
(2.3 $\pm$ 0.9) $\times 10^{-6}$ \\
&
DESK &
0.1 $\pm$ 0.1 &
0.1 $\pm$ 0.1 &
(4.0 $\pm$ 0.1) $\times 10^{4}$ &
(7.0 $\pm$ 0.3) $\times 10^{3}$ &
(3.2 $\pm$ 1.0) $\times 10^{-5}$ &
(1.2 $\pm$ 0.5) $\times 10^{-6}$ \\
\hline
198 &
SEDust &
0.03 $^{0.05}_{0.03} $ &
0.03 $^{0.05}_{0.03} $ &
(6.7 $\pm$ 0.2) $\times 10^{3}$ &
(6.1 $\pm$ 0.2) $\times 10^{3}$ &
(1.5 $\pm$ 0.3) $\times 10^{-6}$ &
(2.6 $\pm$ 0.9) $\times 10^{-6}$  \\
&
DESK &
0.1 $\pm$ 0.1 &
0.1 $\pm$ 0.1 &
(1.1 $\pm$ 0.1) $\times 10^{4}$ &
(7.0 $\pm$ 0.3) $\times 10^{3}$ &
(5.1 $\pm$ 1.5) $\times 10^{-6}$ &
(1.2 $\pm$ 0.5) $\times 10^{-6}$  \\
\hline
203 &
SEDust &
0.28 $\pm$ 0.05 &
0.20 $\pm$ 0.05 &
(8.8 $\pm$ 0.4) $\times 10^{3}$ &
(7.7 $\pm$ 0.3) $\times 10^{3}$ &
(3.9 $\pm$ 1.1) $\times 10^{-6}$ &
(2.4 $\pm$ 0.9) $\times 10^{-6}$ \\
&
DESK &
0.1 $\pm$ 0.1 &
0.7 $\pm$ 0.1 &
(2.6 $\pm$ 0.1) $\times 10^{4}$ &
(9.5 $\pm$ 0.4) $\times 10^{3}$ &
(2.4 $\pm$ 0.7) $\times 10^{-5}$ &
(1.2 $\pm$ 0.5) $\times 10^{-5}$  \\
\hline
\hline
\end{tabular}
\end{center}
\end{table*}

\section{Propagation of the BC$_K$ fit covariance}
\label{app:Propagation}

The bolometric correction is modeled as a quadratic function of color,
\begin{equation}
    BC_K(x) = A\,x^2 + B\,x + C, \qquad x \equiv J-K,
\end{equation}
with best-fit coefficients $(A,B,C)$ and covariance matrix $\Sigma$ obtained
from the nonlinear least-squares fit to the nine O-rich LPVs (Table~2).
The propagated uncertainty on the predicted $BC_K$ at a given color follows
from standard error propagation,
\begin{equation}
    \sigma^2_{BC_K}(x) = \mathbf{g}(x)^{\!\top} \, \Sigma \, \mathbf{g}(x),
    \qquad \mathbf{g}(x) \equiv \left(x^2,\; x,\; 1\right)^{\!\top}.
\end{equation}
Evaluated over the observed color range of the M-type LPVs,
$J-K \in [1.02, 1.28]$~mag, this gives
\begin{equation}
    \sigma_{BC_K} \simeq 0.035\text{--}0.069\ \mathrm{mag},
\end{equation}
substantially smaller than the $50$--$100\%$ fractional uncertainties on the
individual coefficients $A$, $B$, and $C$ owing to their strong covariance.

Since $\log_{10} L = \left[M_{\mathrm{bol},\odot} - BC_K - K + \mu\right]/2.5$,
the corresponding luminosity uncertainty is
\begin{equation}
    \sigma_{\log L} = 0.4\,\sigma_{BC_K} \simeq 0.014\text{--}0.028\ \mathrm{dex}
    \quad (\approx 0.02\ \mathrm{dex}).
\end{equation}
Because the LPV mass-loss rate scales as $\dot{M} \propto L^{3/4}$
(Eq.~2), the calibration-driven uncertainty in the
individual mass-loss rates is
\begin{equation}
    \sigma_{\log \dot{M}} = \tfrac{3}{4}\,\sigma_{\log L}
    \simeq 0.011\text{--}0.021\ \mathrm{dex}
    \quad (\approx 0.01\text{--}0.02\ \mathrm{dex}).
\end{equation}

To propagate this into the \emph{integrated} M-type LPV mass-loss rate, we
performed $10^5$ Monte Carlo realizations, drawing coefficient triplets
$(A,B,C)_k \sim \mathcal{N}\!\left((A,B,C),\,\Sigma\right)$ and applying each
draw \emph{coherently} to every M-type LPV, since the fit coefficients
represent a single, common calibration systematic rather than independent
per-star noise; treating them as independent and adding in quadrature would
artificially shrink the propagated uncertainty on the summed rate. For each
realization $k$ we compute
\begin{equation}
    \dot{M}_{i,k} = \dot{M}_{i,0}\,
    10^{-0.3\left[BC_{K,k}(x_i) - BC_{K,0}(x_i)\right]},
    \qquad
    \dot{M}_{\mathrm{tot},k} = \sum_i \dot{M}_{i,k}.
\end{equation}
The standard deviation of $\dot{M}_{\mathrm{tot},k}$ over the $10^5$
realizations gives the calibration uncertainty,
\begin{equation}
    \sigma_{\mathrm{calib}}(\dot{M}_{\mathrm{LPV,M}})
    \approx 7.5\times10^{-7}\ \mathrm{M_\odot\,yr^{-1}}
    \;\left(\approx 0.1\%\ \mathrm{of\ the\ quoted}\ 6.1\times10^{-4}\
    \mathrm{M_\odot\,yr^{-1}}\ \mathrm{uncertainty}\right).
\end{equation}
This confirms that the BC$_K$ calibration uncertainty is negligible relative
to the total uncertainty already reported for the LPV-based mass-loss
estimate in Table~3, and including it in quadrature does not change the
rounded uncertainties reported there.

\section{Quantifying Spatial Concentration}
\label{app:spatial_metrics}

This section describes the methodology used to quantify the spatial distribution and degree of concentration of stellar populations. The definitions and metrics introduced here are general and are not specific to NGC~6822.

For a given set of sources with equatorial coordinates $(\alpha, \delta)$, a spatial centroid $(\alpha_0, \delta_0)$ is defined as the arithmetic mean of the individual coordinates. The projected radial distance of each source from this centroid is then computed using the small-angle approximation,
\begin{equation}
r = D \sqrt{(\Delta \alpha \cos \delta_0)^2 + (\Delta \delta)^2},
\end{equation}
where $\Delta \alpha = \alpha - \alpha_0$, $\Delta \delta = \delta - \delta_0$, and $D$ is the distance to the system.

To characterize the radial structure of the distribution, sources are grouped into concentric annuli with inner and outer radii $r_{\mathrm{in}}$ and $r_{\mathrm{out}}$. The surface density in each annulus is defined as
\begin{equation}
\Sigma(r) = \frac{N(r)}{\pi \left( r_{\mathrm{out}}^2 - r_{\mathrm{in}}^2 \right)},
\end{equation}
where $N(r)$ is the number of sources within the annulus. A smooth, monotonically declining surface-density profile is characteristic of a centrally concentrated population, while pronounced deviations or multiple peaks may indicate spatial substructure or fragmentation.

%@@@@@@@@@@@@@@@@@@@@@@@@@@@@@@@@@@@@@@@@@@@@@@@@@@@@@@@@@@@@@@@@@@@@@@@@@@@@@@@@@@@@@@@@@@@@

\end{document}